\documentclass[12pt]{article}
\usepackage{amsfonts}
\usepackage[dvips]{color}
\usepackage{psfrag}
\usepackage{feynmp}
\usepackage[pdftex]{graphicx}
\usepackage{braket}
\usepackage{bm}
\usepackage{subfigure}

\usepackage{ulem}
\usepackage{comment}
\includecomment{pdffig}

\usepackage[height=8.5in,width=6.4in]{geometry} 
\usepackage{xparse}
\usepackage{xcolor}
\usepackage{tikz}
\usepackage{cite}
\usepackage[linktocpage]{hyperref}
\usepackage{amssymb, mathtools}
\usepackage[vcentermath]{youngtab}

\allowdisplaybreaks[1]

\Yboxdim{5pt}

\makeatletter
    
    \@addtoreset{equation}{section}
  \makeatother

\def\rem#1{}

\renewcommand{\title}[1]{\vbox{\center\LARGE{#1}}\vspace{5mm}}
\renewcommand{\author}[1]{\vbox{\center\large#1}\vspace{5mm}}

\numberwithin{equation}{section}

\DeclareFontShape{OT1}{cmr}{mx}{n}%
    {<->cmr10}{}

\renewcommand{\hat}{\widehat}
\renewcommand{\tilde}{\widetilde}

\renewcommand{\bar}{\overline}

\allowdisplaybreaks

\begin{document}
\bibliographystyle{utphys}

\begin{titlepage}
\begin{center}
\vspace{5mm}
\hfill {\tt 
}\\
\vspace{20mm}

\title{
\LARGE  
3d Chern--Simons matter theories from generalized Argyres--Douglas theories}
\vspace{7mm}

Takahiro Nishinaka$^{a,b,c}$ and 
Yutaka Yoshida
$^{d,e}$

\vspace{6mm}

\vspace{3mm}
$^a${\small {\it Department of Physics, Graduate School of Science
Osaka Metropolitan University, Osaka 558-8585, Japan}} \\
$^b${\small {\it Nambu Yoichiro Institute of Theoretical and Experimental Physics (NITEP)
Osaka Metropolitan University, Osaka 558-8585, Japan}} \\
$^c${\small {\it  Osaka Central Advanced Mathematical Institute (OCAMI)
Osaka Metropolitan University, Osaka 558-8585, Japan}} \\
$^d${\small {\it Department of Current Legal Study, Faculty of Law Meiji Gakuin University, 1-2-37 
Shirokanedai, Minato-ku, Tokyo 108-8636, Japan}} \\
$^e${\small {\it Institute for Mathematical Informatics, Meiji Gakuin University
1518 Kamikurata-cho, Totsuka-ku, Yokohama 244-8539, Japan}}

\end{center}

\vspace{7mm}
\abstract{
We study 3d $\mathcal{N}=2$ Chern--Simons matter theories describing the
 R-twisted $S^1$-reduction of Argyres--Douglas theories of
 $(A_{M-1},A_{N-1})$ type with $\text{gcd}(M,N)=1$, via a
 recently-proposed 4d/3d correspondence.
In particular, for the
 $(A_2,A_{N-1})$ and $(A_3,A_{N-1})$ theories, we identify
 a series of Chern--Simons matter theories with monopole superpotentials
 consistent with an $\mathcal{N}=4$ supersymmetry
 enhancement in the infrared.  As a by-product,  we also find
 a novel Nahm sum formula for the vacuum character of $(3,8)$ $W_3$
 minimal model, from which we find another Chern--Simons matter
 theory describing the R-twisted $S^1$-reduction of the $(A_2,A_4)$ theory.
}
\vfill

\end{titlepage}

\tableofcontents

\section{Introduction}

It has recently been pointed out that the $U(1)_r$ twisted
compactification of 4d $\mathcal{N}=2$ superconformal field
theories (SCFTs) of Argyres--Douglas (AD)
type \cite{Argyres:1995jj, Argyres:1995xn, Eguchi:1996vu,
Cecotti:2010fi, Xie:2012hs}
give rise to a series of 3d $\mathcal{N}=4$ SCFTs that are realized as the IR fixed points of 3d $\mathcal{N}=2$ Chern--Simons (CS) matter
theories 
\cite{Dedushenko:2023cvd, Dedushenko:2018bpp, ArabiArdehali:2024ysy, Gaiotto:2024ioj,
ArabiArdehali:2024vli}. In particular, the authors of
\cite{Gaiotto:2024ioj} have proposed a general strategy to identify 3d
theories corresponding to a given AD theory in four dimensions. A key
ingredient in their discussions is 
the IR formula for the
Schur index of the 4d AD theories \cite{Cordova:2015nma}, which is written in terms of the
spectrum of 4d BPS states at generic points on the Coulomb
branch. Since this BPS spectrum jumps when a wall of marginal stability
is crossed, there are generally many different 3d $\mathcal{N}=2$ CS
matter theories that are expected to flow to the same $\mathcal{N}=4$ fixed
point in the infrared. Therefore, the wall-crossing phenomena for 4d BPS
states lead to IR dualities for various 3d $\mathcal{N}=2$ CS matter
theories \cite{Gaiotto:2024ioj}. 
When an  AD theory has no flavor symmetry, the resulting 3d $\mathcal{N}=4$ theory at the
fixed point is expected to be rank-zero, in the sense that both its Higgs and Coulomb branches are zero-dimensional.
Supersymmetry enhancement of 3d $\mathcal{N}=2$ CS matter theories flowing
to rank-zero theories in the infrared was first studied in
\cite{Gang:2018huc} and further  developed in \cite{Gang:2021hrd, Gang:2023rei}.

From the viewpoint of the SCFT/VOA correspondence \cite{Beem:2013sza},
the above 3d theories provide a bridge between 4d $\mathcal{N}=2$ SCFTs
and 2d vertex operator algebras (VOAs) \cite{Dedushenko:2023cvd}. Indeed, by applying the general method of \cite{Costello:2018fnz},
one can construct a 2d VOA from a topologically twisted 3d
$\mathcal{N}=4$ theory on a half-space $\mathbb{R}_{\geq 0}\times
\mathbb{C}$. When the H-twist (or A-twist) is applied in the bulk and a
deformed $(0,4)$ boundary condition is imposed, the resulting boundary VOA for
the 3d $\mathcal{N}=4$ SCFTs discussed above is
expected to be identical to the VOA associated with the parent 4d AD
theories in the sense of
\cite{Beem:2013sza}.  Furthermore, one can make contact with these
boundary VOAs from the 3d $\mathcal{N}=2$ CS matter theories in the
ultraviolet. This particularly means that
one can reproduce the vacuum character of the 2d VOAs, or
equivalently the Schur index of the parent 4d $\mathcal{N}=2$ SCFTs, as the half index of the
3d $\mathcal{N}=2$ CS matter theories \cite{Costello:2018fnz,
Gang:2023rei, Gang:2023ggt, Ferrari:2023fez, Baek:2024tuo,
Creutzig:2024ljv, ArabiArdehali:2024ysy, Gaiotto:2024ioj, Gang:2024loa, Kim:2024dxu, Go:2025ixu, Kim:2025klh}.

The proposal of \cite{Gaiotto:2024ioj} has been applied to and tested in
various AD theories of $(A_1, G)$ type with $G=A_N, D_N$ and $E_N$
\cite{Gaiotto:2024ioj, Kim:2024dxu, Go:2025ixu, Kim:2025klh}, where $G$ is a Lie algebra of ADE type. Among other theories, its
application to $(A_1, A_{N-1})$ theories for odd $N$ leads to a series of 3d
$\mathcal{N}=2$ CS matter theories whose half index reproduces the
vacuum character of the $(2,N+2)$ Virasoro minimal models, which are
indeed VOAs associated in the sense of \cite{Beem:2013sza} with
the $(A_1,A_{N-1})$ theories for odd $N$. Given these results, it is
desirable to generalize them 
 to more general AD theories of
$(A_{M-1},A_{N-1})$ type.

In this paper, we apply the proposal of \cite{Gaiotto:2024ioj} to the $(A_{M-1},
A_{N-1})$ theories for coprime $M$ and $N$, for which the corresponding
VOAs in the sense of \cite{Beem:2013sza} are conjectured in \cite{Cordova:2015nma}  to be the
vacuum sector of the $(M,M+N)$ $W_M$
minimal model.
In particular, we identify a series of 3d $\mathcal{N}=2$ CS matter
theories describing the $U(1)_r$ twisted $S^1$-reduction of $(A_2, A_{N-1})$ and $(A_3, A_{N-1})$ theories with $N$
coprime to three and four, respectively. To be more specific, we identify the
gauge group, matter content, mixed CS levels, the charges of matter fields, and
the monopole superpotential that is expected to give rise to an IR
supersymmetry enhancement to $\mathcal{N}=4$. We then confirm that the
half index of these CS matter theories coincide with the Schur index of
the parent 4d theories, $(A_2,A_{N-1})$ and $(A_3,A_{N-1})$.
In addition, for $(A_{M-1},A_{N-1})$
with coprime $M\geq 5$ and $N$, we give a conjectural expression for the
half-monodromy and the Schur index, from which one can read off the
mixed CS levels, the
matter content and the charges of matter fields of the 3d
$\mathcal{N}=2$ CS matter theory.

In addition to the above results, we
also find a novel simple Nahm sum formula for the vacuum character of the
$(3,8)$ $W_3$ minimal model. This can be regarded as a natural generalization of a
similar formula for the $(3,7)$ $W_3$ minimal model discovered in
\cite{Feigin:2005gz}. We read off the 3d $\mathcal{N}=2$ CS matter
theory description for this Nahm sum, which is expected to flow to the
same $\mathcal{N}=4$ SCFT in the infrared as the CS matter theory
we identify for the $(A_2,A_4)$ theory via the method of \cite{Gaiotto:2024ioj}.

The organization of the rest of this paper is the following. In
Sec.~\ref{sec:general}, we describe the general strategy proposed in
\cite{Gaiotto:2024ioj} and study general properties of the half index
and the ellipsoid partition function obtained from the IR formula. In Sec.~\ref{sec:KS}, we
describe how to evaluate the quantum monodromy for the $(A_{M-1},
A_{N-1})$ theories. In Sec.~\ref{sec:A2}, we study $\mathcal{N}=2$ CS
matter theories for
 the $(A_2,A_{N-1})$ theories with $N$ coprime to three, where we  mainly exploit the
method of \cite{Gaiotto:2024ioj} while in Sec.~\ref{subsec:Nahm} we give
an independent analysis on the vacuum character of $(3,8)$ $W_3$ minimal
model. In
Sec.~\ref{sec:A3}, we study CS matter theories for the $(A_3,A_{N-1})$
theories with $N$ coprime to four. We also give in Sec.~\ref{subsec:conjecture} a conjecture on a
general $(A_{M-1}, A_{N-1})$ theory for coprime $M$ and $N$. In Sec.~\ref{sec:summary},
we conclude and discuss future directions. In appendix \ref{app:local},
we summarize localization formulas for the half index, superconformal
index and the ellipsoid partition function of 3d $\mathcal{N}=2$ abelian
CS matter theories. In appendix
\ref{app:derivationIR}, we give a derivation of expressions for the 4d Schur index/3d
ellipsoid partition function corresponding to a quantum monodromy of 4d
BPS states. In appendix
\ref{app:monopole}, we summarize some formulae for charges of monopole operators.

\bigskip

\noindent {\bf Note added:} When this paper was almost finished, the
paper \cite{Kim:2025rog} appeared on arXiv which has a partial overlap with
our discussions in this paper. For some of the $(A_2,A_{M-1})$ and
$(A_3,A_{M-1})$ theories, the authors of \cite{Kim:2025rog} identified
a different 3d CS matter theory from ours, which are expected to be IR
dual to each other as we will discuss in Sec.~\ref{sec:summary}.

 \section{General framework and strategy}
 \label{sec:general}

In
this section, we describe the general strategy proposed in
\cite{Gaiotto:2024ioj} and study general properties of the half index
and the ellipsoid partition function obtained from the IR formula. 

\subsection{Half index from IR formula for 4d Schur index}

Following \cite{Gadde:2011ik,Gadde:2011uv}, the Schur index of a 4d $\mathcal{N}=2$ superconformal field theory is defined by
\begin{equation}
 \mathcal{I}^\text{4d}(q) \equiv \text{Tr}(-1)^Fq^{E-R_{4d }}~,
\end{equation}
where the trace is taken over the space of local operators, and $E$ and
$R_{4d}$ are respectively the
scaling dimension and (the Cartan of) $SU(2)_R$ charge of those
operators.
For the convergence of the index, we assume that $|q|<1$.

The IR formula proposed in \cite{Cordova:2015nma}
is a conjectural formula for the above
Schur index, which is written in terms of the spectrum of BPS states in
a particular chamber on the Coulomb branch. To describe it, 
we first focus on 4d theories whose
Coulomb branch has a special
chamber in which only finitely many BPS states are stable.
We also assume that these stable BPS states are all
hypermultiplets.
Note that the $(A_{M-1}, A_{N-1})$
theories are theories of this type \cite{Alim:2011ae}. The IR formula  then implies
that the Schur index 
is written as \cite{Cordova:2015nma}
\begin{align}
 \mathcal{I}^\text{4d}(q) = (q)^{2r}_{\infty}\,\text{Tr}(S(q)\bar{S(q)})~,\qquad S(q) \equiv
 \prod_{\gamma}^{\curvearrowright}E_q(X_\gamma)~,
\label{eq:4Dindex}
\end{align}
where $(q)_\infty \equiv (q;q)_{\infty}=\prod_{k=1}^{\infty}(1-q^k)$, $r$ is the
dimension of the Coulomb branch\footnote{For the $(A_{M-1},A_{N-1})$
theory with coprime $M$ and $N$,
$r = (M-1)(N-1)/2$.}, $\gamma$ runs over electro-magnetic charges of the stable BPS
hypermultiplets in the chamber discussed above, and $X_\gamma$ is an element of the quantum torus algebra satisfying
\begin{align}
 X_{\gamma_1} X_{\gamma_2} = q^{\frac{\langle
 \gamma_1,\gamma_2\rangle}{2}}X_{\gamma_1+\gamma_2} = q^{\langle
 \gamma_1,\gamma_2\rangle}X_{\gamma_2}X_{\gamma_1}~,
\label{eq:torus}
\end{align}
with $\langle \gamma_1,\gamma_2\rangle$ being the Dirac's pairing of the
electric and magnetic charges.
The $q$-exponential $E_q(X)$ is defined by
\begin{align}
 E_q(X) \equiv \prod_{k=0}^\infty(1+q^{k+\frac{1}{2}}X)^{-1} =
 \sum_{n=0}^\infty \frac{\big(-q^{\frac{1}{2}}X\big)^n}{(q)_n}~,
\label{eq:factorE}
\end{align}
where $(q)_n \equiv (q;q)_{n}= \prod_{k=1}^n(1-q^k)$.
Note that $E_q(X_{\gamma_1})$ and $E_q(X_{\gamma_2})$ do not commute unless
$\langle \gamma_1,\gamma_2\rangle =0$, and therefore the order of
$E_q(X_\gamma)$ must be specified in \eqref{eq:4Dindex}; the order is taken according to the value of $\text{arg}
\, Z(\gamma)$ where $Z(\gamma)$ is the central charge of the 4d
$\mathcal{N}=2$ super-Poincar\'e algebra. The quantity $\bar{S(q)}$ is
obtained from $S(q)$ by replacing all the electro-magnetic charges
$\gamma$ to their charge-conjugates $-\gamma$.
The trace in \eqref{eq:4Dindex} is then defined by\footnote{When the theory has a flavor symmetry,
Eq.~\eqref{eq:trace} is slightly modified.}
\begin{equation}
 \text{Tr} X_\gamma = \delta_{\gamma,0}~.
\label{eq:trace}
\end{equation}
Note that, since one can reduce the number of $X_\gamma$ in each term of
$S(q)\bar{S(q)}$ by using \eqref{eq:torus}, Eq.~\eqref{eq:trace} is sufficient for us to
evaluate the IR formula \eqref{eq:4Dindex}.

As described in detail in Sec.~\ref{sec:generalCS}, by rewriting the IR formula for the Schur index so that it matches the half index of a 3d $\mathcal{N}=2$ CS matter theory, 
one can read off the gauge group, the effective CS levels, and the
matter content of the $S^1$-reduction of the 4d $\mathcal{N}=2$ theory.\footnote{To be more precise, this is a $U(1)_r$-twisted
$S^1$-compactification \cite{Dedushenko:2018bpp, Dedushenko:2023cvd,
ArabiArdehali:2024ysy, Gaiotto:2024ioj} and therefore the factor of
$e^{2\pi i r}$ is introduced in the definition of the 4d index, where
$r$ is the $U(1)_r$ charge. Since AD theories have Coulomb branch
operators of fractional $U(1)_r$ charges, this factor is non-trivial for
these theories.}
Moreover, 
one can  
read off an R-symmetry mixing with topological symmetries.
In order for this CS matter theory to genuinely flow to the twisted reduction of the original 4d $\mathcal{N}=2$ theory, 
one needs to turn on an appropriate monopole superpotential so that the supersymmetry
is enhanced to $\mathcal{N}=4$ in the infrared.
Such a supersymmetry enhancement in RG flows to 3d rank-zero theories was first discussed in \cite{Gang:2018huc}.

\subsection{Ellipsoid partition function from IR formula}
A remarkable observation of \cite{Gaiotto:2024ioj} is that, replacing the $q$-exponentials in the IR formula for the Schur index \eqref{eq:4Dindex} by Faddeev’s non-compact quantum dilogarithms, one obtains the ellipsoid partition function of a 3d $\mathcal{N}=2$ CS matter theory with the same gauge group, CS levels, and matter content:
\begin{align}
 \mathcal{S}^\text{3d}_{\sf b} =
 \text{Tr}({\sf s}_{\sf b}\,\bar{{\sf s}_{\sf b}})~,\qquad {\sf s}_{\sf b} \equiv
 \prod_{\gamma}^{\curvearrowright} \Phi_{\sf b}(x_\gamma)~,
\label{eq:S3part}
\end{align}
up to an overall constant. 
Here, $\Phi_{\sf b}$ is the Faddeev quantum
dilogarithm \cite{Faddeev:1995nb, Faddeev:2000if, Alexandrov:2015xir} (we follow the notation of \cite{Alexandrov:2015xir} here):
\begin{align}
 \Phi_{\sf b}(x) \equiv \exp\left(\frac{1}{4}\int_{\mathbb{R}+i0^+} \frac{e^{-2ixt}}{\sinh({\sf b}t)\sinh({\sf b}^{-1}t)}\frac{dt}{t}\right)=  \prod_{k=0}^\infty\frac{1+ e^{2\pi i
 {\sf b}^2(k+\frac{1}{2})}e^{2\pi {\sf b} x}}{1+e^{-2\pi i
 {\sf b}^{-2}(k+\frac{1}{2})}e^{2\pi {\sf b}^{-1}x}}~,
\label{eq:Faddeev}
\end{align}
and $x_\gamma$ are non-commutative variables such that
\begin{equation}
 [x_{\gamma_1},\,x_{\gamma_2}] = \frac{1}{2\pi i}\langle
  \gamma_1,\gamma_2\rangle~,
\label{eq:xcom}
\end{equation}
and $x_{\gamma_1+\gamma_2} = x_{\gamma_1}+x_{\gamma_2}$. 
It is known that \eqref{eq:Faddeev} can be Fourier-transformed as
\begin{equation}
 \Phi_{\sf b}(x)  =\int dp\, \hat{\Phi}_{\sf b}(p)e^{2\pi i p x}~,
\label{eq:Fourier}
\end{equation}
where 
\begin{equation}
 \hat{\Phi}_{\sf b}(p) = e^{-\pi ip^2}\Phi_{\sf b} \left(\frac{i}{2}({\sf b}+{\sf b}^{-1})-p\right)=e^{-\pi ip^2}
 \prod_{k=0}^\infty \frac{1- e^{2\pi i {\sf b}^2(k+1)}e^{-2\pi {\sf b}p}}{1-e^{-2\pi
 i {\sf b}^{-2}k}e^{-2\pi {\sf b}^{-1}p}}
\label{eq:hatPhi_b}
\end{equation}
up to a $p$-independent constant prefactor. 
Up to an over all constant, the trace in \eqref{eq:S3part} is defined so that
\begin{equation}
 \text{Tr}\, e^{2\pi i \sum_{i}p_i x_{\gamma_i}} = \prod_{i}\delta(p_i)~,
 \label{eq:deltafunc}
\end{equation}
where $\{\gamma_i\}$ is a basis of the electro-magnetic charge
lattice.\footnote{Again, we assume here that the 4d theory has no
flavor symmetry.}

Note that the factor
$\Phi_{\sf b}(\frac{i({\sf b}+{\sf b}^{-1})}{2}-p)$ in \eqref{eq:hatPhi_b} is almost identical to the contribution from a
chiral multiplet to the
localization formula for the ellipsoid partition function, where the
chiral multiplet is assigned with R-charge zero and is coupled to a $U(1)$
vector multiplet whose real scalar is $p$. This implies that
\eqref{eq:S3part} is identified as the ellipsoid partition function of
an $\mathcal{N}=2$ CS matter theory, where $e^{-\pi i
p^2}$ in \eqref{eq:hatPhi_b} is a part of  the  CS level of a
$U(1)$ gauge group corresponding to $p$. 
The precise identification given in the next sub-section 
of \eqref{eq:S3part} with the ellipsoid partition function leads to the 
the same CS matter theory obtained by the Schur index.   
Specifically, we will see that one can read off from the half-index and the ellipsoid partition function that the same linear combination of the reference R-charge and topological $U(1)$ charges is allowed as a symmetry.

\subsection{General formula for 3d CS matter theory from IR formulas}
\label{sec:generalCS}

We now study general properties of the half index and the ellipsoid partition function obtained 
by the IR formulas \eqref{eq:4Dindex} and \eqref{eq:S3part} without flavor symmetry.
To begin with, without restricting ourselves to any particular theory, we consider the following general expression:
\begin{align}
 S(q) &= \prod_{\ell=1}^{L^{\prime}} E_{q}(X_{\sum_{k=1}^{2r} P_{k, \ell } \gamma_k})=
 E_{q}(X_{\sum_{k=1}^{2r} P_{k, 1 } \gamma_k}) \cdots E_{q}(X_{\sum_{k=1}^{2r} P_{k, L^{\prime} } \gamma_k}), 
 \label{eq:halfmd}\\
  {\sf s}_{\sf b} &=\prod_{\ell=1}^{L^{\prime}} \Phi_{\sf b}(x_{\sum_{k=1}^{2r } P_{ k, \ell} \gamma_k})
  =\Phi_{\sf b}(x_{\sum_{k=1}^{2r }P_{ k, 1} \gamma_k}) \cdots \Phi_{\sf b}(x_{\sum_{k=1}^{2r }P_{ k, L^{\prime}} \gamma_k})\,.
  \label{eq:smalls}
 \end{align}
 where $r$ is the dimensions of the Coulomb branch.  $ \gamma_{k}$ for $k=1,\cdots, 2r$ is a basis of the electro-magnetic charges such that
 the matrix defined by the Dirac products $\langle \gamma_i, \gamma_j \rangle$ for $i,j=1,\cdots, 2r$ is non-degenerate. 
  For example, the basis   for $(A_{M-1}, A_{N-1})$ theory is specified as $\gamma^{(i)}_j$ for $i=1,\cdots, M-1$ and 
 $j=1, \cdots, N-1 $ in Figure~\ref{fig:quiver7}. 
 We also assume the rank of $2r \times L^{\prime}$ matrix $P_{ k, l}$ is $2r$.

The Schur index and the $S^{3d}_{\sf b}$ partition function are expressed as
\begin{align}
  \mathcal{I}^\text{4d}  
&
=(q)^{L-N}_{\infty} 
\sum_{n_1, \cdots, n_N \in \mathbb{Z} \atop (n \cdot Q)_1 \ge 0, \cdots, (n \cdot Q)_{L} \ge 0}
\!\!
 \frac{q^{\frac{1}{2}\sum_{a,b=1}^{N} K_{a b} n_{a} n_{b}}(-q^{\frac{1}{2}})^{
 \sum_{a=1}^N \sum_{i=1}^L Q_{a,i} n_{a}  }}
      {\prod_{i=1}^{L}(q)_{\sum_{a=1}^N n_a Q_{a,i}}}
 \,,
 \label{eq:Schur4}
\\
\mathcal{S}^{\rm 3d}_{\sf b}&= 
\int  \prod_{a=1}^{N} d \sigma_a  
 e^{\pi i \sum_{a, b=1}^{N} K_{a b} \sigma_{a} \sigma_{b} } 
 \prod_{i=1}^{L} \widehat{\Phi}_{\sf b} \Bigl(\sum_{a=1}^{N} Q_{a,i} \sigma_{a}\Bigr) \,,
 \label{eq:S3b4}
\end{align}
where $N$ and $L$ are defined by
\begin{align}
L \equiv 2 L^{\prime}\,, \quad N \equiv 2 L^{\prime} -2r\,.
\end{align}
A detailed derivation of the above expressions is presented in appendix~\ref{app:derivationIR}.
Then 
 the IR formulas for Schur index \eqref{eq:Schur4} and $\mathcal{S}^{\rm 3d}_{\sf b}$ \eqref{eq:S3b4} are identified with
 the half index  and the ellipsoid partition function, respectively. It is easy to rewrite the Schur index and  $\mathcal{S}^{\rm 3d}_{\sf b}$  as 
\begin{align}
  \mathcal{I}^\text{4d}  
&=\frac{1}{(q)^{N}_{\infty}} \sum_{n_1, \cdots, n_N \in \mathbb{Z}}
 q^{\frac{1}{2}\sum_{a,b=1}^{N} K_{a b} n_{a} n_{b}} (-q^{\frac{1}{2}})^{- 
 \sum_{a=1}^N \sum_{i=1}^L Q_{a, i} n_{a}  }
 \prod_{i=1}^{L} (q^{1-\sum_{a=1}^N n_a Q_{a, i}} )_{\infty}\,, \\
\mathcal{S}^{\rm 3d}_{\sf b} &= \int \prod_{a=1}^N  d\sigma_{a} 
\exp\Bigl( { \pi i \sum_{a,b=1}^N (K_{a b} -\frac{1}{2} \sum_{i=1}^L Q_{a, i} Q_{b, i}) \sigma_a \sigma_b 
+\frac{\pi}{2}({\sf b}+{\sf b}^{-1}) \sum_{a=1}^N \sum_{i=1}^L Q_{a,i}} \sigma_a \Bigr)
\nonumber \\
& \qquad \times \prod_{i=1}^L
 s_{\sf b} \Bigl( \frac{i}{2}({\sf b}+{\sf b}^{-1})- \sum_a Q_{a, i} \sigma_a \Bigr)\,,
 \label{eq:3dsquashed3}
\end{align}
where $(a)_{\infty} \equiv (a;q)_{\infty}=\prod_{k=0}^{\infty} (1-a q^{k})$, and  $s_{\sf b}$ is the double sine function given by \eqref{eq:doublesine}.
By comparing these expressions with the localization formulas  \eqref{eq:halfindex} and  \eqref{eq:3dsquashed}, we find that the Schur index and $\mathcal{S}^{\rm 3d}_{\sf b}$ agree with the half index and the ellipsoid partition function of a 3d $\mathcal{N}=2$ CS matter theory with the $U(1)^N$ gauge group  coupled to $L$ chiral multiplets of vanishing R-charge, with gauge charges $Q_{a, i}$ for $a=1,\ldots,N$ and $i=1,\ldots,L$. 
The effective gauge  CS level is  $K_{ab}$.

In particular, if the fugacities $x_a$ for the topological $U(1)$ symmetries 
and the Fayet–Iliopoulos (FI) parameters $\xi_a$ are turned off, i.e.\ 
$x_a=1$ and $\xi_a=0$, we find that the gauge–R-symmetry mixed CS level is
\begin{align}
  K_{a R} = -\sum_{i=1}^{L} Q_{a,i}\,.
\end{align}
Equivalently, this can be viewed as turning on background couplings only for a specific linear combination of the topological $U(1)$ symmetries. 
After redefining the R-current so that the mixed CS levels vanish, $K_{aR}=0$, this is implemented as follows.
For the half index, the fugacities $x_a$ for the topological $U(1)$ symmetries are chosen as
\footnote{The appearance of $-q^{\frac{1}{2}}$ instead of $q^{\frac{1}{2}}$ is due to the fact that,
for the integer R-charges, we use 
$\mathrm{Tr}(-1)^{R} q^{J_3+\frac{R}{2}}$ instead of 
$\mathrm{Tr}(-1)^{F} q^{J_3+\frac{R}{2}}$; see footnote~12 of \cite{Dimofte:2017tpi}. 
In our case, all chiral multiplets have R-charge zero, and therefore the integrality condition for R-charges is satisfied.}
\begin{align}
x_{a}=(-q^{\tfrac{1}{2}})^{-\sum_{i=1}^L Q_{a,i}}\,,\qquad K_{a R}=0\,,
\end{align}
while for the ellipsoid partition function the FI parameters are fixed to
\begin{align}
\xi_a =\frac{i}{4}({\sf b}+{\sf b}^{-1})\sum_{i=1}^L Q_{a,i}\,,\qquad K_{a R}=0\,,
\label{eq:FItune}
\end{align}
for $a=1,\cdots, N$.
In the following, we will adopt the convention that $K_{aR}=0$, and correspondingly shift the classical R-charge $R$ as follows:
\begin{align}
R \to R_{\rm shift}\equiv R - \sum_{a=1}^N \sum_{i=1}^L Q_{a, i} J^{(a)}\,.
\label{eq:Rshift}
\end{align}
Here $J^{(a)}$ is the generator of the topological symmetry for the $a$-th $U(1)$ gauge group.
For example,  the combination $\sum_{a,i} Q_{a,i} J^{(a)}$ that enters the R-charge shift for $(A_{M-1}, A_{N-1})$ theories ($M=2,3,4$) is given by
\begin{align}
\sum_{a, i} Q_{a, i}J^{(a)} =
\begin{cases}
2\sum_{i=1}^{N-1} J_{y_i}  & \text{for } (A_1, A_{N-1}) \,,\\
\sum_{i=1}^{N-1}\Bigl( 2(J_{x^{(1)}_i} + J_{x^{(2)}_i}) -  (J_{y_i}+J_{\tilde{y}_i}) \Bigr)& \text{for } (A_2, A_{N-1}) \,,\\[2pt]
\displaystyle
\sum_{i=1}^{N-1} \Bigl( 2\sum_{l=1}^3 J_{x^{(l)}_{i}}+
\sum_{l=1,3}
(J_{y^{(l)}_{i}}+{J}_{\tilde{y}^{(l)}_{i}})
-2(J_{y^{(2)}_{i}}+{J}_{\tilde{y}^{(2)}_{i}})
\Bigr) & \text{for } (A_3, A_{N-1})\,.
\end{cases}
\end{align}
See \eqref{eq:yyyyshift}, \eqref{eq:RIR2}, and \eqref{eq:Rshifta3an} for the detailed definitions.

In the next sub-section, we will review how the above strategy works for the twisted 3d reduction of $(A_1,A_{2N})$ theories, 
focusing in particular on the case $(A_1,A_2)$.

\subsection{Example: review of $(A_1,A_{2N})$ case}
\label{sec:A1A2}

In this section, we review the CS matter theories identified in
\cite{Gaiotto:2024ioj} as twisted $S^1$-compactifications of the $(A_1,A_{2N})$
theories, by applying the procedure  above.

\subsubsection{$(A_1,A_2)$ theory}

Let us start with the $(A_1,A_2)$
theory. For $(A_1,A_2)$, the IR formula \eqref{eq:4Dindex} is written as
\begin{equation}
 \mathcal{I}^\text{4d} = \text{Tr}\Big(E_q(X_{\gamma_1})E_q(X_{\gamma_2})E_q(X_{-\gamma_1})E_q(X_{-\gamma_2})\Big)~,
\end{equation}
with $\langle\gamma_1, \gamma_2 \rangle = 1$. This can be evaluated as
\cite[Eq.~(4.10)]{Cordova:2015nma}
\begin{equation}
 \mathcal{I}^\text{4d}(q) = (q;q)_\infty^2 \sum_{n_1,n_2=0}^\infty
 \frac{q^{n_1n_2+n_1+n_2}}{(q;q)_{n_1}^2(q;q)_{n_2}^2}~.
\label{eq:Schur-A1A2}
\end{equation}
It was found in \cite{Cordova:2015nma} that this expression coincides with the
(normalized) vacuum character of the $(2,5)$ Virasoro minimal model.

According to the conjecture of \cite{Gaiotto:2024ioj}, the ellipsoid
 partition function of the 
 $U(1)_r$-twisted compactification of the $(A_1,A_2)$ theory can be
 expressed as
\begin{equation}
 \mathcal{S}_{\sf b}^\text{3d} = \text{Tr}\Big(
 \Phi_{\sf b}(x_{\gamma_1})\Phi_{\sf b}(x_{\gamma_2}) \Phi_{\sf b}(x_{-\gamma_1})\Phi_{\sf b}(x_{-\gamma_2}) \Big)~.
\end{equation}
Using the commutation relation $[x_{\gamma_1},x_{\gamma_2}] = \frac{1}{2\pi i}\langle
 \gamma_1,\gamma_2\rangle = \frac{1}{2\pi i}$ and \eqref{eq:Fourier}, we find
\begin{equation}
 \mathcal{S}_{\sf b}^\text{3d} =\int dp_1dp_2
 \left(\hat{\Phi}_{\sf b}(p_1)\right)^2\left(\hat{\Phi}_{\sf b}(p_2)\right)^2e^{2\pi
 i p_1p_2}~,
\label{eq:ZS3-A1A2}
\end{equation}
up to an constant prefactor.

Note that \eqref{eq:ZS3-A1A2} is equivalent to the localization formula for
the ellipsoid partition function of 3d $\mathcal{N}=2$ $U(1)^2$
Chern-Simons gauge theory coupled to four chiral multiplets. In
particular, each $\hat{\Phi}_{\sf b}(p_i)$ corresponds to a chiral multiplet
coupled to a $U(1)$ vector multiplet whose real scalar is $p_i$. The
last factor $e^{2\pi i p_1p_2}$ stands for a non-trivial mixed
Chern--Simons level between the two $U(1)$ gauge groups. Indeed, this
$U(1)^2$ Chern--Simons matter theory with four chiral multiplets is
characterized by the following two matrices:
\begin{align}
 K = \left(
\begin{array}{cc}
0 & 1\\
1 & 0\\
\end{array}
\right)~,\qquad Q = \left(
\begin{array}{cccc}
1 & 1 & 0 & 0 \\
0 & 0 & 1 & 1 \\
\end{array}
\right)~,
\label{eq:KI-A1A2}
\end{align}
where $K$ represents the effective CS levels, and $Q$ represents the
gauge charges of the four chiral multiplets.
This CS matter theory is expected to flow in the
infrared to an $\mathcal{N}=4$ fixed point obtained by the twisted
$S^1$-compactification of the
$(A_1,A_2)$ theory.

Having identified the above CS matter theory, one can recover
the Schur index of the original 4d $(A_1,A_2)$ theory by considering the
half index. 
Indeed, the half index of the above $U(1)^2$ CS matter theory
under the $(\mathcal{D},D_c)$ boundary condition is evaluated as
\begin{align}
 I\!\!I_{(\mathcal{D}, D_c)} = \sum_{n_1,n_2=0}^\infty
 q^{n_1n_2}(y_1)^{n_1}(y_2)^{n_2}\frac{(q^{1-n_1};q)_\infty^2(q^{1-n_2};q)_\infty^2}{(q;q)_\infty^2}~,
\end{align}
where $n_1$ and $n_2$ are monopole charges for the two $U(1)$ gauge
groups, and $y_1$ and $y_2$ are fugacities for the topological $U(1)^2$
symmetry.
One can rewrite the above expression as
\begin{align}
 I\!\!I_{(\mathcal{D}, D_c)} = (q;q)_\infty^2\sum_{n_1,n_2=0}^\infty \frac{q^{n_1n_2}(y_1)^{-n_1}(y_2)^{-n_2}}{(q;q)_{n_1}^2(q;q)_{n_2}^2}~.
\end{align}
We see that, by setting
\begin{align}
 y_1 = y_2 = q^{-1}~,
\label{eq:substitution}
\end{align}
the above half index coincides with the following expression for the
Schur index \eqref{eq:Schur-A1A2} of the $(A_1,A_2)$ theory. 

Here, the constraint \eqref{eq:substitution} suggests that the
topological $U(1)^2$ symmetry and the R-symmetry is broken by a superpotential including
monopole operators. Indeed, without turning on such a superpotential,
the 3d theory has a too large global symmetry to be
identified as the 3d reduction of $(A_1,A_2)$. As identified in
\cite{Gaiotto:2024ioj}, the correct monopole superpotential is the sum
of the following three gauge-invariant (dressed) monopole operators:\footnote{Note that, there is also a gauge invariant monopole operator $\tilde
\varphi_2 V_{(-1,0)}$ in addition to \eqref{eq:superpot_A1A2}. However,
including it breaks the $U(1)_A$ symmetry that we will discuss below,
and therefore we do not include it here.}
\begin{equation}
 \varphi_1 V_{(0,-1)}~,\qquad \tilde \varphi_1 V_{(0,-1)}~,\qquad
  \varphi_2 V_{(-1,0)}~.
\label{eq:superpot_A1A2}
\end{equation}
Here, $\varphi_i$ and $\tilde \varphi_i$ are two chiral multiplets which
have charge one under the $i$-th $U(1)$ gauge group and charge zero
under the other $U(1)$, and $V_{(n_1,n_2)}$ are (bare) monopole operators with
monopole charge $(n_1,n_2)$.

One can check that \eqref{eq:substitution} is consistent with the above
monopole superpotential. 
As explained in sub-section~\ref{sec:generalCS},
we see that  the replacement \eqref{eq:substitution} is equivalent to
redefining the R-charge as
\begin{equation}
 R \to R - 2\left(J_{y_1} + J_{y_2} \right)~,
\end{equation}
where $J_{y_i}$ is the topological charge corresponding to the fugacity
$y_i$.
Since all the chiral multiplets have zero R-charge before this
shift, so do the monopole operators. Therefore the monopole
superpotential \eqref{eq:superpot_A1A2} breaks the original R-charge.
However, after the above shift of R-charge,
all the monopole operators listed in \eqref{eq:superpot_A1A2} have
R-charge two. This means that, while the original R-charge $R$ and
the topological charges $J_{x_i}$ are both broken by the
monopole superpotential \eqref{eq:superpot_A1A2}, the linear
combination 
\begin{equation}
R_\text{shift} = R-2(J_{y_1} + J_{y_2})
\label{eq:shifta1a2}
\end{equation} 
is preserved, which is then
identified as a new R-charge of the theory deformed by the
superpotential.  
The condition \eqref{eq:substitution} precisely replaces the original R-charge
used in the definition of the half index with the one preserved by the
superpotential \eqref{eq:superpot_A1A2}.

We also see that adding the the monopole superpotential 
\eqref{eq:superpot_A1A2} leads to the correct global symmetry for the twisted 3d reduction of the
$(A_1,A_2)$ theory. Indeed, before including the monopole
superpotential, the 3d theory has $U(1)^2$
topological symmetry as well as $U(1)^2$ flavor symmetry rotating the
phase of the
chiral multiplets. In total, the 3d theory has $U(1)^4$ $\mathcal{N}=2$
flavor symmetry
before the superpotential deformation. The
superpotential deformation by \eqref{eq:superpot_A1A2} breaks this $U(1)^4$ symmetry to
a single $U(1)$ symmetry, which we denote by
$U(1)_A$.\footnote{Specifically, this $U(1)_A$ rotates the phase of $\tilde
\varphi_2$.} Therefore, the 3d theory has global
$U(1)_A\times U(1)_{R_\text{shift}}$ symmetry, which is expected to be enhanced to the 3d
$\mathcal{N}=4$ R-symmetry, $SO(4)_R$, in the
infrared.\footnote{The superconformal R-charge of the IR fixed point is
a non-trivial linear combination of the $U(1)_R$ charge and $U(1)_A$ charge.} Unless there is
an accidental global symmetry, the IR fixed point has no
continuous global symmetry commuting with the 3d $\mathcal{N}=4$
superconformal symmetry, which is the correct global symmetry for the 3d
reduction of the $(A_1,A_2)$ theory. This is in the same spirit as
\cite{Gang:2018huc}, and indeed it was shown in \cite{Gaiotto:2024ioj}
that the above CS matter theory is dual to
a theory discovered in \cite{Gang:2018huc}.

\subsubsection{$(A_1,A_{2N})$ theories}

It is straightforward to generalize the above discussion to the
$(A_1,A_{2N})$ theories. For $(A_1,A_{2N})$, the IR formula
\eqref{eq:4Dindex} is given by
\begin{align}
 \mathcal{I}^\text{4d}(q) = (q;q)_\infty^{2N}\text{Tr}\Big(
 \prod_{i:\text{odd}}E_q(X_{\gamma_i})\prod_{j:\text{even}}E_q(X_{\gamma_j})\prod_{k:\text{odd}}E_q(X_{-\gamma_k})\prod_{\ell:\text{even}}E_q(X_{-\gamma_\ell})\Big)~,
\label{eq:index_A1A2n}
\end{align}
where the Dirac's pairing is 
\begin{align}
 \langle \gamma_i, \gamma_j\rangle  = (-1)^{i+1}\delta_{i+1,j} - (-1)^{i}\delta_{i,j+1}~.
\end{align}
Note that $E_q(X_{\gamma_i})$ and
$E_q(X_{\gamma_j})$ commute with each other if $i$ and $j$ are both even
or both odd. The formula \eqref{eq:index_A1A2n} can be evaluated as 
\begin{align}
 \mathcal{I}^\text{4d} =
 (q;q)_\infty^{2N}\sum_{n_1,\cdots,n_{2N}=0}^\infty
 \frac{q^{\sum_{i=1}^{2N-1}n_in_{i+1} +
 \sum_{i=1}^{2N}n_i}}{\prod_{i=1}^{2N}\left[(q;q)_{n_i}\right]^2}~,
\label{eq:Schur_A1A2n}
\end{align}
which coincides with the vacuum character of the $(2,2N+3)$ Virasoro
minimal model \cite{Cordova:2015nma}.

By replacing $E_q(X_\gamma)$ in \eqref{eq:index_A1A2n} with
$\Phi_{\sf b}(x_\gamma)$, one can identify the ellipsoid partition function \eqref{eq:S3part} of
the twisted $S^1$-compactification of the $(A_1,A_{2N})$ theory. The
result is expressed as 
\begin{align}
 \mathcal{S}^\text{3d}_{\sf b} = \int \prod_{i=1}^{2N} \left(
 dp_i\Big(\hat{\Phi}_{\sf b}(p_i)\Big)^2\right)e^{2\pi i \sum_{k=1}^{2N-1}p_kp_{k+1}}~,
\end{align}
up to a constant prefactor.
This can be identified as the ellipsoid partition function of
3d $\mathcal{N}=2$ $U(1)^{2N}$ Chern--Simons
theory coupled to $4N$ chiral multiplets
\cite[Eq.~(5.12)]{Gaiotto:2024ioj}. Its (effective) mixed Chern--Simons levels
and the gauge charges of the chiral multiplets are respectively
encoded in the following $K$ and $Q$ matrices:
\begin{align}
 K_{ij} = \delta_{i,j+1} + \delta_{i,j-1}~,\qquad Q = {\bf 1}_{2N}
 \otimes (1,1) = \left(
\begin{array}{cccccccc}
1 & 1 & &&&& \\
& & 1 & 1 &  & &&\\
&&&& \ddots & \ddots &&\\
&&&&&& 1 & 1\\
\end{array}
\right)~.
\end{align}

The half index of the above CS-matter theory with the $(\mathcal{D},D_c)$ boundary
condition imposed is
written as
\begin{equation}
 I\!\!I_{(\mathcal{D}, D_c)} 
 =  (q;q)_\infty^{2N}\sum_{n_1,\cdots,n_{2N}=0}^\infty \left(\prod_{i=1}^{2N}(y_i)^{-n_i}\right)\frac{q^{\sum_{i=1}^{2N-1}n_in_{i+1}}}{\prod_{i=1}^{2N}\left[(q;q)_{n_i}\right]^2}~,
\end{equation}
where $y_1,\cdots, y_{2N}$ are fugacities for the topological $U(1)^{2N}$ symmetry. 
By setting
\begin{align}
y_1 = y_2 = \cdots = y_{2N} = q^{-1}~, 
\label{eq:yyyy}
\end{align}
the above half index is identical to the Schur index \eqref{eq:Schur_A1A2n} of the
$(A_1,A_{2N})$ theory. Eqs. \eqref{eq:yyyy} means the R-charge is shifted as 
\begin{align}
R \to R_{\rm shift}\equiv R- 2\sum_{i=1}^{2N} J_{y_i}~, 
\label{eq:yyyyshift}
\end{align}
where $J_{y_i}$ is the topological charge for $i$-the $U(1)$ gauge group.

The condition \eqref{eq:yyyy} implies that the $U(1)^{2N}$ topological
symmetry is broken by some monopole superpotential. One can identify
the correct superpotential as \cite[Eq.~(5.14)]{Gaiotto:2024ioj}
\begin{equation}
 W = \sum_{i=0}^{2N-1} \varphi_i V_{i+1}\varphi_{i+2} +
  \sum_{i=0}^{2N-2}\tilde \varphi_{i}V_{i+1}\tilde \varphi_{i+2}~,
\end{equation}
where $V_i$ is the monopole operator that has magnetic flux $-1$ for the
$i$-th $U(1)$ gauge group and vanishing magnetic flux for all the other
$U(1)$, the fields $\varphi_i$ and $\tilde \varphi_i$ are chiral multiplets that have
charge $+1$ under the $i$-th $U(1)$ group and are neutral under all the
other $U(1)$, and we set $\varphi_0 = \tilde \varphi_0 =
\varphi_{2N+1}=1$.

\section{Quantum monodromy of $(A_{M-1},A_{N-1})$}
\label{sec:KS}

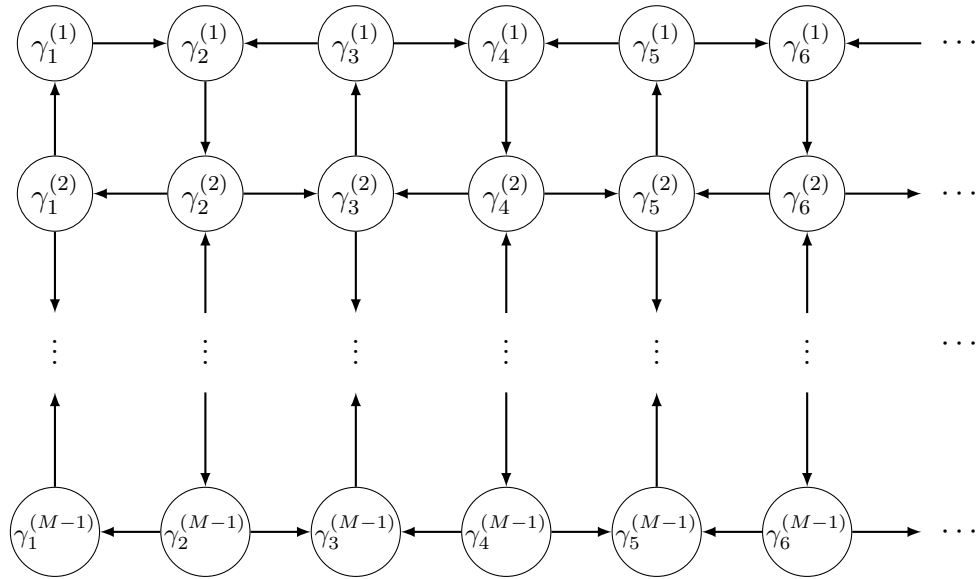
\begin{figure}[thb]
\centering
 \begin{tikzpicture}[vec/.style={thick,-latex,draw=black},gauge/.style={circle,draw=black,inner sep=0pt,minimum size=10mm},flavor/.style={rectangle,draw=black,inner sep=0pt,minimum size=10mm},auto]
  \node[gauge] (1) at (0,0) {$\gamma_{1}^{(1)}$};
  \node[gauge] (2) at (2,0) {$\gamma_{2}^{(1)}$};
  \node[gauge] (3) at (4,0) {$\gamma_{3}^{(1)}$};
  \node[gauge] (4) at (6,0) {$\gamma_{4}^{(1)}$};
  \node[gauge] (5) at (8,0) {$\gamma_{5}^{(1)}$};
  \node[gauge] (6) at (10,0) {$\gamma_{6}^{(1)}$};
  \node (7) at (12,0) {\;$\cdots$};

  \node[gauge] (a) at (0,-2) {$\gamma_{1}^{(2)}$};
  \node[gauge] (b) at (2,-2) {$\gamma_{2}^{(2)}$};
  \node[gauge] (c) at (4,-2) {$\gamma_{3}^{(2)}$};
  \node[gauge] (d) at (6,-2) {$\gamma_{4}^{(2)}$};
  \node[gauge] (e) at (8,-2) {$\gamma_{5}^{(2)}$};
  \node[gauge] (f) at (10,-2) {$\gamma_{6}^{(2)}$};
  \node (g) at (12,-2) {\;$\cdots$};

  \node (A) at (0,-4) {$\vdots$};
  \node (B) at (2,-4) {$\vdots$};
  \node (C) at (4,-4) {$\vdots$};
  \node (D) at (6,-4) {$\vdots$};
  \node (E) at (8,-4) {$\vdots$};
  \node (F) at (10,-4) {$\vdots$};
  \node (G) at (12,-4) {\;$\cdots$};

  \node (A2) at (0,-4.5) {};
  \node (B2) at (2,-4.5) {};
  \node (C2) at (4,-4.5) {};
  \node (D2) at (6,-4.5) {};
  \node (E2) at (8,-4.5) {};
  \node (F2) at (10,-4.5) {};
  \node (G2) at (12,-4.5) {};

  \node[gauge] (AA) at (0,-6.5) {\footnotesize $\gamma_{1}^{(M-1)}$};
  \node[gauge] (BB) at (2,-6.5) {\footnotesize $\gamma_{2}^{(M-1)}$};
  \node[gauge] (CC) at (4,-6.5) {\footnotesize $\gamma_{3}^{(M-1)}$};
  \node[gauge] (DD) at (6,-6.5) {\footnotesize $\gamma_{4}^{(M-1)}$};
  \node[gauge] (EE) at (8,-6.5) {\footnotesize $\gamma_{5}^{(M-1)}$};
  \node[gauge] (FF) at (10,-6.5) {\footnotesize $\gamma_{6}^{(M-1)}$};
  \node (GG) at (12,-6.5) {\;$\cdots$};

  \draw[vec] (1) to (2);
  \draw[vec] (3) to (4);
  \draw[vec] (5) to (6);
  \draw[vec] (3) to (2);
  \draw[vec] (5) to (4);
  \draw[vec] (7) to (6);

  \draw[vec] (b) to (a);
  \draw[vec] (d) to (c);
  \draw[vec] (f) to (e);
  \draw[vec] (b) to (c);
  \draw[vec] (d) to (e);
  \draw[vec] (f) to (g);

  \draw[vec] (BB) to (AA);
  \draw[vec] (BB) to (CC);
  \draw[vec] (DD) to (CC);
  \draw[vec] (DD) to (EE);
  \draw[vec] (FF) to (EE);
  \draw[vec] (FF) to (GG);

  \draw[vec] (a) to (1);
  \draw[vec] (a) to (A);
  \draw[vec] (c) to (3);
  \draw[vec] (c) to (C);
  \draw[vec] (e) to (5);
  \draw[vec] (e) to (E);

  \draw[vec] (2) to (b);
  \draw[vec] (B) to (b);
  \draw[vec] (4) to (d);
  \draw[vec] (D) to (d);
  \draw[vec] (6) to (f);
  \draw[vec] (F) to (f);

\draw[vec] (AA) to (A2);
\draw[vec] (B2) to (BB);
\draw[vec] (CC) to (C2);
\draw[vec] (D2) to (DD);
\draw[vec] (EE) to (E2);
\draw[vec] (F2) to (FF);

 \end{tikzpicture}
\caption{The BPS quiver for the $(A_{M-1},A_{N-1})$ theory for odd
 $M$. 
 When $M$ is even, the orientation of the arrows attached to the nodes in the bottom row are reversed. The diagram has $(M-1)$ rows and $(N-1)$
 columns, and therefore $(M-1)(N-1)$ gauge nodes. The charge shown
 inside each circle stands for the primitive charge associated with
 the corresponding node. }
\label{fig:quiver7}
\end{figure}

In this section, we describe how to evaluate the quantum monodromy $S(q)\bar{S(q)}$ for
the $(A_{M-1},A_{N-1})$ theory, following \cite{Cecotti:2010fi,
Xie:2012gd}. We will
use this half-monodromy to study the 3d CS matter theories corresponding
to these theories in the next two sections.

The
$(A_{M-1}, A_{N-1})$ theories have a special chamber on
the Coulomb branch in which all the stable BPS states are
hypermultiplets. In that chamber, the half-monodromy $S(q)$ is expressed as
\begin{equation}
 S(q) \equiv \prod_{\gamma}^{\curvearrowright}E_{\gamma}(q)~,
\end{equation}
where $\gamma$ runs over the electro-magnetic charges of the stable BPS
hypermultiplets whose central charge phase $\text{arg} Z$ satisfies $0
\leq \text{arg} Z < \pi$.\footnote{BPS states whose central charge phase
satisfies $\pi\leq \text{arg} Z<2\pi$ are charge conjugates of these BPS
states.} 
To evaluate the half-monodromy, one needs to identify the charge spectrum of the
stable BPS hypermultiplets and the order of their central charge
phases. Below, we will explain how to identify it for the
$(A_{M-1},A_{N-1})$ theory by using the its BPS quiver diagram shown in
Fig~\ref{fig:quiver7}.

First,
recall that a quiver mutation $\mu_i$ at a node $i$
       associated with electro-magnetic charge $\gamma_i$ is the
       following operation:
\begin{enumerate}
 \item For every node $j$ to which an arrow comes from the node $i$, add
       $\gamma_i$ to the corresponding electro-magnetic charge, say,
       $\gamma_j$. Therefore, the mutation $\mu_i$ induces the
       replacement $\gamma_j \to \mu_i(\gamma_j) \equiv \gamma_j + \gamma_i$ if there is an arrow
       from $i$ to $j$ before the mutation.

 \item Flip the sign of the the charge associated with the node $i$,
       i.e., $\gamma_i\to \mu_i(\gamma_i) \equiv  -\gamma_i$.
 \item Flip the direction of all arrows originating from or ending at
       the node $i$.
 \item Add an extra arrow corresponding to the ``meson'' in the language
       of the
       Seiberg-duality. That is, if the quiver before the mutation
       contains an arrow $f$ from a node $j$ to
       the node $i$ and an arrow $g$ from the node $i$
       node to a node $k$, then the mutation $\mu_i$ add to the quiver diagram
       an extra arrow from $j$
      to $k$. In
       this operation, we might cancel an existing arrow
       with the newly-added one if they are in mutually opposite
       directions. (In what follows, all the mesonic arrows are indeed
       canceled out.)
\end{enumerate}

With the above definition of the quiver mutation, the half-monodromy $S(q)$ is
identified as follows:
\begin{enumerate}
 \item Let us denote by $\Gamma_i$ the charge associated with a node
       $i$ at the beginning.
 \item Find a series 
\begin{equation}
\mu \equiv \mu_{i_n} \circ \mu_{i_{n-1}}\circ \cdots
       \circ \mu_{i_1} 
\label{eq:mu0}
\end{equation}
 of quiver mutations that maps the quiver
       diagram to itself but flips the sign of the electro-magnetic charge at
       every node. 
 \item  Define
\begin{equation}
 \Gamma^{(k)}\equiv \big( \mu_{i_{k-1}}\circ \cdots \circ
  \mu_{i_1}\big)(\Gamma_{i_k})~.
\end{equation}
Namely, the charge $\Gamma^{(k)}$ is the charge associated with the node $i_k$ {\it
       just before} the action of $\mu_{i_k}$ in \eqref{eq:mu}.

 \item Given the above $\Gamma^{(1)},\,\Gamma^{(2)},\,\cdots,\Gamma^{(n)}$ associated with \eqref{eq:mu0}, the half-monodromy $S(q)$
 is identified as
\begin{equation}
 S(q) = E_{\Gamma^{(n)}}(q)E_{\Gamma^{{(n-1)}}}(q)\cdots E_{\Gamma^{(1)}}(q)~.
\end{equation}
\end{enumerate}

The most non-trivial task in the above procedure is
to find a chain of mutations \eqref{eq:mu0} that preserves the quiver
diagram except that the charge associated with every node receives a
sign flip. For $(A_{M-1},A_{N-1})$,
two expressions for such
$\mu$ have been discussed in \cite[Sec. 8.3]{Cecotti:2010fi} and \cite[page 68]{Xie:2012gd}. That is, 
\begin{equation}
 \mu = \left(\mu_{+-}\right)^{M}
\label{eq:mu1}
\end{equation}
and
\begin{equation}
 \mu = \left(\mu_{-+}\right)^{N}~.
\label{eq:mu2}
\end{equation}
Here, $\mu_{+-}$ is the composition of the mutations at all the nodes
which are {\it sinks} in the horizontal direction and {\it sources} in
the vertical directions. 
Similarly, $\mu_{-+}$ is the composition of the mutations
at all the nodes which are {\it sources} in the horizontal direction and
{\it sinks} in the vertical directions. Note that the mutations included in
a single $\mu_{+-}$ or $\mu_{-+}$ are all commutative.
While one can use either \eqref{eq:mu1} or \eqref{eq:mu2} to identify the half-monodromy,
it is practically easier to
use $\eqref{eq:mu1}$ when $M<N$, and therefore we will use \eqref{eq:mu1} in this paper.

\begin{figure}
\centering
 \begin{tikzpicture}[vec/.style={thick,-latex,draw=black},gauge/.style={circle,draw=black,inner sep=0pt,minimum size=8mm},flavor/.style={rectangle,draw=black,inner sep=0pt,minimum size=10mm},auto]
  \node[gauge] (1) at (0,0) {$\gamma_1^{(1)}$};
  \node[gauge] (2) at (2,0) {$\gamma_2^{(1)}$};
  \node[gauge] (3) at (4,0) {$\gamma_3^{(1)}$};
  \node[gauge] (4) at (0,-2) {$\gamma^{(2)}_1$};
  \node[gauge] (5) at (2,-2) {$\gamma^{(2)}_2$};
  \node[gauge] (6) at (4,-2) {$\gamma^{(2)}_3$};
  \draw[vec] (1) to (2);
  \draw[vec] (3) to (2);
  \draw[vec] (4) to (1);
  \draw[vec] (5) to (4);
  \draw[vec] (5) to (6);
  \draw[vec] (2) to (5);
  \draw[vec] (6) to (3);
 \end{tikzpicture}
\caption{The BPS quiver for the $(A_2,A_3)$ theory. The charge shown
 inside each circle stands for the primitive charge associated with
 the corresponding node.}
\label{fig:quiver1}
\end{figure}
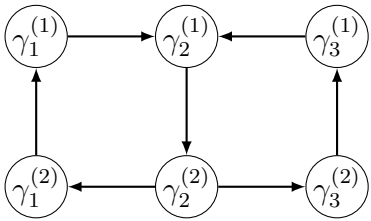

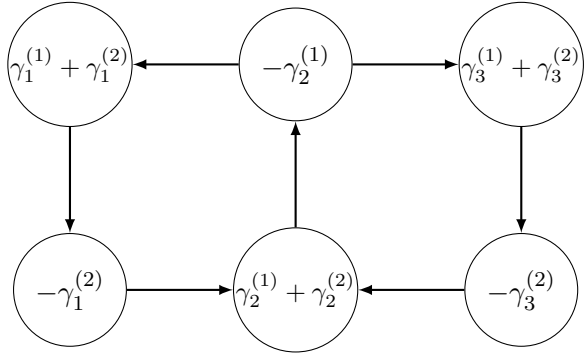
\begin{figure}
\centering
 \begin{tikzpicture}[vec/.style={thick,-latex,draw=black},gauge/.style={circle,draw=black,inner sep=0pt,minimum size=15mm},flavor/.style={rectangle,draw=black,inner sep=0pt,minimum size=10mm},auto]
  \node[gauge] (1) at (0,0) {\footnotesize $\gamma_1^{(1)} +\gamma^{(2)}_1$};
  \node[gauge] (2) at (3,0) {\small $-\gamma^{(1)}_2$};
  \node[gauge] (3) at (6,0) {\footnotesize $\gamma^{(1)}_3+\gamma^{(2)}_3$};
  \node[gauge] (4) at (0,-3) {\small $-\gamma^{(2)}_1$};
  \node[gauge] (5) at (3,-3) {\footnotesize $\gamma^{(1)}_2+\gamma^{(2)}_2$};
  \node[gauge] (6) at (6,-3) {\small $-\gamma^{(2)}_3$};
  \draw[vec] (2) to (1);
  \draw[vec] (2) to (3);
  \draw[vec] (1) to (4);
  \draw[vec] (4) to (5);
  \draw[vec] (6) to (5);
  \draw[vec] (5) to (2);
  \draw[vec] (3) to (6);
 \end{tikzpicture}
\caption{The BPS quiver after one operation of $\mu_{+-}$}
\label{fig:quiver4}
\end{figure}

Let us demonstrate how the above procedure works for
       $(A_2,A_3)$. The relevant BPS quiver is shown in
       Fig.~\ref{fig:quiver1}.
 The original charges associated with the
       quiver nodes are $\gamma^{(1)}_i$ and $\gamma^{(2)}_i$ for $i=1,2$ and
       $3$. The chain of mutations we use is \eqref{eq:mu1} for $M=3$:
\begin{equation}
 \mu = \mu_{+-}\circ \mu_{+-}\circ \mu_{+-}~.
\label{eq:mu}
\end{equation}
Note that at each step of $\mu_{+-}$ the quiver nodes are associated
with different electro-magnetic charges, and therefore the above three
$\mu_{+-}$ give rise to different factors of $E_\gamma(q)$. The first
(and therefore the rightmost)
$\mu_{+-}$ is the composition of the mutations at $\gamma^{(2)}_1,\gamma^{(2)}_3$
and $\gamma^{(1)}_2$, which gives rise to the factor
\begin{equation}
 E_{\gamma^{(1)}_2}(q)E_{\gamma^{(2)}_1}(q)E_{\gamma^{(2)}_3}(q)
\end{equation}
in $S(q)$.
 Note that these three factors of $E_\gamma(q)$ commute with each other since
the electro-magnetic charges $\gamma^{(1)}_2,\gamma^{(2)}_1$ and $\gamma^{(2)}_3$ have
mutually vanishing Dirac's pairing. After this operation of $\mu_{+-}$, the
BPS quiver is now of the form in Fig.~\ref{fig:quiver4}. Note that all
 the ``mesonic'' arrows induced by the mutations are canceled out.

\begin{figure}
\centering
 \begin{tikzpicture}[vec/.style={thick,-latex,draw=black},gauge/.style={circle,draw=black,inner sep=0pt,minimum size=15mm},flavor/.style={rectangle,draw=black,inner sep=0pt,minimum size=10mm},auto]
  \node[gauge] (1) at (0,0) {\footnotesize $-\gamma^{(1)}_1 -\gamma^{(2)}_1$};
  \node[gauge] (2) at (3,0) {\small $\gamma^{(2)}_2$};
  \node[gauge] (3) at (6,0) {\footnotesize $-\gamma^{(1)}_3-\gamma^{(2)}_3$};
  \node[gauge] (4) at (0,-3) {\small $\gamma^{(1)}_1$};
  \node[gauge] (5) at (3,-3) {\footnotesize $-\gamma^{(1)}_2-\gamma^{(2)}_2$};
  \node[gauge] (6) at (6,-3) {\small $\gamma^{(1)}_3$};
  \draw[vec] (1) to (2);
  \draw[vec] (3) to (2);
  \draw[vec] (4) to (1);
  \draw[vec] (5) to (4);
  \draw[vec] (5) to (6);
  \draw[vec] (2) to (5);
  \draw[vec] (6) to (3);
 \end{tikzpicture}
\caption{The BPS quiver after the second operation of $\mu_{+-}$}
\label{fig:quiver5}
\end{figure}
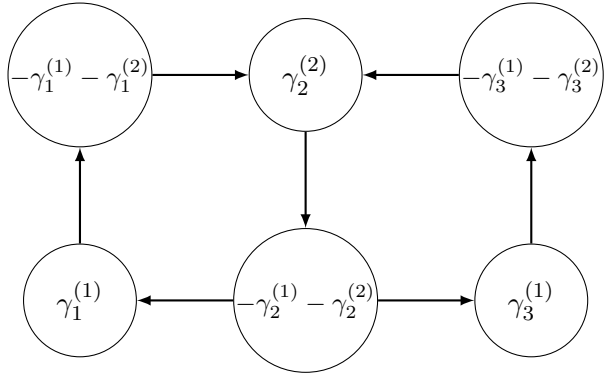

The second $\mu_{+-}$ in \eqref{eq:mu} is then the composition of the
three mutations at the nodes corresponding to
$\gamma^{(1)}_1+\gamma^{(2)}_1,\,\gamma^{(1)}_2+\gamma^{(2)}_2$ and $\gamma^{(1)}_3+\gamma^{(2)}_3$ in Fig.~\ref{fig:quiver4}. This
gives rise to the following factor in $S(q)$:
\begin{equation}
 E_{\gamma^{(1)}_1 +
  \gamma^{(2)}_1}(q)E_{\gamma^{(1)}_2+\gamma^{(2)}_2}(q)E_{\gamma^{(1)}_3+\gamma^{(2)}_3}(q)~.
\label{eq:EEE2}
\end{equation}
We see again that these three factors commute with each other, and
therefore the ordering does not matter in the expression
\eqref{eq:EEE2}.
After this second operation of $\mu_{+-}$, the BPS quiver is now of the form in Fig.~\ref{fig:quiver5}.

It is now clear that the third $\mu_{+-}$ is the composition of the
mutations at the nodes corresponding to $\gamma^{(1)}_1,\,\gamma^{(1)}_3$ and
$\gamma^{(2)}_2$. It gives rise to the factor 
\begin{equation}
 E_{\gamma^{(1)}_1}(q)E_{\gamma^{(1)}_3}(q)E_{\gamma^{(2)}_2}(q)
\end{equation}
in $S(q)$. Again, these three factors of $E_\gamma(q)$ are mutually
commutative. We see that, after this third operation of $\mu_{+-}$, the
BPS quiver now comes back to the original one shown in Fig.~\ref{fig:quiver1},
with $\gamma^{(j)}_i$ replaced by $-\gamma^{(j)}_i$.

From the above computations, the half-monodromy $S(q)$ for the
$(A_2,A_3)$ theory is read off as
\begin{equation}
 S(q) = \Big(E_{\gamma^{(1)}_1}(q)E_{\gamma^{(1)}_3}(q)E_{\gamma^{(2)}_2}(q) \Big)\Big(E_{\gamma^{(1)}_1+\gamma^{(2)}_1}(q)E_{\gamma^{(1)}_2+\gamma^{(2)}_2}(q)E_{\gamma^{(1)}_3+\gamma^{(2)}_3}(q)\Big)\Big(E_{\gamma^{(1)}_2}(q)E_{\gamma^{(2)}_1}(q)E_{\gamma^{(2)}_3}(q)\Big)~.
\end{equation}

\begin{figure}
\centering
 \begin{tikzpicture}[vec/.style={thick,-latex,draw=black},gauge/.style={circle,draw=black,inner sep=0pt,minimum size=9mm},flavor/.style={rectangle,draw=black,inner sep=0pt,minimum size=10mm},auto]
  \node[gauge] (1) at (0,0) {$\gamma^{(1)}_1$};
  \node[gauge] (2) at (2,0) {$\gamma^{(1)}_2$};
  \node[gauge] (3) at (4,0) {$\gamma^{(1)}_3$};
  \node[gauge] (4) at (6,0) {$\gamma^{(1)}_4$};
  \node[gauge] (5) at (8,0) {$\gamma^{(1)}_5$};
  \node[gauge] (6) at (10,0) {$\gamma^{(1)}_6$};
  \node (7) at (12,0) {\;$\cdots$};

  \node[gauge] (a) at (0,-2) {$\gamma^{(2)}_1$};
  \node[gauge] (b) at (2,-2) {$\gamma^{(2)}_2$};
  \node[gauge] (c) at (4,-2) {$\gamma^{(2)}_3$};
  \node[gauge] (d) at (6,-2) {$\gamma^{(2)}_4$};
  \node[gauge] (e) at (8,-2) {$\gamma^{(2)}_5$};
  \node[gauge] (f) at (10,-2) {$\gamma^{(2)}_6$};
  \node (g) at (12,-2) {\;$\cdots$};

  \draw[vec] (1) to (2);
  \draw[vec] (3) to (4);
  \draw[vec] (5) to (6);
  \draw[vec] (3) to (2);
  \draw[vec] (5) to (4);
  \draw[vec] (7) to (6);

  \draw[vec] (b) to (a);
  \draw[vec] (d) to (c);
  \draw[vec] (f) to (e);
  \draw[vec] (b) to (c);
  \draw[vec] (d) to (e);
  \draw[vec] (f) to (g);

  \draw[vec] (a) to (1);
  \draw[vec] (c) to (3);
  \draw[vec] (e) to (5);

  \draw[vec] (2) to (b);
  \draw[vec] (4) to (d);
  \draw[vec] (6) to (f);

 \end{tikzpicture}
\caption{The BPS quiver for the $(A_2,A_{N-1})$ theory. The charge shown
 inside each circle stands for the primitive charge associated with
 the corresponding node.}
\label{fig:quiver16}
\end{figure}
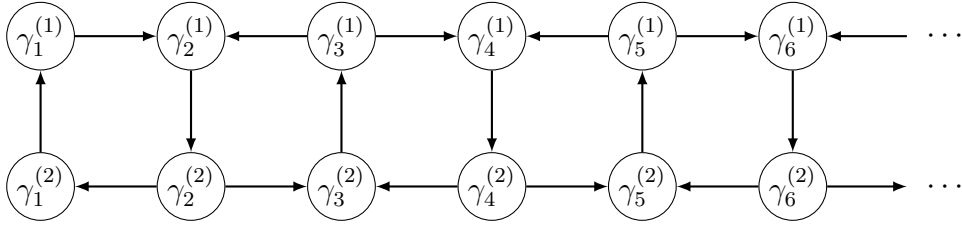

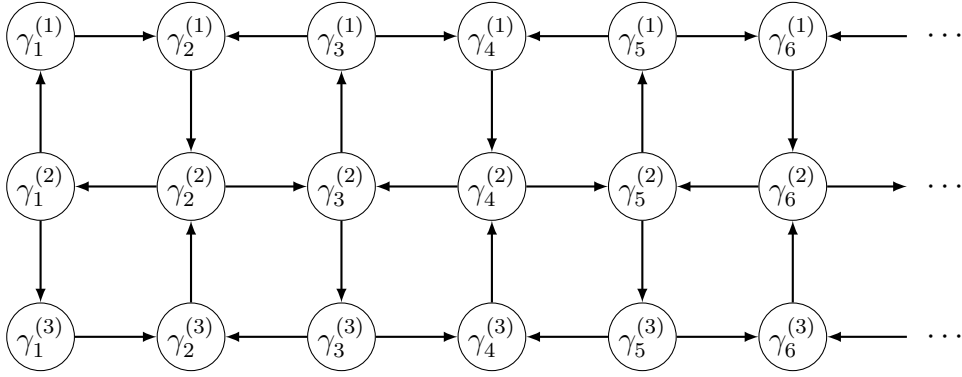
\begin{figure}
\centering
 \begin{tikzpicture}[vec/.style={thick,-latex,draw=black},gauge/.style={circle,draw=black,inner sep=0pt,minimum size=9mm},flavor/.style={rectangle,draw=black,inner sep=0pt,minimum size=10mm},auto]
  \node[gauge] (1) at (0,0) {$\gamma^{(1)}_1$};
  \node[gauge] (2) at (2,0) {$\gamma^{(1)}_2$};
  \node[gauge] (3) at (4,0) {$\gamma^{(1)}_3$};
  \node[gauge] (4) at (6,0) {$\gamma^{(1)}_4$};
  \node[gauge] (5) at (8,0) {$\gamma^{(1)}_5$};
  \node[gauge] (6) at (10,0) {$\gamma^{(1)}_6$};
  \node (7) at (12,0) {\;$\cdots$};

  \node[gauge] (a) at (0,-2) {$\gamma^{(2)}_1$};
  \node[gauge] (b) at (2,-2) {$\gamma^{(2)}_2$};
  \node[gauge] (c) at (4,-2) {$\gamma^{(2)}_3$};
  \node[gauge] (d) at (6,-2) {$\gamma^{(2)}_4$};
  \node[gauge] (e) at (8,-2) {$\gamma^{(2)}_5$};
  \node[gauge] (f) at (10,-2) {$\gamma^{(2)}_6$};
  \node (g) at (12,-2) {\;$\cdots$};

  \node[gauge] (A) at (0,-4) {$\gamma^{(3)}_1$};
  \node[gauge] (B) at (2,-4) {$\gamma^{(3)}_2$};
  \node[gauge] (C) at (4,-4) {$\gamma^{(3)}_3$};
  \node[gauge] (D) at (6,-4) {$\gamma^{(3)}_4$};
  \node[gauge] (E) at (8,-4) {$\gamma^{(3)}_5$};
  \node[gauge] (F) at (10,-4) {$\gamma^{(3)}_6$};
  \node (G) at (12,-4) {\;$\cdots$};

  \draw[vec] (1) to (2);
  \draw[vec] (3) to (4);
  \draw[vec] (5) to (6);
  \draw[vec] (3) to (2);
  \draw[vec] (5) to (4);
  \draw[vec] (7) to (6);

  \draw[vec] (b) to (a);
  \draw[vec] (d) to (c);
  \draw[vec] (f) to (e);
  \draw[vec] (b) to (c);
  \draw[vec] (d) to (e);
  \draw[vec] (f) to (g);

  \draw[vec] (A) to (B);
  \draw[vec] (C) to (D);
  \draw[vec] (E) to (F);
  \draw[vec] (C) to (B);
  \draw[vec] (E) to (D);
  \draw[vec] (G) to (F);
 
  \draw[vec] (a) to (1);
  \draw[vec] (a) to (A);
  \draw[vec] (c) to (3);
  \draw[vec] (c) to (C);
  \draw[vec] (e) to (5);
  \draw[vec] (e) to (E);

  \draw[vec] (2) to (b);
  \draw[vec] (B) to (b);
  \draw[vec] (4) to (d);
  \draw[vec] (D) to (d);
  \draw[vec] (6) to (f);
  \draw[vec] (F) to (f);

 \end{tikzpicture}
\caption{The BPS quiver for the $(A_3,A_{N-1})$ theory. The charge shown
 inside each circle stands for the primitive charge associated with
 the corresponding node.}
\label{fig:quiver6}
\end{figure}

One can generalize this procedure to the
$(A_{M-1},A_{N-1})$ theory. For instance, for the $(A_2,A_{N-1})$
theory, the half-monodromy is identified as
\begin{equation}
 S_{(A_2,A_{N-1})}(q) = \prod_{i:\text{odd}}E_q(X_{\gamma^{(1)}_i})\prod_{j:\text{even}}E_q(X_{\gamma^{(2)}_j})
\prod_{i=1}^{N-1}E_q(X_{\gamma^{(1)}_i +
 \gamma^{(2)}_i})\prod_{i:\text{even}}E_q(X_{\gamma^{(1)}_i})\prod_{j:\text{odd}}E_q(X_{\gamma^{(2)}_j})~,
\label{eq:S-A2AN}
\end{equation}
where the electro-magnetic charges are labeled as in Fig.~\ref{fig:quiver16}.
Similarly, for the $(A_3,A_{N-1})$ theory, the
half-monodromy $S(q)$ is identified as
\begin{align}
 S_{(A_3,A_{N-1})}(q) &= \Big(\prod_{\substack{i:\text{even}\\ j:\text{odd}}}E_{\gamma^{(1)}_j}(q)E_{\gamma^{(2)}_i}(q)E_{\gamma^{(3)}_j}(q)\Big)\Big(\prod_{\substack{i:\text{even}\\
 j:\text{odd}}}E_{\gamma^{(1)}_i+\gamma^{(2)}_i}(q)E_{\gamma^{(1)}_j+\gamma^{(2)}_j+\gamma^{(3)}_j}(q)E_{\gamma^{(2)}_i+\gamma^{(3)}_i}(q)\Big)
\nonumber\\
&\qquad \times\Big(\prod_{\substack{i:\text{even}\\
 j:\text{odd}}}E_{\gamma^{(1)}_j+\gamma^{(2)}_j}(q)E_{\gamma^{(1)}_i+\gamma^{(2)}_i+\gamma^{(3)}_i}(q)E_{\gamma^{(2)}_j+\gamma^{(3)}_j}(q)\Big)\Big(\prod_{\substack{i:\text{even}\\
 j:\text{odd}}}E_{\gamma^{(1)}_i}(q)E_{\gamma^{(2)}_j}(q)E_{\gamma^{(3)}_i}(q)\Big)~,
\label{eq:S-A3AN}
\end{align}
where the electro-magnetic charges are assigned as in
Fig.~\ref{fig:quiver6}.
We will use \eqref{eq:S-A2AN} and \eqref{eq:S-A3AN} in the following
sections to identify the
CS-matter theory corresponding to the twisted compactification of the
$(A_2,A_{N-1})$ and $(A_3,A_{N-1})$ theories.

\section{$(A_2,A_{N-1})$ theory}
\label{sec:A2}

\subsection{Schur index for $(A_2, A_{N-1})$}
\label{sec:schurA2An}
In this section, we focus on the $(A_2,A_{N-1})$ theory for $N$ coprime to $3$.
The BPS quiver has $2(N-1)$ nodes, corresponding to the charges $\gamma^{(1)}_i$ and $\gamma^{(2)}_i$ for
$i=1,2,\cdots,N-1$ such that
\begin{align}
 \langle \gamma^{(1)}_i,\gamma^{(1)}_{i+1}\rangle = (-1)^{i+1}~,\qquad \langle
 \gamma^{(2)}_i,\gamma^{(2)}_{i+1}\rangle = (-1)^i~,\qquad \langle
 \gamma^{(1)}_i,\gamma^{(2)}_i\rangle = (-1)^i~.
\end{align}
Below, we first evaluate the Schur index of this theory via the
IR formula \cite{Cordova:2015nma}, and then study the CS matter theory corresponding to the
twisted 3d reduction of $(A_2,A_{N-1})$.

The 4d Schur index of $(A_2,A_{N-1})$ is evaluated as
\begin{align}
 \mathcal{I}^\text{4d} =
 (q;q)^{2(N-1)}_{\infty}\text{Tr}\left(S(q)\overline{{S}(q)}\right)~,
\label{eq:4Dindex-A2A3N}
\end{align}
where $S(q)$ is given by \eqref{eq:S-A2AN}, and $\bar{S(q)}$ is obtained
by replacing $X_\gamma$ with $X_{-\gamma}$ for all $\gamma$.
By explicit computations, the index \eqref{eq:S-A2AN} can be
evaluated as\footnote{This expression will also be discussed in a separate
paper \cite{Nishinaka:toappear} by S.~Tanigawa and the first-named author. The first-named author thanks S.~Tanigawa for this separate
collaboration.}
\begin{align}
S(q)&= \sum_{i=1}^{N-1}\sum_{n^{(1)}_{i}, n^{(2)}_{i}=0}^{\infty}
\sum_{k_{i}=0}^{{\rm min}(n^{(1)}_{i},n^{(2)}_{i})}   
(q^{\frac{1}{2}} )^{ \sum_{i=1}^{N-2}(n^{(1)}_{i} n^{(1)}_{i+1} + n^{(2)}_{i} n^{(2)}_{i+1}-2 k_{i} k_{i+1}) 
+\sum_{i=1}^{N-1}(   k^2_{i} -n^{(1)}_{i} n^{(2)}_{i} )}
\nonumber \\
&  
\qquad \times \frac{(-q^{\frac{1}{2}})^{  \sum_{i=1}^{N-1} \Bigl(  
n^{(1)}_{i} +  n^{(2)}_{i}  - k_{i} \Bigr) }
 }{
\prod_{i=1}^{N-1} (q)_{n^{(1)}_i- k_i} (q)_{n^{(2)}_i -k_i}  
 (q)_{k_i}
 }
 X_{\sum_{i=1}^{N-1}\sum_{j=1}^2n_i^{(j)}  \gamma_{i}^{(j)}  
}
\end{align}
Plugging this into \eqref{eq:4Dindex-A2A3N}, we find 
\begin{align}
\mathcal{I}^\text{4d} &=  (q)_\infty^{2(N-1)}\sum_{n^{(1)}_{ i},n^{(2)}_{i}=0}^\infty\sum_{k_i,\tilde{k}_i=0}^{\text{min}(n^{(1)}_{i}, n^{(2)}_{i})}q^{\sum_{i=1}^{N-2}(n^{(1)}_{i}  n^{(1)}_{i+1}+n^{(2)}_{i} n^{(2)}_{i+1}-k_ik_{i+1}-\tilde{k}_i\tilde{k}_{i+1})
 - \sum_{i=1}^{N-1}(n^{(1)}_i n^{(2)}_i-\frac{1}{2}k_i^2-\frac{1}{2}\tilde{k}_i^2)}
\nonumber\\
&\qquad \times \frac{\prod_{i=1}^{N-1}q^{n^{(1)}_i}q^{n^{(2)}_i}(-q^{-\frac{1}{2}})^{k_i}(-q^{-\frac{1}{2}})^{\tilde{k}_i}
}{(q)_{n^{(1)}_i-k_i}(q)_{n^{(2)}_i-k_i}(q)_{k_i}(q)_{n^{(1)}_i-\tilde{k}_i}(q)_{n^{(2)}_i-\tilde{k}_i}(q)_{\tilde{k}_i}}~.
\label{eq:4Dindex-A2A3M-2}
\end{align}

It was conjectured in \cite{Cordova:2015nma} that this Schur index
coincides with the vacuum character of $(3,N+3)$ $W_3$ minimal model via
the SCFT/VOA correspondence. To check this conjecture, we can try to
expand \eqref{eq:4Dindex} in powers of $q$ and evaluate the first
several terms.
However, one difficulty here is that the expression
\eqref{eq:4Dindex-A2A3M-2} involves a conditionally-convergent sum, as
in the cases discussed in \cite{Cordova:2016uwk}.
If we truncate the sum over $n_i^{(j)}$ by
introducing a cut-off as proposed in \cite{Cordova:2016uwk}, we reproduce the first few terms of the vacuum
character of the $(3,N+3)$ minimal model.
A more careful treatment of this ill-defined expression
for the Schur index will be discussed in \cite{Nishinaka:toappear}.

\subsection{Half index and CS matter theory}
\begin{table}[thb]
\begin{center}
\begin{tabular}{c | c c c cc c }
& $\phi^{(1)}_{i}$  & $\phi^{(2)}_{i}$ & $\tilde{\phi}^{(1)}_{i}$  & $\tilde{\phi}^{(2)}_{i}$ & $\varphi_{i}$ & $\tilde{\varphi}_{i}$   \\ \hline
$U(1)_{x^{(1)}_{i}}$ &  $1$  & $0$  & $1$  & $0$  & $0$ & 0\\ 
$U(1)_{x^{(2)}_{i}}$ & $0$   & $1$  & $0$  & $1$  & $0$&   0 \\ 
$U(1)_{y_i}$             & $-1$  & $-1$ & $0$  & $0$  & $1$ & 0 \\ 
$U(1)_{\tilde{y}_{i}}$ & $0$  & $0$ & $-1$  & $-1$ & $0$&  1 \\ 
\end{tabular} 
\caption{The gauge charge assignment in the $i$-th sector, $i = 1,\dots,N-1$. 
The chiral multiplets in the $i$-th sector carry no charge under the gauge groups of the other sectors.
}
\label{tabel:chargeA2An}
\end{center}
\end{table}

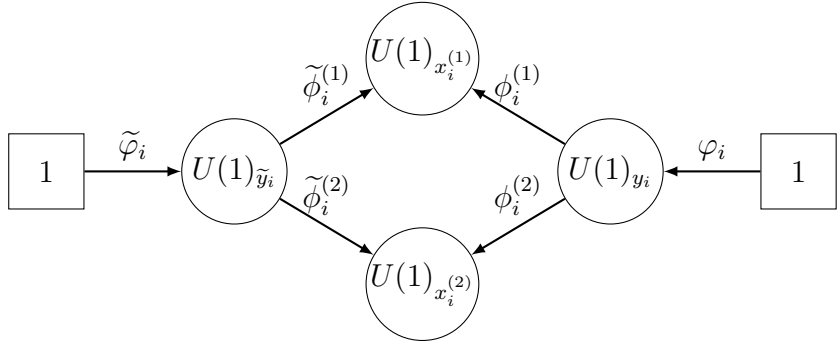
\begin{figure}
\centering
 \begin{tikzpicture}[vec/.style={thick,-latex,draw=black},gauge/.style={circle,draw=black,inner sep=0pt,minimum size=14mm},flavor/.style={rectangle,draw=black,inner sep=0pt,minimum size=10mm},auto]
  \node[gauge] (1) at (0,1.5) {$U(1)_{x^{(1)}_i}$};
  \node[gauge] (2) at (0,-1.5) {$U(1)_{x^{(2)}_i}$};
  \node[gauge] (3) at (2.5,0) {$U(1)_{y_i}$};
  \node[gauge] (4) at (-2.5,0) {$U(1)_{\tilde{y}_i}$};
  \node[flavor] (5) at (5,0) {$1$};
  \node[flavor] (6) at (-5,0) {$1$};
  \draw[vec] (6) to node[above]{$\tilde \varphi_i$} (4);
  \draw[vec] (4) to node[above]{$\tilde \phi^{(1)}_i$} (1);
  \draw[vec] (4) to node[above]{$\tilde \phi^{(2)}_i$} (2);
  \draw[vec] (5) to node[above]{$\varphi_i$}(3);
  \draw[vec] (3) to node[above]{$\phi^{(1)}_i$} (1);
  \draw[vec] (3) to node[above]{$\phi^{(2)}_i$} (2);
 \end{tikzpicture}
\bigskip
\caption{The quiver diagram for the $i$-th set of gauge and matter
 multiplets included in the CS matter theory for $(A_2,A_{N-1})$ (for
 $N$ coprime to $3$). Each circle stands for a $U(1)$ gauge group. Each arrow connecting two circles stands for a
 bifundamental chiral multiplet, and each arrow from a box to a circle
 stands for a chiral multiplet charged under the $U(1)$ corresponding to
 the circle. The subscript of $U(1)$ in each circle stands for the
 fugacity for the corresponding topological symmetry. This diagram can be read off from \eqref{eq:4Dindex-A2A3M-2}. }
\label{fig:quiver2}
\end{figure}

As in the $(A_1,A_2)$ case reviewed in Section~\ref{sec:A1A2}, we determine the gauge group and matter content of a 3d
$\mathcal{N}=2$ CS matter theory in such a way that its half index reproduces the Schur index \eqref{eq:4Dindex-A2A3M-2}.
We consider a CS matter theory with gauge group $U(1)^{4(N-1)}$ coupled to $6(N-1)$ chiral multiplets. 
The field content and the corresponding quiver diagram are summarized in Table~\ref{tabel:chargeA2An} and in Figure~\ref{fig:quiver2}.
We denote the gauge group as
\begin{align}
  U(1)^{4(N-1)}
  = \prod_{i=1}^{N-1} U(1)_{x^{(1)}_i}
    \times \prod_{i=1}^{N-1} U(1)_{x^{(2)}_i}
    \times \prod_{i=1}^{N-1} U(1)_{y_i}
    \times \prod_{i=1}^{N-1} U(1)_{\tilde{y}_i}\,,
\end{align}
and label the chiral multiplets by
$\phi^{(1)}_i, \phi^{(2)}_i, \tilde{\phi}^{(1)}_i, \tilde{\phi}^{(2)}_i, \varphi_i, \tilde{\varphi}_i$ (for $i = 1,\ldots,N-1$), as depicted in Figure~\ref{fig:quiver2}. 
We assume that the reference R-charges of all these chiral multiplets are zero.

We choose the gauge Chern-Simons levels to be
\begin{align}
  K_{ab} \equiv \tilde{K}_{ab} + \tilde{K}_{ba}\,,
\end{align}
with
\begin{align}
\tilde{K}_{ab} \equiv
\begin{cases}
  1 & \text{if } a = b - 1,\quad b = 1,\ldots,2N-1,\\[2pt]
  -1 & \text{if } a = b - 1,\quad b = 2N-1,\ldots,4(N-1),\\[2pt]
  -1 & \text{if } a = b - N + 1,\quad b = 1,\ldots,N-1,\\[2pt]
  \dfrac{1}{2} & \text{if } a = b,\quad b = 2N-1,\ldots,4(N-1),\\[2pt]
  0 & \text{otherwise}.
\end{cases}
\end{align}
Using the localization formula summarized in Appendix~\ref{app:local}, the half index of this theory takes the following form:
\begin{align}
 I\!\!I_{(\mathcal{D}, D_c)} &= \frac{1}{(q)_{\infty}^{4(N-1)}} \sum_{n^{(1)}_i,n^{(2)}_i,k_i,\tilde{k}_i\in \mathbb{Z}}
 q^{\sum_{i=1}^{N-2}(n^{(1)}_in^{(1)}_{i+1}+n^{(2)}_in^{(2)}_{i+1}-
 k_ik_{i+1}-\tilde{k}_i\tilde{k}_{i+1}) - \sum_{i=1}^{N-1}\left(n^{(1)}_in^{(2)}_i -
 \frac{k_i^2+\tilde{k}_i^2}{2}\right)} \nonumber \\
 & \quad \times 
 \Bigl(\prod_{i=1}^{N-1}(x^{(1)}_i)^{n^{(1)}_i}(x^{(2)}_i)^{n^{(2)}_i}(y_i)^{k_i}(\tilde{y}_i)^{\tilde{k}_i}\Bigr)
\nonumber\\
&\quad \times
 \prod_{i=1}^{N-1}(q^{1-(n^{(1)}_i-k_i)})_\infty(q^{1-(n^{(2)}_i-k_i)})_\infty(q^{1-(n^{(1)}_i-\tilde{k}_i)})_\infty(q^{1-(n^{(2)}_i-\tilde{k}_i)})_\infty(q^{1-k_i})_\infty(q^{1-\tilde{k}_i})_\infty
\nonumber\\
 &= (q)_\infty^{2(N-1)}\sum_{n^{(1)}_i, n^{(2)}_i=0}^\infty\sum_{k_i,\tilde{k}_i=0}^{\text{min}(n^{(1)}_i,n^{(2)}_i)}
 q^{\sum_{i=1}^{N-2}(n^{(1)}_in^{(1)}_{i+1}+n^{(2)}_in^{(2)}_{i+1}-
 k_ik_{i+1}-\tilde{k}_i\tilde{k}_{i+1}) - \sum_{i=1}^{N-1}
\left(n^{(1)}_i n^{(2)}_i -
 \frac{k_i^2+\tilde{k}_i^2}{2}\right)}
\nonumber\\
&\qquad \times
 \prod_{i=1}^{N-1}\frac{(x^{(1)}_i)^{-n^{(1)}_i}(x^{(2)}_i)^{-n^{(2)}_i}(y_i)^{-k_i}(\tilde{y}_i)^{-\tilde{k}_i}}
 {(q)_{n^{(1)}_i-k_i}(q)_{n^{(2)}_i-k_i}(q)_{n^{(1)}_i-\tilde{k}_i}(q)_{n^{(2)}_i-\tilde{k}_i}(q)_{k_i}(q)_{\tilde{k}_i}}\,,
\label{eq:half-A2A3M}
\end{align}
where ${x}^{(1)}_i,x^{(2)}_i,y_i$ and $\tilde{y}_i$ are fugacities for the topological
symmetry associated with the gauge group $U(1)_{x^{(1)}_i}, U(1)_{x^{(2)}_i},  U(1)_{y_i},  U(1)_{\tilde{y}_i} $ .

Comparing \eqref{eq:half-A2A3M} with \eqref{eq:4Dindex-A2A3M-2}, we find
that these two expressions coincide with each other if
\begin{align}
 x^{(1)}_i = x^{(2)}_i = q^{-1}~,\qquad y_i = \tilde{y}_i = - q^{\frac{1}{2}}~.
\label{eq:cond}
\end{align}
These conditions must be interpreted in terms of a monopole
superpotential. Namely, we need to identify the most general expression
for the monopole superpotential consistent with \eqref{eq:cond}.

\subsection{Monopole superpotential}
\begin{table}[thb]
\begin{center}
\begin{tabular}{c | c c c cc c }
& $V^{(1)}_{i}$  & $V^{(2)}_{i}$ & $V_{i}$  & $\tilde{V}^{(1)}_{i}$ & $\tilde{V}^{(2)}_{i}$ & $\tilde{V}_{i}$   \\ \hline
$U(1)_{x^{(1)}_{i}}$ &  $-1$  & $0$  & $0$  & $-1$  & $0$ & 0\\ 
$U(1)_{x^{(2)}_{i}}$ & $0$   & $-1$  & $0$  & $0$  & $-1$&   0 \\ 
$U(1)_{y_i}$           & $-1$  & $-1$ & $1$  & $0$  & $0$ & 0 \\ 
$U(1)_{\tilde{y}_{i}}$ & $0$  & $0$ & $0$  & $-1$ & $-1$&   $1$\\ 
\end{tabular} 
\caption{The magnetic charge for monopole operators in the $i$-th sector, $i = 1,\dots,N-1$. 
The monopole operators in the $i$-th sector carry no charge under the gauge groups of the other sectors.
}
\label{table:chargeA2An}
\end{center}
\end{table}

The non-vanishing fugacities \eqref{eq:cond} for the topological
symmetries imply that the
correct R-symmetry is a linear
combination of the 
reference R-symmetry and the topological symmetries. This
suggests that a superpotential involving monopole operators is turned
on.
From the formula for the gauge charge of  monopole operator given in appendix~\ref{app:monopole}, 
we find that the following six types of dressed monopole operators satisfy the gauge-invariance condition 
$Q_a[\mathcal{O}_{(n,m)}]=0$:
\begin{align}
&\phi^{(1)}_{i-1} V^{(1)}_i \phi^{(1)}_{i+1},  \;\; \phi^{(2)}_{i-1} V^{(2)}_i \phi^{(2)}_{i+1}, \;\; \varphi_{i-1} V_i \varphi_{i+1}, \nonumber \\[2mm]
&\tilde{\phi}^{(1)}_{i-1} \tilde{V}^{(1)}_i \tilde{\phi}^{(1)}_{i+1}, \;\; \tilde{\phi}^{(2)}_{i-1} \tilde{V}^{(2)}_i \tilde{\phi}^{(2)}_{i+1},  \;\;
\tilde{\varphi}_{i-1} \tilde{V}_i \tilde{\varphi}_{i+1},
\label{eq:dressmono}
\end{align}
with $i=1,\cdots,N-1$. Here we set $\phi^{(1)}_{0} =\phi^{(1)}_{N}=\phi^{(2)}_{0} =\phi^{(2)}_{N} =\tilde{\varphi}_{0}=\tilde{\varphi}_{N}=1$.
The magnetic charges of the monopole operators $V^{(l)}_i, \tilde{V}^{(l)}_i$ and 
$V_i, \tilde{V}_i$ are listed in Table~\ref{table:chargeA2An}.

Note that, when the above terms are added to the superpotential, the 
reference
R-symmetry is broken since the monopole operators are charged under the
R-symmetry \eqref{eq:RchargeV}. However, the following linear
combination of the reference R-symmetry and topological symmetries is preserved:
\begin{align}
R_{\rm shift}=  R  -2\sum_{i=1}^{N-1}(J_{x^{(1)}_i} + J_{x^{(2)}_i}) + \sum_{i=1}^{N-1} (J_{y_i}+J_{\tilde{y}_i}) ~,
\label{eq:RIR2}
\end{align}
where $J_{x^{(l)}_i}$ $(l=1,2)$,  $J_{y_i}$ and $J_{\tilde{y}_i}$ are topological charges associated with
$U(1)$ gauge groups corresponding to $x^{(l)}_i$ $(l=1,2)$, $y_i$ and $\tilde{y}_i$, respectively.
 Indeed, each term of \eqref{eq:dressmono} has precisely charge
two for this linear combination, which means that \eqref{eq:RIR2}
can be identified as the preserved R-charge of the theory perturbed by
the superpotential.

It is then natural to consider the following term as a superpotential term:
\begin{align}
  &\sum_{i=1}^{N-1} \Bigl(  {\phi}^{(1)}_{i-1} {V}^{(1)}_i  \phi^{(1)}_{i+1}  +
  {\phi}^{(2)}_{i-1} {V}^{(2)}_i  \phi^{(2)}_{i+1}+
 \tilde{\phi}^{(1)}_{i-1}\tilde{V}_i^{(1)}\tilde{\phi}^{(1)}_{i+1}
 \nonumber \\
 & \qquad \qquad
 +
 \tilde{\phi}^{(2)}_{i-1}\tilde{V}_i^{(2)}\tilde{\phi}^{(2)}_{i+1}
 +
  \varphi_{i-1} {V}_i \varphi_{i+1}
 +
 \tilde{\varphi}_{i-1}\tilde{V}_i\tilde{\varphi}_{i+1}\Bigr)~.
\end{align} 
However, in order to  preserve a single $U(1)$ flavor symmetry, which will be studied in the next subsection, 
we slightly modify the above terms and instead include the following terms in the superpotential:
\begin{align}
W &=
 \sum_{i=1}^{N-1} \Bigl(  {\phi}^{(1)}_{i-1} {V}^{(1)}_i  \phi^{(1)}_{i+1}  +
  {\phi}^{(2)}_{i-1} {V}^{(2)}_i  \phi^{(2)}_{i+1}+
 \tilde{\phi}^{(1)}_{i-1}\tilde{V}_i^{(1)}\tilde{\phi}^{(1)}_{i+1}
 \nonumber \\
 & \qquad \qquad
 +
  \varphi_{i-1} {V}_i \varphi_{i+1}
 +
 \tilde{\varphi}_{i-1}\tilde{V}_i\tilde{\varphi}_{i+1}\Bigr)
 +\sum_{i=1}^{N-2} \tilde{\phi}^{(2)}_{i-1}\tilde{V}_i^{(2)}\tilde{\phi}^{(2)}_{i+1}~.
\label{eq:superpot3}
\end{align} 

Now we study the  choice of fugacities.  
Identifying \eqref{eq:RIR2} as the  R-charge precisely corresponds to turning on
\begin{align}
  x^{(1)}_i = x^{(2)}_i = q^{-1} \,, \qquad
  y_i = \tilde{y}_i = - q^{\frac{1}{2}} \,.
  \label{eq:fugacity1}
\end{align}

\subsection{Residual global symmetry}
\label{sec:ressymA2An}
Before turning on the monopole superpotential, the CS matter theory 
 has a $U(1)^{4(N-1)}$ topological symmetry and a $U(1)^{2(N-1)}$ flavor symmetry. 
On the other hand, since this theory is expected to flow in the IR to 3d SCFT obtained by twisted compactification of the $(A_2,A_{N-1})$ theory, 
its genuine UV flavor symmetry  should reduce to a single $U(1)_A$. 
In particular, the  $U(1)_A$ and  $U(1)$ R-symmetry in the 3d $\mathcal{N}=2$ theory is  identified with linear combinations of 
the maximal torus of the  $SU(2)_H \times SU(2)_C$ R-symmetry at the 3d $\mathcal{N}=4$ superconformal fixed point. 
In the following we will show that, once the superpotential \eqref{eq:superpot3} is turned on, the global symmetry of the CS 
matter theory is indeed broken down to a single $U(1)$, in agreement with this expectation.

Let us first describe how the $U(1)^{2(N-1)}$ flavor symmetry acts on the fields. 
Up to gauge equivalence,
the $U(1)^{2(N-1)}$ flavor symmetry acts on the chiral multiplets as
\begin{equation}
\tilde \phi_k^{(l)} \to e^{i\zeta_k^{(l)}}\tilde \phi_k^{(l)}\qquad (l
 = 1,2)~,
\end{equation}
with all the other chiral multiplets kept fixed. It also acts on the
monopole operators as
\begin{align}
 V_k^{(1)} &\to e^{-\frac{i}{2}\zeta_k^{(1)}} V_k^{(1)}~,\qquad V_k^{(2)}\to
 e^{-\frac{i}{2}\zeta^{(2)}_k}V_k^{(2)}~,
\\
\tilde V_k^{(1)} &\to e^{-\frac{i}{2}\zeta_k^{(2)}}\tilde V_k^{(1)}~,\qquad
 \tilde V_k^{(2)}\to e^{-\frac{i}{2}\zeta_k^{(1)}}\tilde V_k^{(2)}~,
\\
\tilde V_k &\to e^{-\frac{i}{2}(\zeta_k^{(1)} + \zeta^{(2)}_k)} \tilde V_k~,
\end{align}
with $V_k$ kept fixed.
On the other hand, the topological $U(1)^{4(N-1)}$ symmetry acts only on
the monopole operators as
\begin{align}
V_k^{(1)} \to e^{i(-\alpha^{(1)}_k -\beta_k)}V_k^{(1)}~&,\qquad V_k^{(2)}
 \to e^{i\left(-\alpha^{(2)}_k - \beta_k\right)}V_k^{(2)}~,
\\
 \tilde V_k^{(1)} \to
  e^{i\left(-\alpha^{(1)}_k - \tilde\beta_k\right)} \tilde V_k^{(1)}~&,\qquad
 \tilde V_k^{(2)} \to e^{i\left(-\alpha^{(2)}_k-\tilde \beta_k\right)}\tilde V_k^{(2)}~,
\\
 V_k \to e^{i\beta_k}V_k~&,\qquad \tilde V_k \to e^{i\tilde \beta_k}\tilde{V}_k~,
\end{align}
where $e^{i\alpha^{(1)}_k},\,e^{i\alpha^{(2)}_k},\,e^{i\beta_k}$ and $e^{i\tilde
\beta_k}$ are topological $U(1)$ associated with the $U(1)$ gauge groups
corresponding to $x_k^{(1)},\,x_k^{(2)},\,y_k$ and $\tilde y_k$, respectively.

Under the combined $U(1)^{6(N-1)}$ global symmetry, the superpotential
term \eqref{eq:superpot3} transform as
\begin{align}
 W &\to \sum_{k=1}^{N-1} \left(e^{i\beta_k}\varphi_{k-1}V_{k}\,\varphi_{k+1} +
e^{i\left(-\frac{\zeta^{(1)}_k +\zeta^{(2)}_k }{2}+\tilde
 \beta_k\right)} \tilde{\varphi}_{k-1}\tilde V_k\,\tilde{\varphi}_{k+1}
 \right)
\nonumber \\
 &+ \sum_{k=1}^{N-1} \left(
 e^{i\left(-\frac{\zeta_k^{(1)}}{2}-\alpha^{(1)}_k - \beta_k\right)}\phi_{k-1}^{(1)} V_k^{(1)}\phi_{k+1}^{(1)}  +
e^{i\left(-\frac{\zeta^{(2)}_k}{2}-\alpha^{(2)}_k
 -\beta_k\right)} \phi_{k-1}^{(2)}V_k^{(2)}\phi_{k+1}^{(2)}
 \right)
\nonumber \\
 &+  \sum_{k=1}^{N-1} e^{i\left(\zeta_{k-1}^{(1)}+\zeta^{(1)}_{k+1}-\frac{\zeta^{(2)}_k}{2}-\alpha^{(1)}_k - \tilde \beta_k\right)}\tilde{\phi}^{(1)}_{k-1} \tilde{V}_k^{(1)} \tilde \phi^{(1)}_{k+1}  +
 \sum_{k=1}^{N-2}e^{i\left(\zeta^{(2)}_{k-1} +
 \zeta^{(2)}_{k+1}-\frac{\zeta^{(1)}_k}{2}-\alpha^{(2)}_k - \tilde
 \beta_k\right)}\tilde \phi^{(2)}_{k-1}\tilde{V}_k^{(2)}\tilde \phi^{(2)}_{k+1}~,
\end{align}
where we defined $\zeta^{(l)}_0 = \zeta^{(l)}_N=0$ for
convenience.
For these superpotential terms to be invariant, we need to impose
\begin{align}
 \beta_k &= 0~,\qquad \tilde \beta_k =
 \frac{\zeta_k^{(1)}+\zeta^{(2)}_k}{2}~,\qquad \alpha^{(l)}_k =
 -\frac{\zeta_k^{(l)}}{2}~,
\label{eq:ggab}
\\
\zeta^{(2)}_k &= \zeta^{(1)}_{k-1}+\zeta^{(1)}_{k+1}~,
\label{eq:theta-zeta}
\end{align}
for $1\leq k \leq N-1$, and
\begin{equation}
 \zeta_k^{(1)} = \zeta^{(2)}_{k-1}
 + \zeta^{(2)}_{k+1}\label{eq:theta-zeta2}
\end{equation}
for $1\leq k\leq N-2$.
From \eqref{eq:theta-zeta} and \eqref{eq:theta-zeta2} for $k=1$ and $k=2$, we see that
\begin{equation}
\zeta_2^{(1)} = \zeta^{(2)}_1~,\qquad \zeta^{(2)}_2 = \zeta^{(1)}_1~,\qquad \zeta^{(1)}_3 =
 \zeta^{(2)}_3 = 0~.
\label{eq:init_cond}
\end{equation}
Furthermore, for $2\leq k\leq N-2$, Eqs.~\eqref{eq:theta-zeta} and
\eqref{eq:theta-zeta2} imply that
\begin{equation}
 \zeta_{k+2}^{(l)} = -\zeta_k^{(l)} -\zeta_{k-2}^{(l)}\qquad (l=1,2)~.
\label{eq:general_cond}
\end{equation}
The constraints \eqref{eq:init_cond} and \eqref{eq:general_cond}
uniquely fix $\zeta_k^{(1)}$ and $\zeta^{(2)}_k$ in terms of $\zeta^{(1)}_1$ and
$\zeta^{(2)}_1$ as
\begin{equation}
 \zeta_k^{(1)}= \left\{
\begin{array}{l}
 (-1)^{n+1}\zeta_1^{(1)}\qquad \text{when} \qquad k = 3n-2 \quad \text{for}
  \quad n\in \mathbb{N}\\
(-1)^{n+1}\zeta_1^{(2)}\qquad \text{when}\qquad k=3n-1 \quad \text{for}
 \quad n\in \mathbb{N}\\
0 \qquad \text{when}\qquad k=3n \quad \text{for}\quad n\in \mathbb{N}\\
\end{array}
\right.~,
\end{equation}
and
\begin{equation}
 \zeta_k^{(2)}= \left\{
\begin{array}{l}
 (-1)^{n+1}\zeta^{(2)}_1\qquad \text{when} \qquad k = 3n-2 \quad \text{for}
  \quad n\in \mathbb{N}\\
(-1)^{n+1}\zeta^{(1)}_1\qquad \text{when}\qquad k=3n-1 \quad \text{for}
 \quad n\in \mathbb{N}\\
0 \qquad \text{when}\qquad k=3n \quad \text{for}\quad n\in \mathbb{N}\\
\end{array}
\right.~.
\end{equation}
Since $\alpha^{(l)}_k,\beta_k$ and $\tilde \beta_k$ are constrained
by \eqref{eq:ggab}, all the parameters are now fixed in terms of
$\zeta^{(1)}_1$ and $\theta^{(2)}_1$.

Finally, the constraint \eqref{eq:theta-zeta} for $k=N-1$ implies 
\begin{equation}
 \zeta^{(1)}_{N-1} = \zeta^{(1)}_{N-2}~.
\label{eq:final-const}
\end{equation}
Since $N$ is coprime to $3$, either $N-1$ or $N-2$ is an integer
multiple of $3$. Therefore \eqref{eq:final-const} set one of $\zeta^{(1)}_1$
and $\zeta^{(2)}_1$ to zero. Therefore, there is only one global $U(1)$ symmetry
preserved by the superpontential $W$
in addition to the $U(1)$ R-symmetry. This global symmetry is identified
with the $U(1)_A$ symmetry.

Note that, if the second sum in \eqref{eq:superpot3} ran over
$k=1,2,\cdots,N-1$, Eq.~\eqref{eq:theta-zeta2} for $k=N-1$
gives rise to an extra constraint
$\zeta_{N-1}^{(1)} = \zeta_{N-2}^{(2)}$, which removes the above $U(1)_A$
symmetry. Therefore, to preserve the $U(1)_A$ symmetry, we take the second sum
in \eqref{eq:superpot3} to run over $k=1,2,\cdots,N-2$.

\subsection{Example: $(A_2,A_1)$}
When 3d $\mathcal{N}=2$ supersymmetry is enhanced to $\mathcal{N}=4$, 
the superconformal index typically has an expansion of the form \cite{Gang:2018huc}:
\begin{align}
  I_{\rm SCI} \;=\; 1 - q - \left(\eta + \frac{1}{\eta}\right) q^{\frac{3}{2}} + \mathcal{O}(q^2)~,
  \label{eq:enhance}
\end{align}
where $\eta$ is the fugacity for the $U(1)_A$ symmetry and $q$ is the fugacity for the superconformal R-charge.
We will show that the IR supersymmetry enhancement can be seen from the superconformal index for 
the $N=2$ case, i.e.\ for the $U(1)^4$ CS matter theory associated with $(A_2,A_1)$. 
This provides nontrivial evidence that the 3d CS matter theory indeed flows to the twisted reduction of the corresponding AD theory.

The mixed CS levels $K_{ab}$ and electric charges $Q_{a i}$ 
of the matter chiral multiplets are characterized by 
\begin{align}
 K = \left(
\begin{array}{cccc}
0 & -1 & 0 & 0\\
-1 & 0 & 0 & 0\\
0 & 0 & 1 & 0 \\
0 & 0 & 0 & 1 \\
\end{array}
\right)~,\qquad Q =\left(
\begin{array}{cccccc}
1 & 1 & 0 & 0 & 0 &0\\
0 & 0 & 1 & 1 &0 &0\\
-1& 0 & -1 & 0 &1 &0\\
0 & -1 & 0 & -1 & 0 & 1\\
\end{array}
\right)~,
\end{align}
where the matter chiral multiplets are ordered as $(\phi^{(1)}_1,\tilde
\phi^{(1)}_1, \phi^{(2)}_1,\tilde{\phi}^{(2)}_1,\varphi_1,\tilde{\varphi}_1)$.
The monopole superpotential is
\begin{align}
 W &= V_1 + \tilde{V}_1 + V_1^{(1)} + V_1^{(2)} + \tilde V_1^{(1)}~.
\end{align}
The global symmetry preserved by this superpotential is read off from
\eqref{eq:ggab} and \eqref{eq:theta-zeta} as
\begin{align}
 \tilde{\beta}_1 &= \frac{\zeta^{(1)}_1}{2}~, \quad \alpha^{(1)}_1 = -\frac{\zeta^{(1)}_1}{2}~, \quad
 \beta_1=\zeta^{(2)}_1=\alpha^{(2)}_1 =\zeta^{(2)}_1=0~.
\end{align}
This residual symmetry non-trivially act on the 
\begin{align}
 \phi^{(1)}_1 \to e^{i\zeta_1}\phi^{(1)}_1~,
\end{align}
where all the other chiral multiplets are kept invariant.
This induces the following transformation of the monopole operators:
\begin{align}
V_1 \to V_1\,,\;\; \tilde{V}_1\to e^{-\frac{i}{2} \zeta^{(1)}_1}\tilde{V}_1\,,\;\;  V_1^{(1)} 
\to e^{-\frac{i}{2} \zeta^{(1)}_1 }V_1^{(1)}\,,\;\; \tilde V_1^{(1)} \to
 \tilde V_1^{(1)}\,,\;\; V_1^{(2)} \to  V_1^{(2)}~.
\end{align}
Therefore, by using the localization formula \eqref{eq:SCI3}, the superconformal index of this theory is evaluated as\footnote{As in the case of the half index, the trace is defined using $(-1)^R$ instead of $(-1)^F$.}

\begin{align}
 I_{\rm SCI} &= \sum_{i=1}^2 \sum_{n_i,k_i \in \mathbb{Z}} \oint
\prod_{i=1}^2 \frac{d s_i dz_i}{(2\pi i)^2 s_i  z_i}
 q^{-n_1-n_2}\left(-q^{\frac{1}{2}}\right)^{k_1 + 
 k_2} e^{-i\zeta^{(1)}_1n_2} s_1^{-n_2} s_2^{-n_1} z_1^{k_1} z_2^{ k_2}
\nonumber\\
&\qquad \times
\prod_{i,j=1}^2 \left(-q^{\frac{1}{2}}\frac{z_j}{s_i}e^{-i\zeta^{(1)}_1 \delta_{i,1} \delta_{j,1}} \right)^{\frac{n_i-k_j +
 |n_i-k_j|}{2}}\frac{\left(\frac{z_j}{s_i} e^{-i\zeta^{(1)}_1 \delta_{i,1} \delta_{j,1}}q^{1+\frac{|n_i-k_j|}{2}};q\right)_\infty}{(\frac{s_i}{z_j}e^{i\zeta^{(1)}_1 \delta_{i,1} \delta_{j,1}} q^{\frac{|n_i-k_j|}{2}};q)_\infty}
 \nonumber \\
& \qquad  \times
\prod_{i=1}^2 \left(-q^{\frac{1}{2}}\frac{1}{z_i}\right)^{\frac{k_i+|k_i|}{2}}
\frac{\left(\frac{1}{z_i}q^{1+\frac{|k_i|}{2}};q\right)_\infty}{\left(z_iq^{\frac{|k_i|}{2}};q\right)_\infty
 }\,,
\label{eq:I-A2A1}
\end{align}
where $q^{-n_1-n_2}\left(-q^{\frac{1}{2}}\right)^{k_1 + k_2}$ reflects the fact that the classical R-charge is mixed with the
topological symmetry, and $e^{i\zeta^{(1)}_1}$ is the
fugacity for $U(1)_A$. According to Mathematica computations, when we set
\begin{align}
 e^{i\zeta^{(1)}_1} = -q^{-\frac{1}{2}}\eta~,
 \label{eq:rescfuga}
\end{align}
we find 
\begin{align}
 I_{\rm SCI} = 1 - q - \left(\eta + \frac{1}{\eta}\right)q^{\frac{3}{2}} - 2q^2 + \mathcal{O}(q^{\frac{5}{2}})~,
\end{align}
which reproduces the behavior \eqref{eq:enhance},  and  also the right answer for the $(A_2,A_1) = (A_1,A_2)$.

\subsection{Nahm sum formula for  $(3,8)$ $W_3$ minimal model.}
\label{subsec:Nahm}

In \cite{Gang:2023rei} the authors constructed a 3d CS matter theory whose infrared dynamics exhibits an enhancement to $\mathcal{N}=4$ superconformal symmetry.
Interestingly, the half index of this theory takes the form of a Nahm  sum representation of the vacuum character of 
the $(2,2N+3)$ Virasoro minimal model. 
Since  3d $\mathcal{N}=4$ supersymmetric  theory admit a topological twist that produces  VOA 
on a 2d boundary \cite{Costello:2018fnz}, this observation suggests that the boundary VOA for 
IR limit of the CS matter theory is the Virasoro minimal model.

On the other hand, as reviewed in Section~\ref{sec:A1A2}, the authors of \cite{Gaiotto:2024ioj} constructed  another CS matter theory
based the Schur index of $(A_1, A_{2N})$ theory. Since the Schur index agrees with the vacuum character of the corresponding VOA \cite{Cordova:2015nma}, by construction, 
the half index reproduces the vacuum character of the $(2,2N+3)$ Virasoro minimal model. 
These two theories    flow to the same IR fixed point
which obtained by twisted reduction of $(A_1, A_{2N})$ theory \cite{Gaiotto:2024ioj, ArabiArdehali:2024ysy}. 

In the previous sections we have considered extending the correspondence between the $(A_1,A_{2N})$ theories,
or equivalently the Virasoro minimal models, and CS matter theories to the more general $(A_{M-1},A_{N-1})$ theories, or equivalently the $W_N$ minimal models. 
In particular, up to this point we have mainly pursued the latter approach \cite{Gaiotto:2024ioj}, 
constructing 3d CS matter theories on the basis of the IR formula for the Schur index of $(A_{M-1},A_{N-1})$ AD theory.

By analogy with the Virasoro case, it is then natural to ask whether one can construct 3d CS matter theories whose half indices 
directly reproduce Nahm sum expressions for the vacuum characters of the $W_N$ minimal models. 
However, a universal Nahm sum formula for the $W_N$ minimal models, valid for arbitrary coprime $(M,N)$, is not known. 
In what follows we therefore focus on the $(M,M+N)=(3,8)$ $W_3$ minimal model, for which we have found a new  Nahm sum expression for the vacuum character or equivalently the Schur index for $(A_2,A_4)$ theory.\footnote{For the $(M,M+N)=(3,7)$ $W_3$ minimal model, a Nahm sum formula was proven in \cite{Feigin:2005gz}, and the corresponding CS matter theory was studied very recently in \cite{Kim:2025klh}.}

First, let us recall the definition of a Nahm sum. 
It is a $q$-series of the form
\begin{align}
 \sum_{n_1,\ldots,n_N=0}^{\infty} 
 \frac{q^{\frac{1}{2}\sum_{a,b=1}^{N}  K_{a b} n_a n_b + \sum_{a=1}^N B_a n_a}}
      {\prod_{a=1}^{N} (q;q)_{n_a}} \,,
 \label{eq:nahm}
\end{align}
where $K_{ab}$ is an $N\times N$ matrix and $B_a$ is an $N$-dimensional vector.
For $N=5$, by choosing $K$ and $B$ as specified below:
\begin{align}
\frac{1}{2}K =
\begin{pmatrix}
1 & 1 & 1 & 1 & 0 \\
1 & 2 & 2 & 2 & \tfrac{1}{2} \\
1 & 2 & 3 & 3 & 1 \\
1 & 2 & 3 & 4 & \tfrac{3}{2} \\
0 & \tfrac{1}{2} & 1 & \tfrac{3}{2} & 1
\end{pmatrix}, 
\qquad
B =(1,2,3,4,2)\,,
\label{eq:CSlevel2}
\end{align}  
 we have found experimentally that the resulting Nahm sum agrees with the vacuum character \cite{andrews1998a2b} of the $(3,8)$ $W_3$ minimal model up to 
 a high order in $q$:\footnote{After uploading the first version to arXiv, we are informed by S. O. Warnaar that \eqref{eq:nahm} can be derived from Eq.~(7.6) of \cite{Warnaar:2023a2BaileyTree}. We are grateful to him for this remark.}
\begin{align}
 &\sum_{n_1,\ldots,n_5=0}^{\infty} 
 \frac{q^{\frac{1}{2}\sum_{a,b=1}^{5}  K_{a b} n_a n_b + \sum_{a=1}^4 a n_a+2n_5}}
      {\prod_{a=1}^{5} (q;q)_{n_a}}
 \nonumber \\
 &    \qquad  =
 1 + q^2 + 2 q^3 + 3 q^4 + 4 q^5 + 7 q^6 + 8 q^7 + 14 q^8 + 18 q^9 + 
 26 q^{10} + 34 q^{11}   +\cdots\,.
 \label{eq:nahm}
\end{align}

\subsubsection*{CS matter theory}
Once  the Nahm sum \eqref{eq:nahm} is given,
it is straightforward to read off the gauge group, CS levels, and chiral multiplet gauge charges of the corresponding CS matter theory. 
The gauge group is $U(1)^5$, the CS level matrix is given by \eqref{eq:CSlevel2}, and the five chiral multiplets $\{\phi_i\}_{i=1}^5$ carry diagonal gauge charges $Q_{a, i} = \delta_{ai}$ for  $a, i=1,\cdots, 5$ with the R-charge zero.
Without superpotential, the half index with the $(\mathcal{D}, D_c)$ boundary condition is written as
\begin{align}
I\! \! I_{(\mathcal{D}, D_c)}
&= \frac{1}{(q;q)^5_{\infty}} \sum_{n_1,\ldots,n_5 \in \mathbb{Z}} 
 q^{\frac{1}{2}\sum_{a,b=1}^{5}  K_{a b} n_a n_b }  \prod_{a=1}^5 x^{n_a}_a  
  (q^{1-n_a};q)_{\infty}
 \,,
\nonumber \\
&= \sum_{n_1,\ldots,n_5=0}^{\infty} 
 \frac{q^{\frac{1}{2} \sum_{a,b=1}^{5} K_{a b} n_a n_b } } 
      {\prod_{a=1}^{5} (q;q)_{n_a}} \Bigl( \prod_{a=1}^5 x^{-n_a}_a \Bigr) \,,
 \label{eq:nahmA2A4}
\end{align}
where $K_{ab}$ is given by \eqref{eq:CSlevel2} and $x_a$ $(a=1,\cdots, 5)$ is the the fugacity for the topological symmetry 
for the $a$-th $U(1)$ gauge group. Next we introduce a superpotential deformation that breaks the $U(1)^5$ topological symmetry down to
a single $U(1)$ generated by a  linear combination of the topological symmetries.
From the formulas \eqref{eq:gaugeV} and \eqref{eq:RchargeV}, we find that 
the following superpotential is gauge invariant:
\begin{align}
  W
  = V_{m^{(1)}} + V_{m^{(2)}} + V_{m^{(3)}}+\phi_5 V_{m^{(4)}} \,,
\end{align}
where the magnetic charges of the bare monopole operators $V_{m^{(l)}}$ for $l=1,2,3,4$ are given by
\begin{equation}
\begin{aligned}
  m^{(1)} &= (-1, 2, -1, 0, 0)\,, \\
  m^{(2)} &= (0, -1, 2, -1, 0)\,, \\
  m^{(3)} &= (-1, 0, 1, 2, -2)\,, \\
  m^{(4)} &= (2, -1, 0, 0, 0)\,.
\end{aligned}
\label{eq:m_vectors}
\end{equation}
The following linear combination of the topological $U(1)$ symmetries:
\begin{align}
T \equiv  \sum_{a=1}^4 a\, J^{(a)} + 2 J^{(5)} \,,
\label{eq:topnahm}
\end{align}
leaves the superpotential invariant and the single  remained global symmetry.
Here $J^{(a)}$ denotes the generator of the topological $U(1)$ symmetry associated with the $a$-th gauge group.
Then the fugacities for the $U(1)^5$ topological symmetries is restricted to 
\begin{align}
  x_a = \eta^{a} \quad (a = 1,2,3,4), 
  \qquad x_5 = \eta^{2} \,.
\end{align}
If we take   $\eta=q^{-2}$, the half index, which means the R-charge is shifted by $R \to R_{\rm shift}= R - 2 T$ in the definition of the half index,
\eqref{eq:nahmA2A4}  reproduces the Nahm sum formula \eqref{eq:nahm} for the vacuum character. 

Next we give the identification between the $U(1)$ R-charge and topological charges in \eqref{eq:RchargeV} and \eqref{eq:topnahm} 
and the Cartan generators of the IR $\mathcal{N}=4$ $SU(2)_H \times SU(2)_C$ 
R-symmetry as follows.
Since the half index for a 3d $\mathcal{N}=2$ theory with the $(\mathcal{D}, D_{c})$ boundary condition is engineered to 
reproduce \eqref{eq:nahm}, i.e.\ the  Schur index and hence the vacuum character of the corresponding VOA, the half index also agrees with the vacuum character of the VOA  that appears in the 4d SCFT/VOA correspondence. 
On the other hand, the H-twisted (a.k.a. A-twisted) half index for 3d $\mathcal{N}=4$ theories agrees with the vacuum character of 
the VOA that appears at the 2d boundary of spacetime. Thus, when a 3d $\mathcal{N}=2$ theory flows in the IR to a 3d $\mathcal{N}=4$ SCFT \cite{Costello:2018fnz},
the half index of the 3d $\mathcal{N}=2$ theory should be identified with the H-twisted half index $I\!\!I^{H}$ of the IR $\mathcal{N}=4$ SCFT \cite{Gaiotto:2024ioj}:
\begin{align}
I\! \! I_{(\mathcal{D}, D_c)}(q)=I\! \! I^{H}(q) \equiv \mathrm{Tr}(-1)^{J_C+J_{H}} q^{J_3+J_{H}}\,.
\end{align}
Here $J_{H}$ and $J_{C}$ are the Cartan generators of the $SU(2)_H \times SU(2)_C$ R-symmetry. 
Then the linear combination of topological charges $T$, the $\mathcal{N}=2$ R-charge $R$ and an $\mathcal{N}=4$ R-charge $J_H$  are related to
\begin{align}
R-2 T=2 J_{H}\,.
\end{align}
To determine this relation more precisely, we consider a refinement of the 3d $\mathcal{N}=4$ twisted half index by including the fugacity for $J_H-J_C$:
\begin{align}
I\! \! 
I^{H}(q,t) \equiv \mathrm{Tr}(-1)^{J_C+J_{H}} q^{J_3+J_H} t^{J_C-J_H}\,.
\label{eq:refinedhalf}
\end{align} 
Recently, it was proposed \cite{Kim:2025klh} that this refined  H-twisted index, $I\! \!I^{H}(q,t)$ agrees with the Macdonald index $\mathcal{I}^{\rm 4d}(q,t)$ of the corresponding 4d theory:
\begin{align}
I\!\!I^{H}(q, t)=\mathcal{I}^{\rm 4d} (q, t)
\end{align}
where
\begin{align}
\mathcal{I}^{\rm 4d} (q, t)& \equiv \mathrm{Tr}(-1)^{F} q^{E-R_{\rm 4d}} t^{R_{\rm 4d}-r_{\rm 4d}}\,.
\label{eq:macd}
\end{align}
Here $R_{\rm 4d}$ and $r_{\rm 4d}$ are the generators of 4d $SU(2) \times U(1)$ R-symmetry.
Note that, when $t=1$, \eqref{eq:refinedhalf} and \eqref{eq:macd} reduce to the H-twisted index and the Schur index, respectively.

The 3d $\mathcal{N}=2$ theory has a single $U(1)$ global symmetry generated by \eqref{eq:topnahm}, and this symmetry 
should be identified with the one generated by $J_C - J_H$. Indeed, if the fugacity is chosen to be $\eta=(qt)^{-1}$,
we find that $I\! \! I_{(\mathcal{D}, D_c)}$ reproduces the Macdonald index for the $(A_2, A_4) \simeq (A_1,E_{8})$ Argyres--Douglas theory\footnote{A similar 
refinement of the Nahm sum for $(3,7)$ $W_3$ minimal model was  proposed in \cite{Foda:2019guo}.}:
\begin{align}
I\! \! I_{(\mathcal{D}, D_c)}
&= \sum_{n_1,\ldots,n_5=0}^{\infty} 
 \frac{q^{\frac{1}{2} \sum_{a,b=1}^{5} K_{a b} n_a n_b } } 
      {\prod_{a=1}^{5} (q;q)_{n_a}} (q t)^{\sum_{a=1}^4a n_a +2 n_5} \nonumber \\
&=1+ t q^2+
   \left(t+ t^2\right)q^3+
   \left(t+ 2 t^2\right)q^4+
   \left(t+2t^2+ t^3\right)q^5
   + \left(t+3 t^2+ 3t^3\right)q^6+\cdots
\end{align}
from which we obtain the identification of generators
$2 T = J_C - J_{H}$.
Therefore we have detected the following relation between the UV $U(1)$ charges and the IR ones:
\begin{align}
R= J_{H}+J_C\,, \quad 2 T= J_C-J_{H}.
\end{align}

Finally, we compute the superconformal index:
\begin{align}
I_{\rm SCI}
&= \sum_{n_a \in \mathbb{Z}} \sum_{a=1}^5
\oint \prod_{a=1}^4 \frac{d z_a}{2\pi {\rm i} z_a}\,
(-q^{-\frac{1}{2}} \eta)^{\sum_{a=1}^4 a n_a + 2 n_5}
\prod_{a,b=1}^4 z_{a}^{K_{ab} n_b}
\nonumber \\
&\qquad \times \prod_{a=1}^5
(-q^{-\frac{1}{2}} z_a)^{-\frac{n_a+|n_a|}{2}}
\frac{(z^{-1}_a q^{1+\frac{|n_a|}{2}};q)_{\infty}}{(z_a q^{\frac{|n_a|}{2}};q)_{\infty}}
\nonumber \\
&= 1 - q
  + \left( 1 + \eta^2 + \frac{1}{\eta^2} \right) q^{2}
  + \left( 3 \eta + \frac{3}{\eta} \right) q^{\frac{5}{2}}
  + \cdots \,.
\end{align}
This expansion agrees with the superconformal index obtained in \cite{Kim:2025klh} from a $U(1)^3$ CS matter theory associated with an alternative expression for the vacuum character of the $(3,8)$ $W_3$ minimal model.

\section{From $(A_3,A_{N-1})$ to a conjectural formula for $(A_{M-1},A_{N-1})$}
\label{sec:A3}

In this section we determine, by following the same procedure as in the previous section for the $(A_2,A_{N-1})$ case,
a 3d $\mathcal{N}=2$ CS matter theory that flows to the 3d $\mathcal{N}=4$ SCFT associated with the twisted $S^1$ reduction of the $(A_3,A_{N-1})$ theory with ${\rm gcd}(4,N)=1$.

\subsection{Schur index for $(A_{3}, A_{N-1})$}
From the analysis in the section \ref{sec:KS}, the half monodromy operator for $(A_3,A_{N-1})$ theory is  given by \eqref{eq:S-A3AN}.
Again the Schur index \eqref{eq:4Dindex} is evaluated in terms of the commutation relation of quantum torus algebra, the series expansion of 
$q$-exponentials and the Dirac product.

 \begin{align}
S(q)&=\Bigl( \prod_{j:{\rm odd} \atop i:{\rm even}} E_q(X_{\gamma^{(1)}_j}) E_q(X_{\gamma^{(2)}_i}) E_q(X_{\gamma^{(3)}_j}) \Bigr)
\Bigl( \prod_{j:{\rm odd} \atop i:{\rm even}} E_q(X_{\gamma^{(1)}_i+\gamma^{(2)}_i}) E_q(X_{\gamma^{(1)}_j+\gamma^{(2)}_j+\gamma^{(3)}_j}) E_q(X_{\gamma^{(2)}_i+\gamma^{(3)}_i}) \Bigr) \nonumber \\
& \times \Bigl( \prod_{j:{\rm odd} \atop i:{\rm even}} E_q(X_{\gamma^{(1)}_j+\gamma^{(2)}_j}) E_q(X_{\gamma^{(1)}_i+\gamma^{(2)}_i+\gamma^{(3)}_i}) E_q(X_{\gamma^{(2)}_j+\gamma^{(3)}_j}) \Bigr)
\Bigl( \prod_{j:{\rm odd} \atop i:{\rm even}} E_q(X_{\gamma^{(1)}_i}) E_q(X_{\gamma^{(2)}_j}) E_q(X_{\gamma^{(3)}_i}) \Bigr)
\end{align}
The Dirac product of charge vectors satisfy the following relations:
\begin{align}
\langle \gamma^{(1)}_i, \gamma^{(1)}_j \rangle&=\langle \gamma^{(3)}_i, \gamma^{(3)}_j \rangle=(-1)^{i+1}  (\delta_{i+1,j} + \delta_{i,j+1}), \\
\langle \gamma^{(2)}_i, \gamma^{(2)}_j \rangle&=(-1)^{i}  (\delta_{i+1,j} + \delta_{i,j+1}), \\
\langle \gamma^{(1)}_i, \gamma^{(2)}_j \rangle&=(-1)^{i}  \delta_{i,j}\,, \\
\langle \gamma^{(2)}_i, \gamma^{(3)}_j \rangle&=(-1)^{i+1}  \delta_{i,j}\,. 
\end{align}
After an elementary but somewhat involved computation, we obtain the following expression of $S(q)$ and  
 the Schur index:
 \begin{align}
S(q) 
&=
\sum_{i=1}^{N-1} \sum_{n^{(1)}_{i}, n^{(2)}_{i}, n^{(3)}_{i}=0}^{\infty} \sum_{k^{(1)}_{i}, k^{(2)}_{i}, k^{(3)}_{i} \in D} 
 q^{\frac{A}{2}} (-q^{\frac{1}{2}})^{\sum_{i=1}^{N-1}( \sum_{l=1}^3 n^{(l)}_{i}-k^{(1)}_{i} -2k^{(2)}_{i} - k^{(3)}_{i} )}
 \nonumber \\
 & \times \frac{1}
 { \prod_{i=1}^{N-1} (q)_{n^{(1)}_{i}-k^{(1)}_{i}-k^{(2)}_{i}}  (q)_{n^{(2)}_{i}-k^{(1)}_{i}- k^{(2)}_{i} - k^{(3)}_{i} }
 (q)_{n^{(3)}_{i}-k^{(2)}_{i} -k^{(3)}_{i}} \prod_{l=1}^3(q)_{ k^{(l)}_{i}   }}
  X_{\sum_{l=1}^{3}\sum_{i=1}^{N-1} n^{(l)}_{i} \gamma^{(l)}_{i}}
\end{align}
and 
\begin{align}
\mathcal{I}^{\rm 4d}
&=(q)_{\infty}^{3(N-1)}
\sum_{i=1}^{N-1} \sum_{l=1}^3 \sum_{n^{(l)}_{i} =0}^{\infty} \sum_{k^{(l)}_{i} \in D}
\sum_{\tilde{k}^{(l)}_{i} \in \tilde{D}} q^{\frac{A+\tilde{A}}{2}} (-q^{\frac{1}{2}})^{\sum_{i=1}^{N-1}( 2 \sum_{l=1}^3 n^{(l)}_{i}-k^{(1)}_{i} -2 k^{(2)}_{i} - k^{(3)}_{i}
 -\tilde{k}^{(1)}_{i} -2 \tilde{k}^{(2)}_{i} - \tilde{k}^{(3)}_{i}) }
\nonumber \\
&\qquad \times
 \frac{1}{ \prod_{i=1}^{N-1} (q)_{n^{(1)}_{i}-k^{(1)}_{i}-k^{(2)}_{i}}  
 (q)_{n^{(2)}_{i}- k^{(1)}_{i} - k^{(2)}_{i} - k^{(3)}_{i}} (q)_{n^{(3)}_{i}-k^{(2)}_{i} - k^{(3)}_{i}} 
 \prod_{l=1}^3 (q)_{k^{(l)}_{i}}} 
 \nonumber \\
 & \qquad \times \frac{1}
 {
\prod_{i=1}^{N-1}  (q)_{n^{(1)}_{i}-\tilde{k}^{(1)}_{i}-\tilde{k}^{(2)}_{i}}  (q)_{n^{(2)}_{i}- \tilde{k}^{(1)}_{i} - \tilde{k}^{(2)}_{i} - \tilde{k}^{(3)}_{i}} (q)_{n^{(3)}_{i}-\tilde{k}^{(2)}_{i} - \tilde{k}^{(3)}_{i}}
 \prod_{l=1}^3 ( q)_{\tilde{k}^{(l)}_{i}}}
 \label{eq:A3AnSchur}
 \end{align}
Here $A$ and $\tilde{A}$ are defined by
\begin{align}
A & \equiv \sum_{i=1}^{N-2} \Bigl( \sum_{l=1}^3 \Bigl[ n^{(l)}_{i} n^{(l)}_{i+1} - 2 k^{(l)}_{i} k^{(l)}_{i+1}  \Bigr] -2\sum_{l=1}^2 \Bigl[ k^{(l)}_i k^{(l+1)}_{i+1}+ k^{(l+1)}_i k^{(l)}_{i+1}\Bigr]\Bigr)
\nonumber \\
&\qquad
+ 
+\sum_{i=1}^{N-1} \Bigl( -\sum_{l=1}^2 n^{(l)}_i n^{(l+1)}_{i}
+ \sum_{1 \le l < l^{\prime} \le 3}  k^{(l)}_i k^{(l^{\prime})}_i   
+(k^{(1)}_i)^2+2(k^{(2)}_i)^2+(k^{(3)}_i)^2
\Bigr)\,,\\
\tilde{A}& \equiv A|_{k^{(l)}_i \to \tilde{k}^{(l)}_i }\,,
\end{align}
and $D$,  $\tilde{D}$ are  defined by
\begin{align}
 D& \equiv \{ ( k^{(1)}_i, k^{(2)}_i, k^{(3)}_i) \in \mathbb{Z}^{3N}_{\ge 0} |  
k^{(1)}_i + k^{(2)}_i \le  n^{(1)}_i, 
k^{(1)}_i + k^{(2)}_i + k^{(3)}_i \le n^{(2)}_i, k^{(2)}_i + k^{(3)}_i \le n^{(3)}_i \}, \\   
 \tilde{D}& \equiv D |_{k^{(l)}_i \to \tilde{k}^{(l)}_i}\,.
\end{align}
Note that \eqref{eq:A3AnSchur} should agree with the vacuum character of the 
$(4,N+4)$ $W_{4}$ minimal model. As mentioned in the end of sub-section~\ref{sec:schurA2An}, 
we simply introduce by hand a cutoff in the sum over magnetic charges in the ill-defined expression \eqref{eq:A3AnSchur}, 
and observe that the resulting expression \eqref{eq:A3AnSchur} for $N=2,4$ reproduces the first few terms of the vacuum character of the $W_{4}$ minimal model.

\subsection{Half index and CS matter theory}
As before, we determine the gauge group and matter content of a 3d
$\mathcal{N}=2$ CS matter theory in such a way that its half index reproduces the Schur index \eqref{eq:4Dindex-A2A3M-2}.
We consider a CS matter theory with gauge group $U(1)^{9(N-1)}$ coupled to $12(N-1)$ chiral multiplets. 
The field content and the corresponding quiver diagram are summarized in Table~\ref{table:chiralA3An} and in Figure~\ref{fig:quiver2}.
We denote the gauge group as
\begin{align}
  U(1)^{9(N-1)}
  = \prod_{l=1}^3 \prod_{i=1}^{N-1} U(1)_{x^{(l)}_i}
    \times \prod_{l=1}^3 \prod_{i=1}^{N-1} U(1)_{y^{(l)}_i}
    \times \prod_{l=1}^3 \prod_{i=1}^{N-1} U(1)_{\tilde{y}^{(l)}_i}\,,
\end{align}
and label the $12(N-1)$ chiral multiplets by
$\phi^{(l)}_i, \tilde{\phi}^{(l)}_i, \varphi^{(l)}_i, \tilde{\varphi}^{(l)}_i$ (for $l=1,2,3$ and $i=1,\ldots,N-1$). 
We assume that the UV (or reference) R-charges of all these chiral multiplets are zero.

The gauge Chern--Simons levels $K_{ab}$ are chosen to be
\begin{align}
  K_{ab} \equiv \tilde{K}_{ab} + \tilde{K}^T_{ab}\,,
\end{align}
with
\begin{align}
\tilde{K}_{ab} \equiv
\left\{
\begin{array}{ll}
  \delta_{a, b-1} & \text{if } a=(N-1)(l-1)+i,\; i=1,\cdots, N-2,\; l=1,2,3\,,\\[2pt]
  -\delta_{a, b-1} & \text{if } a=(N-1)l+i,\; i=1,\cdots, N-2,\; l=3,4,5,6,7,8\,,\\[2pt]
  -\delta_{a, b-N} & \text{if } a=(N-1)l+i,\; i=1,\cdots, N-2,\; l=3,4,6,7\,,\\[2pt]
  -\delta_{a-N, b} & \text{if } a=(N-1)l+i+1,\; i=1,\cdots, N-2,\; l=3,4,6,7\,,\\[2pt]
  \dfrac{1}{2}\delta_{a, b} & \text{if } a=(N-1)l+i,\; i=1,\cdots, N-1,\; l=3,5,6,8\,,\\[2pt]
  \delta_{a, b} & \text{if } a=(N-1)l+i,\; i=1,\cdots, N-1,\; l=4,7\,,\\[2pt]
  -\delta_{a, b+(N-1)} & \text{if } a=(N-1)+i,\; i=1,\cdots, N-1\,,\\[2pt]
  \delta_{a, b+(N-1)} & \text{if } a=(N-1)l+i,\; i=1,\cdots, N-1,\; l=3,4,6,7\,,\\[2pt]
  \delta_{a, b+2(N-1)} & \text{if } a=(N-1)l+i,\; i=1,\cdots, N-1,\; l=3,6\,,\\[2pt]
  0 & \text{otherwise}. 
\end{array}
\right.
\label{eq:CSlevelA3An}
\end{align}
Note that the CS level satisfies  the following 
relation:
\begin{align}
\sum_{a, b=1}^{9(N-1)} K_{a b} \mathfrak{n}_a \mathfrak{n}_b = A+\tilde{A}\,.
\label{eq:CSA3An}
\end{align}
Here we regard $\mathfrak{n}=(\mathfrak{n}_1, \cdots, \mathfrak{n}_{9(N-1)}) \in \mathbb{Z}^{9(N-1)}$ as the vector obtained by concatenating
the nine $(N-1)$-component vectors $n^{(l)}:=(n^{(l)}_1,\dots,n^{(l)}_{N-1}),
k^{(l)}:= (k^{(l)}_1,\dots,k^{(l)}_{N-1}),
\tilde{k}^{(l)}:=(\tilde{k}^{(l)}_1,\dots,\tilde{k}^{(l)}_{N-1}), l=1,2,3,$ in the following order:
$   
n^{(1)}\,, n^{(2)}\,, n^{(3)}\,, k^{(1)}\,, k^{(2)}\,, k^{(3)}\,, \tilde{k}^{(1)}\,, \tilde{k}^{(2)}\,, \tilde{k}^{(3)}\,.
$
Then  the half index of $U(1)^{9(N-1)}$ 3d CS matter theory with the chiral multiplets depicted by Table \ref{table:chiralA3An}  and the CS level $K_{ab}$ \eqref{eq:CSlevelA3An} is evaluated as  
\begin{align}
  I\! \! I_{(\mathcal{D}, D_c)}&=\frac{1}{(q)_{\infty}^{9(N-1)}}
\sum_{i=1}^{N-1} \sum_{l=1}^3 \sum_{n^{(l)}_{i}, k^{(l)}_{i}, \tilde{k}^{(l)}_{i} \in \mathbb{Z}}
q^{\frac{A+\tilde{A}}{2}} \Bigl( \prod_{l=1}^3 \prod_{i=1}^{N-1}(x^{(l)}_{i})^{n^{(l)}_{i}} (y^{(l)}_{i})^{k^{(l)}_{i}}
(\tilde{y}^{(l)}_{i})^{\tilde{k}^{(l)}_{i}} \Bigr)
 \nonumber \\
 & \times
 \Bigl( \prod_{i=1}^{N-1} (q^{1-(n^{(1)}_{i}-k^{(1)}_{i}-k^{(2)}_{i})})_{\infty}
 (q^{1-(n^{(2)}_{i}- k^{(1)}_{i} - k^{(2)}_{i} - k^{(3)}_{i})})_{\infty}
  (q^{1-(n^{(3)}_{i}-k^{(2)}_{i} - k^{(3)}_{i})})_{\infty} \prod_{l=1}^3 (q^{1-k^{(l)}_{i}})_{\infty} \Bigr)
 \nonumber \\
 &  \times 
\Bigl( \prod_{i=1}^{N-1} (q^{1-(n^{(1)}_{i}-\tilde{k}^{(1)}_{i}-\tilde{k}^{(2)}_{i})})_{\infty}  
  (q^{1-(n^{(2)}_{i}- \tilde{k}^{(1)}_{i} - \tilde{k}^{(2)}_{i} - \tilde{k}^{(3)}_{i})})_{\infty} 
  (q^{1-(n^{(3)}_{i}-\tilde{k}^{(2)}_{i} - \tilde{k}^{(3)}_{i})})_{\infty}
  \prod_{l=1}^3 (q^{1-\tilde{k}^{(l)}_{i}})_{\infty} \Bigr)
  \nonumber \\
  &=(q)_{\infty}^{3(N-1)}
\sum_{i=1}^{N-1} \sum_{l=1}^3 \sum_{n^{(l)}_{i} =0}^{\infty} \sum_{k^{(l)}_{i} \in D}
\sum_{\tilde{k}^{(l)}_{i} \in \tilde{D}} q^{\frac{A+\tilde{A}}{2}} 
\Bigl( \prod_{l=1}^3 \prod_{i=1}^{N-1}(x^{(l)}_{i})^{-n^{(l)}_{i}} (y^{(l)}_{i})^{-k^{(l)}_{i}} (\tilde{y}^{(l)}_{i})^{-\tilde{k}^{(l)}_{i}} \Bigr)
\nonumber \\
& \times
 \frac{1}{ \prod_{i=1}^{N-1} (q)_{n^{(1)}_{\alpha}-k^{(1)}_{i}-k^{(2)}_{i}}  (q)_{n^{(2)}_{i}- k^{(1)}_{i} - k^{(2)}_{i} - k^{(3)}_{i}} ( q)_{n^{(3)}_{i}-k^{(2)}_{i} - k^{(3)}_{i}} \prod_{l=1}^3 (q)_{k^{(l)}_{i}}}  
 \nonumber \\
 & \times \frac{1}
 {\prod_{i=1}^{N-1}
  (q)_{n^{(1)}_{i}-\tilde{k}^{(1)}_{i}-\tilde{k}^{(2)}_{i}}  (q)_{n^{(2)}_{i}- \tilde{k}^{(1)}_{i} - \tilde{k}^{(2)}_{i} - \tilde{k}^{(3)}_{i}} (q; q)_{n^{(3)}_{i}-\tilde{k}^{(2)}_{i} - \tilde{k}^{(3)}_{i}}
 \prod_{l=1}^3 ( q)_{\tilde{k}^{(l)}_{i}}}\,.
 \label{eq:A3Anhalfind1}
 \end{align}
If the fugacities are chosen to the following value:
\begin{align}
x^{(l)}_{i} = q^{-1}, \quad y^{(l)}_i = \tilde{y}^{(l)}_i   = -q^{-\frac{1}{2}}, 
\label{eq:topspecA3An}
\end{align}
the half index \eqref{eq:A3Anhalfind1}  reproduces the Schur index for $(A_{3}, A_{N-1})$ theory \eqref{eq:A3AnSchur}.
Next we will write down  possible gauge invariant superpotential term,  which impose the  specialization of the fugacities \eqref{eq:topspecA3An}.

\subsection{Monopole superpotential  }

\begin{table}[thb]
\begin{center}
\begin{tabular}{c | ccc ccc ccc ccc }
& $\phi^{(1)}_{i}$  & $\phi^{(2)}_{i}$ & $\phi^{(3)}_{i}$  & $\tilde{\phi}^{(1)}_{i}$  & $\tilde{\phi}^{(2)}_{i}$ & $\tilde{\phi}^{(3)}_{i}$ & 
$\varphi^{(1)}_{i}$&  $\varphi^{(2)}_{i}$&  $\varphi^{(3)}_{i}$&  $\tilde{\varphi}^{(1)}_{i}$&  $\tilde{\varphi}^{(2)}_{i}$&  $\tilde{\varphi}^{(3)}_{i}$ \\ \hline
$U(1)_{x^{(1)}_{i}}$ & $1$  & $0$ & $0$            & $1$  & $0$ & $0$  & $0$  & $0$ & $0$ & $0$ & $0$ & $0$\\ 
$U(1)_{x^{(2)}_{i}}$ & $0$  & $1$ & $0$            & $0$ & $1$  & $0$  &$0$   & $0$ & $0$ & $0$ & $0$ & $0$\\ 
$U(1)_{x^{(3)}_{i}}$ & $0$  & $0$ & $1$             & $0$  & $0$ & $1$ &$0$   & $0$ & $0$ & $0$ & $0$ & $0$\\ 
$U(1)_{y^{(1)}_{i}}$ & $-1$  & $-1$ & $0$           & $0$ & $0$ & $0$  &$1$  & $0$ & $0$ & $0$ & $0$ & $0$\\ 
$U(1)_{y^{(2)}_{i}}$ & $-1$  & $-1$ & $-1$          & $0$ & $0$ & $0$  &$0$   & $1$ & $0$ & $0$ & $0$ & $0$\\ 
$U(1)_{y^{(3)}_{i}}$ & $0$  & $-1$ & $-1$          & $0$ & $0$ &  $0$ &$0$   & $0$ & $1$ & $0$ & $0$ & $0$\\ 
$U(1)_{\tilde{y}^{(1)}_{i}}$ & $0$  & $0$ & $0$  & $-1$ & $-1$ & $0$ &$0$  & $0$ & $0$ & $1$ & $0$ & $0$\\ 
$U(1)_{\tilde{y}^{(2)}_{i}}$ & $0$  & $0$ & $0$   & $-1$ & $-1$ & $-1$&$0$  & $0$ & $0$ & $0$ & $1$ & $0$ \\ 
$U(1)_{\tilde{y}^{(3)}_{i}}$ & $0$  & $0$ & $0$  & $0$ & $-1$ & $-1$ &$0$  & $0$ & $0$ & $0$ & $0$ & $1$ \\ 
\end{tabular} 
\caption{ The charge assignment for $i$-th set of gauge and matter multiplets in the CS matter theory associated with 
$(A_3,A_{N-1})$. The charge assignment  can be read off from the Schur index \eqref{eq:A3AnSchur}.
}
\label{table:chiralA3An}
\end{center}
\end{table}

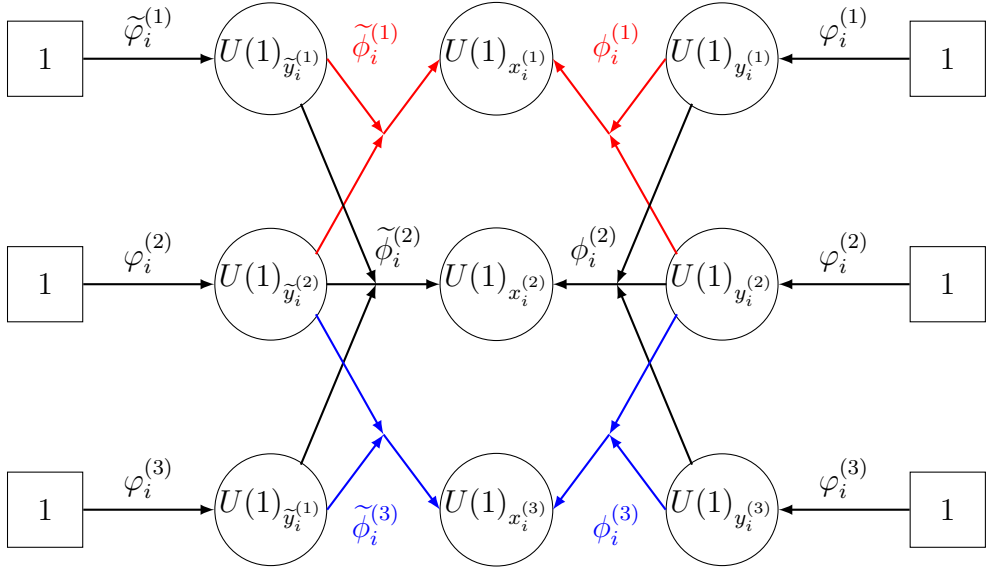
\begin{figure}[thb]
\centering
 \begin{tikzpicture}[vec/.style={thick,-latex,draw=black},gauge/.style={circle,draw=black,inner sep=0pt,minimum size=14mm},flavor/.style={rectangle,draw=black,inner sep=0pt,minimum size=10mm},auto]
  \node[gauge] (1) at (0,1.5) {$U(1)_{x^{(1)}_i}$};
  \node[gauge] (2) at (0,-1.5) {$U(1)_{x^{(2)}_i}$};
  \node[gauge] (3) at (0,-4.5) {$U(1)_{x^{(3)}_i}$};
  \node[gauge] (4) at (3,1.5) {$U(1)_{y^{(1)}_i}$};
  \node[gauge] (5) at (3,-1.5) {$U(1)_{y^{(2)}_i}$};
  \node[gauge] (6) at (3,-4.5) {$U(1)_{y^{(3)}_i}$};
  \node[gauge] (7) at (-3,1.5) {$U(1)_{\tilde{y}^{(1)}_i}$};
 \node[gauge] (8) at (-3,-1.5) {$U(1)_{\tilde{y}^{(2)}_i}$};
 \node[gauge] (9) at (-3,-4.5) {$U(1)_{\tilde{y}^{(1)}_i}$};
  \node[flavor] (10) at (6,1.5) {$1$};
  \node[flavor] (11) at (6,-1.5) {$1$};
  \node[flavor] (12) at (6,-4.5) {$1$};
  \node[flavor] (13) at (-6,1.5) {$1$};
  \node[flavor] (14) at (-6,-1.5) {$1$};
  \node[flavor] (15) at (-6,-4.5) {$1$};
%
  \draw[vec] (10) to node[above]{$\varphi^{(1)}_i$} (4);
  \draw[vec] (11) to node[above]{$\varphi^{(2)}_i$} (5);
  \draw[vec] (12) to node[above]{$\varphi^{(3)}_i$} (6);
  \draw[vec] (13) to node[above]{$\tilde{\varphi}^{(1)}_i$} (7);
  \draw[vec] (14) to node[above]{$\varphi^{(2)}_i$} (8);
  \draw[vec] (15) to node[above]{$\varphi^{(3)}_i$} (9);
    \draw[vec,red] (1.5, 0.5) -- (0.75, 1.5);
    \draw[vec,red] (2.25, 1.5) -- (1.5, 0.5);
    \draw[vec,red] (2.4, -1.1) -- (1.5, 0.5);
    \node[red] at (1.6,1.7) {$\phi^{(1)}_i$};
    \draw[vec,red] (-1.5, 0.5) -- (-0.75, 1.5);
    \draw[vec,red] (-2.25, 1.5) -- (-1.5, 0.5);
    \draw[vec,red] (-2.4, -1.1) -- (-1.5, 0.5);
    \node[red] at (-1.6,1.7) {$\tilde{\phi}^{(1)}_i$};
    \draw[vec] (2.6, 0.9) -- (1.6, -1.5);
    \draw[vec] (2.26, -1.5) -- (0.75, -1.5);
      \draw[vec] (2.6, -3.9) -- (1.6, -1.5);
    \node at (1.3,-1) {$\phi^{(2)}_i$};
    \draw[vec] (-2.6, 0.9) -- (-1.6, -1.5);
    \draw[vec] (-2.26, -1.5) -- (-0.75, -1.5);
      \draw[vec] (-2.6, -3.9) -- (-1.6, -1.5);
         \node at (-1.3,-1) {$\tilde{\phi}^{(2)}_i$};

    \draw[vec, blue] (1.5, -3.5) -- (0.75, -4.5);
    \draw[vec, blue] (2.4, -1.9) -- (1.5, -3.5);
    \draw[vec, blue] (2.25, -4.5) -- (1.5, -3.5);
    \node[blue] at (1.6,-4.7) {$\phi^{(3)}_i$};
    \draw[vec, blue] (-1.5, -3.5) -- (-0.75, -4.5);
    \draw[vec, blue] (-2.4, -1.9) -- (-1.5, -3.5);
    \draw[vec, blue] (-2.25, -4.5) -- (-1.5, -3.5);
    \node[blue] at (-1.6,-4.7) {$\tilde{\phi}^{(3)}_i$};
 \end{tikzpicture}
\bigskip
\caption{The quiver diagram for the $i$-th set of gauge and matter
 multiplets included in the CS matter theory for $(A_3,A_{N-1})$ for ${\rm gcd}(4,N)=1$. Each circle stands for a $U(1)$ gauge group. 
 Each junction of three or four arrows represents a chiral multiplet, with the color indicating which junctions correspond to the same chiral multiplet.
Each arrow from a box to a circle
 stands for a chiral multiplet $\varphi^{(l)}_i$ (resp. $\tilde{\varphi}^{(l)}_i$) charged under the $U(1)_{y^{(l)}_i}$ 
 (resp. $U(1)_{\tilde{y}^{(l)}_i}$) corresponding to
 the circle. The subscript of $U(1)$ in each circle stands for the
 fugacity for the corresponding topological symmetry.
 An incoming (outgoing) arrow attached to a circle indicates that the corresponding chiral multiplet has gauge charge $+1$ ($-1$) under the $U(1)$ gauge group associated with that circle.
 }
\label{fig:quiver3}
\end{figure}

\begin{table}[h]
\begin{center}
\begin{tabular}{c | ccc ccc ccc ccc }
& $V^{(1)}_{i}$  & $V^{(2)}_{i}$ & $V^{(3)}_{i}$  & $V^{(4)}_{i}$  & $V^{(5)}_{i}$ & $V^{(6)}_{i}$ &  $\tilde{V}^{(1)}_{i}$&  $\tilde{V}^{(2)}_{i}$&  $\tilde{V}^{(3)}_{i}$&  $\tilde{V}^{(4)}_{i}$&  $\tilde{V}^{(5)}_{i}$&  $\tilde{V}^{(6)}_{i}$ \\ \hline
$U(1)_{x^{(1)}_{i}}$ & $-1$  & $0$ & $0$            & $0$  & $0$ & $0$  & $-1$  & $0$ & $0$ & $0$ & $0$ & $0$\\ 
$U(1)_{x^{(2)}_{i}}$ & $0$  & $-1$ & $0$            & $0$ & $0$  & $0$  &$0$   & $-1$ & $0$ & $0$ & $0$ & $0$\\ 
$U(1)_{x^{(3)}_{i}}$ & $0$  & $0$ & $-1$             & $0$  & $0$ & $0$ &$0$   & $0$ & $-1$ & $0$ & $0$ & $0$\\ 
$U(1)_{y^{(1)}_{i}}$ & $-1$  & $0$ & $0$           & $0$ & $1$ & $-1$  &$0$  & $0$ & $0$ & $0$ & $0$ & $0$\\ 
$U(1)_{y^{(2)}_{i}}$ & $0$  & $-1$ & $0$          & $1$ & $-1$ & $1$  &$0$   & $0$ & $0$ & $0$ & $0$ & $0$\\ 
$U(1)_{y^{(3)}_{i}}$ & $0$  & $0$ & $-1$          & $-1$ & $1$ &  $0$ &$0$   & $0$ & $0$ & $0$ & $0$ & $0$\\ 
$U(1)_{\tilde{y}^{(1)}_{i}}$ & $0$  & $0$ & $0$  & $0$ & $0$ & $0$ &$-1$  & $0$ & $0$ & $0$ & $1$ & $-1$\\ 
$U(1)_{\tilde{y}^{(2)}_{i}}$ & $0$  & $0$ & $0$   & $0$ & $0$ & $0$&$0$  & $-1$ & $0$ & $1$ & $-1$ & $1$ \\ 
$U(1)_{\tilde{y}^{(3)}_{i}}$ & $0$  & $0$ & $0$  & $0$ & $0$ & $0$ &$0$  & $0$ & $-1$ & $-1$ & $1$ & $0$ \\ 
\end{tabular} 
\caption{ The monopole charge  
for the $i$-th set of monopole operators included in the CS matter theory for $(A_3,A_{N-1})$ for ${\rm gcd}(4,N)=1$.
}
\label{tabel:monopoleA3An}
\end{center}
\end{table}

We consider monopole operators $V^{(l)}_{i}$ and $\tilde{V}^{(l)}_{i}$ for $l=1,2,3$ and $i=1,\cdots, N-1$, depicted by Table \ref{tabel:monopoleA3An}. The gauge charge and  R-charge are again computed by the formulas \eqref{eq:gaugeV} and \eqref{eq:RchargeV}.
Then we find that the following dressed monopole operators are 
gauge invariant:
\begin{align}
&\phi^{(l)}_{i-1} V^{(l)}_i \phi^{(l)}_{i+1}, \quad   
\tilde{\phi}^{(l)}_{i-1} \tilde{V}^{(l)}_i \tilde{\phi}^{(l)}_{i+1}, \quad  l=1,2,3 \, \text{ and } \, i=1,2,\cdots,N-1\,.
\label{eq:dressmonoA3An}
\end{align}
Here we set  $\phi^{(l)}_i=\tilde{\phi}^{(l)}_i=\varphi^{(l)}_i=\tilde{\varphi}^{(l)}_i=1$ for $i=0,N$.

We define a shifted $U(1)$ R-symmetry as the following linear combination of the reference R-charge and the generators of the topological $U(1)$ symmetries:
\begin{align}
R_{\rm shift}=R -\sum_{i=1}^{N-1} \Bigl[ 2\sum_{l=1}^3 J_{x^{(l)}_{i}}
-\sum_{l=1,3}
(J_{y^{(l)}_{i}}+{J}_{\tilde{y}^{(l)}_{i}})
-2(J_{y^{(2)}_{i}}+{J}_{\tilde{y}^{(2)_{i}}})
\Bigr]\,.
\label{eq:Rshifta3an}
\end{align}
Here $J_{x^{(l)}_{i}}, {J}_{{y}^{(l)}_{i}}, {J}_{\tilde{y}^{(l)}_{i}}$ are the generator of topological symmetries associated 
with gauge groups $U(1)_{x^{(l)}_{i}}, U(1)_{{y}^{(l)}_{i}}, U(1)_{\tilde{y}^{(l)}_{i}}$, respectively.
With respect to this assignment, each term in \eqref{eq:dressmonoA3An} carries charge $2$. 
Thus the gauge-invariant dressed monopole operators \eqref{eq:dressmonoA3An} are natural candidates for superpotential terms. 
In the next section we will show that 
appropriate linear combinations of these operators that are compatible with the single $U(1)$ flavor symmetry.

\subsection{Residual symmetry}

As in the analysis of Section~\ref{sec:ressymA2An} for the $(A_2, A_{N-1})$ case, we will show that 
the $3(N-1)$ flavor symmetries are broken down to a single $U(1)$  by
turning on an appropriate monopole superpotential.

Up to gauge equivalence, the $U(1)^{3(N-1)}$ flavor symmetry acts on the
chiral multiplets as
\begin{equation}
 \tilde \phi_k^{(l)} \to e^{i\zeta^{(l)}_k} \tilde \phi^{(l)}_k
\end{equation}
for $l=1,2,3$, with all the other chiral multiplets kept fixed. Under
this transformation, the monopole operators transform as
\begin{align}
 V_k^{(1)} &\to e^{-\frac{i\zeta^{(1)}_k}{2}}V_\alpha^{(1)}~,\qquad
  \tilde V^{(1)}_\alpha \to e^{-\frac{i\zeta^{(2)}_\alpha}{2}}\tilde
  V_k^{(1)}~,
\\
V_k^{(2)} &\to
 e^{-\frac{i\zeta^{(2)}_k}{2}}V_k^{(2)}~,\qquad \tilde
 V_k^{(2)} \to e^{-\frac{i(\zeta^{(1)}_k +
 \zeta^{(3)}_k)}{2}}\tilde V_k^{(2)}~,
\\
V_k^{(3)} &\to
 e^{-\frac{i\zeta_k^{(3)}}{2}}V_k^{(3)}~,\qquad \tilde
 V_k^{(3)} \to e^{-\frac{i\zeta_k^{(2)}}{2}}\tilde
 V_k^{(3)}~,
\\
\tilde V_k^{(4)} &\to e^{-\frac{i\zeta^{(1)}_k}{2}}\tilde
 V_k^{(4)}~,\qquad \tilde V_k^{(5)} \to
 e^{-\frac{i\zeta_k^{(2)}}{2}}\tilde V_k^{(5)}~,\qquad \tilde V_k^{(6)} \to
 e^{-\frac{i\zeta_k^{(3)}}{2}}\tilde V_k^{(6)}~,
\end{align}
with $V_k^{(4)},V_k^{(5)}$ and $V_k^{(6)}$ kept fixed.

On the other hand, the $U(1)^{9(N-1)}$ topological symmetry acts on the
monopole operators as
\begin{equation}
V_k^{(l)} \to e^{-i(\alpha^{(l)}_k+\beta^{(l)}_k)}V_k^{(l)}~,\qquad
  \tilde V_k^{(l)} \to e^{-i(\alpha^{(l)}_k+\tilde \beta^{(l)}_k)}\tilde
 V_k^{(l)}~,
\end{equation}
for $l=1,2,3$ and
\begin{align}
V_k^{(4)} &\to e^{+i(\beta^{(2)}_k-\beta^{(3)}_k)}V_k^{(4)}~,\qquad
 \tilde V_k^{(4)} \to e^{+i(\tilde \beta^{(2)}_k-\tilde \beta^{(3)}_k)}\tilde
 V_k^{(4)}~,
\\
V_k^{(5)} &\to
 e^{+i(\beta^{(1)}_k-\beta^{(2)}_k+\beta^{(3)}_k)}V_k^{(5)}~,\qquad \tilde
 V_k^{(5)} \to e^{+i(\tilde \beta^{(1)}_k - \tilde \beta^{(2)}_k  + \tilde
 \beta^{(3)}_k)}\tilde V_k^{(5)}~,
\\
V_k^{(6)} &\to e^{+i(-\beta^{(1)}_k+\beta^{(2)}_k)}V_k^{(6)}~,\qquad
 \tilde V_k^{(6)} \to e^{+i(-\tilde \beta^{(1)}_k + \tilde
 \beta^{(2)}_k)}\tilde V_k^{(6)}~,
\end{align}
where $\alpha_k^{(l)}$ and $\beta_k^{(l)}$ are phase rotations
corresponding to the topological symmetry.

The superpontential terms we consider here are the following:
\begin{align}
W &= W_1 + W_2~,
\\[1mm]
 W_1 &=
 \sum_{l=1}^3\sum_{k=1}^{N-1}\left(\phi^{(l)}_{k-1}V_k^{(l)}\phi_{k+1}^{(l)}
 + \tilde \phi_{k-1}^{(l)} \tilde V_k^{(l)} \tilde
 \phi_{k+1}^{(l)} \right)~,
\label{eq:M=4_W1}
\\
W_2 &=
 \sum_{l=1}^3\sum_{k=1}^{N-1}\left(\varphi_{k-1}^{(l)}V_k^{(l+3)}\varphi^{(l)}_{k+1}
 + \tilde \varphi_{k-1}^{(l)}\tilde V_k^{(l+3)}\tilde \varphi_{k+1}^{(l)}\right)~,
\end{align}
where we defined $\phi_0^{(l)} = \tilde \phi_0^{(l)} = \varphi_0^{(l)}
= \tilde \varphi_0^{(l)} = \phi_{N}^{(l)} = \tilde \phi_N^{(l)} =
\varphi_N^{(l)} = \tilde \varphi_N^{(l)} = 0$.
Under the flavor and topological symmetry transformations, they
transform as
\begin{align}
 W_1 &\to
 \sum_{l=1}^3\sum_{k=1}^{N-1}\left(e^{i\left(-\frac{\zeta_k^{(l)}}{2}
 - \alpha^{(l)}_k -
 \beta_k^{(l)}\right)}\phi^{(l)}_{k-1}V_k^{(l)}\phi_{k+1}^{(l)}\right)
\nonumber\\
&\qquad 
 + e^{i\left(\zeta_{k-1}^{(1)}+\zeta_{k+1}^{(1)}-\frac{ \zeta_k^{(2)}}{2}
 -  \alpha^{(1)}_k - \tilde \beta_k^{(1)}\right)}\tilde \phi_{k-1}^{(1)} \tilde V_k^{(1)} \tilde
 \phi_{k+1}^{(1)}
\nonumber\\
&\qquad 
 + e^{i\left(\zeta_{k-1}^{(2)}+\zeta_{k+1}^{(2)}-\frac{ \zeta_k^{(1)}+\zeta_k^{(3)}}{2}
 - \alpha^{(2)}_k - \tilde \beta_k^{(2)}\right)}\tilde \phi_{k-1}^{(2)} \tilde V_k^{(2)} \tilde
 \phi_{k+1}^{(2)}
\nonumber\\
&\qquad 
+e^{i\left(\zeta_{k-1}^{(3)}+\zeta_{k+1}^{(3)}-\frac{ \zeta_k^{(2)}}{2}
 -  \alpha^{(3)}_k - \tilde \beta_k^{(3)}\right)}\tilde \phi_{k-1}^{(3)} \tilde V_k^{(3)} \tilde
 \phi_{k+1}^{(3)}~,
\\[2mm]
W_2 &\to  \sum_{k=1}^{N-1}\left( e^{i(\beta_k^{(2)} -\beta_k^{(3)})}\varphi_{k-1}^{(1)}V_k^{(4)}\varphi^{(1)}_{k+1}
 + e^{i(-\frac{\zeta_k^{(1)}}{2} + \tilde \beta_k^{(2)} - \tilde
 \beta_k^{(3)})}\tilde \varphi_{k-1}^{(1)}\tilde
 V_k^{(4)}\tilde \varphi_{k+1}^{(1)}\right)
\nonumber\\
&\qquad + \sum_{k=1}^{N-1}\left( e^{i(\beta_k^{(1)} -\beta_k^{(2)}+\beta_k^{(3)})}\varphi_{k-1}^{(2)}V_k^{(5)}\varphi^{(2)}_{k+1}
 + e^{i(-\frac{\zeta_k^{(2)}}{2} + \tilde \beta_k^{(1)} - \tilde
 \beta_k^{(2)}+\tilde \beta_k^{(3)})}\tilde \varphi_{k-1}^{(2)}\tilde
 V_k^{(5)}\tilde \varphi_{k+1}^{(2)}\right)
\nonumber\\
& \qquad + \sum_{k=1}^{N-1}\left( e^{i(-\beta_k^{(1)} +\beta_k^{(2)})}\varphi_{k-1}^{(3)}V_k^{(6)}\varphi^{(3)}_{k+1}
 + e^{i(-\frac{\zeta_k^{(3)}}{2} - \tilde \beta_k^{(1)} + \tilde
 \beta_k^{(2)})}\tilde \varphi_{k-1}^{(3)}\tilde
 V_k^{(6)}\tilde \varphi_{k+1}^{(3)}\right)~.
\end{align}
For these superpotential terms to be invariant, we need to
impose
\begin{align}
 \alpha_k^{(l)} &=
 -\frac{\zeta_k^{(l)}}{2}~,\qquad \beta_k^{(l)} = 0~,
\\
 \tilde \beta_k^{(1)} 
&= 
 \frac{\zeta_k^{(1)} + \zeta_k^{(2)}}{2}~,\qquad  \tilde \beta_k^{(2)} =
 \frac{\zeta_k^{(1)}+\zeta_k^{(2)}+\zeta_k^{(3)}}{2}~,
\qquad \tilde
 \beta_k^{(3)} 
= \frac{\zeta_k^{(2)}+\zeta_k^{(3)}}{2}~,
\\[2mm]
0&
 =
 \zeta_{k-1}^{(1)} + \zeta_{k+1}^{(1)}
 -\zeta_k^{(2)}~,\qquad 
0
 =
 \zeta_{k-1}^{(3)} + \zeta_{k+1}^{(3)} - \zeta_k^{(2)}
~,\qquad
0
 =
 \zeta_{k-1}^{(2)} + \zeta_{k+1}^{(2)} -
 \zeta_k^{(1)}-\zeta_k^{(3)}~.
\label{eq:zeta}
\end{align}
Note that all the parameters except for $\zeta_k^{(l)}$ are fixed
at this stage.
Then the constraints \eqref{eq:zeta} for
$k=1,2,\cdots,N-2$ imply that, for $k=1,2,\cdots,N-2$,
\begin{align}
 \zeta_k^{(1)}&=\left\{
\begin{array}{l}
\pm \zeta_1^{(1)} \qquad \text{when}\qquad k=8n\pm 1 \quad \text{for} \quad
 n\in \mathbb{N}
\\[2mm]
\pm \zeta_1^{(2)}\qquad \text{when}\qquad k=8n\pm 2 \quad \text{for} \quad
 n\in \mathbb{N}
\\[2mm]
\pm \zeta_1^{(3)}\qquad \text{when}\qquad k=8n\pm 3 \quad \text{for} \quad
 n\in \mathbb{N}
\\[2mm]
0 \qquad \text{when}\qquad k=4n \quad \text{for} \quad
 n\in \mathbb{N}
\end{array}
\right.~,
\\[5mm]
 \zeta_k^{(2)}&=\left\{
\begin{array}{l}
\pm \zeta_1^{(2)} \qquad \text{when}\qquad k=8n\pm 1 \quad \text{for} \quad
 n\in \mathbb{N}
\\[2mm]
\pm\left(\zeta_1^{(1)}+\zeta_1^{(3)}\right)\qquad \text{when}\qquad
k=8n\pm 2 \quad \text{for} \quad
 n\in \mathbb{N}
\\[2mm]
\pm \zeta_1^{(2)}\qquad \text{when}\qquad k=8n\pm 3 \quad \text{for} \quad
 n\in \mathbb{N}
\\[2mm]
0 \qquad \text{when}\qquad k=4n \quad \text{for} \quad
 n\in \mathbb{N}
\end{array}
\right.~,
\\[5mm]
 \zeta_k^{(3)}&=\left\{
\begin{array}{l}
\pm \zeta_1^{(3)} \qquad \text{when}\qquad k=8n\pm 1 \quad \text{for} \quad
 n\in \mathbb{N}
\\[2mm]
\pm\zeta_1^{(2)}\qquad \text{when}\qquad
k=8n\pm 2 \quad \text{for} \quad
 n\in \mathbb{N}
\\[2mm]
\pm \zeta_1^{(1)}\qquad \text{when}\qquad k=8n\pm 3 \quad \text{for} \quad
 n\in \mathbb{N}
\\[2mm]
0 \qquad \text{when}\qquad k=4n \quad \text{for} \quad
 n\in \mathbb{N}
\end{array}
\right.~.
\end{align}
Note that all the parameters are now fixed in terms of
\begin{equation}
 \zeta_1^{(1)}~,\qquad \zeta_1^{(2)}~,\qquad \zeta_1^{(3)}~.
\label{eq:3DOF}
\end{equation}
Finally, the constraints \eqref{eq:zeta}
for $k=N-1$ imply 
\begin{equation}
\zeta_{N-2}^{(1)}= \zeta_{N-2}^{(3)} = \zeta_{N-1}^{(2)}~,\qquad
 \zeta_{N-2}^{(2)} = \zeta_{N-1}^{(1)} + \zeta_{N-1}^{(3)}~.
\label{eq:last-c}
\end{equation}

Since $N$ is coprime to $4$, $N$ is an odd integer. For odd $N$,
imposing both of the two constraints in \eqref{eq:last-c} implies 
\begin{equation}
 \zeta_1^{(1)} = \zeta_1^{(2)} = \zeta_1^{(3)} =0~,
\end{equation}
which means no (non R-symmetric) global symmetry exists. To preserve a
(non R-symmetric) $U(1)$ global symmetry, one can replace
\eqref{eq:M=4_W1} with, for instance,
\begin{equation}
 W_1 =
 \sum_{k=1}^{N-1}\left(\sum_{l=1}^3\phi^{(l)}_{k-1}V_k^{(l)}\phi_{k+1}^{(l)}
 + \sum_{l=1}^2\tilde \phi_{k-1}^{(l)} \tilde V_k^{(l)} \tilde
 \phi_{k+1}^{(l)} \right) +\sum_{k=1}^{N-2}\tilde
 \phi_{k-1}^{(3)}\tilde V_k^{(3)}\tilde \phi_{k+1}^{(3)}~.
\end{equation}
This replacement leaves one of the three degrees of freedom
\eqref{eq:3DOF} unfixed, leading to a (non R-symmetric) $U(1)$ global
symmetry.

\subsection{Conjecture for $(A_{M-1},A_{N-1})$}
\label{subsec:conjecture}

Here, we briefly discuss the generalization of our discussions to
a general  $(A_{M-1}, A_{N-1})$ theory for coprime $M$ and $N$.
In particular, we conjecture that the half-monodromy of the theory is written as
\begin{equation}
 S(q) = \sum_{a_{i,j}=0}^\infty\sum_{k^{B,T}_j=0}^\infty
  \frac{q^{\frac{A}{2}}(-q^{\frac{1}{2}})^{\sum_{i=1}^{M-1}\sum_{j=1}^{N-1}a_{i,j}-\sum_{j=1}^{N-1}\sum_{1\leq
  B< T\leq
  M-1}(T-B)k^{B,T}_j}}{ \Big(\prod_{i=1}^{M-1}\prod_{j=1}^{N-1}(q)_{\hat{a}_{i,j} }  \Big)\Big( \prod_{j=1}^{N-1}\prod_{1\leq B< T\leq
  M-1}(q)_{k^{B,T}_j}\Big)} X_{\sum_{i=1}^{M-1}\sum_{j=1}^{N-1}
  a_{i,j}\gamma^{(i)}_{j} }~,
\label{eq:S-AMAN}
\end{equation}
where $i,B$ and $T$ runs over $1,2,\cdots, M-1$ under the constraint
that $B<T$, and $j$ runs over $1,2,\cdots, N-1$. In the above
expression, we also used
\begin{align}
 \hat{a}_{i,j} &= a_{i,j}-
  \sum_{\substack{1\leq B<T\leq M-1 \\ (B\leq i\leq T)}}k_j^{B,T}~,
\\[3mm]
 A &= -\sum_{i=1}^{M-2}\sum_{j=1}^{N-1}a_{i,j}a_{i+1,j}
 +\sum_{i=1}^{N-1}\sum_{j=1}^{N-2}a_{i,j}a_{i,j+1} 
\nonumber\\
&\qquad
 +\sum_{j=1}^{N-1}\sum_{B_1,B_2=1}^{M-2}\sum_{T_1=B_1+1}^{M-1}\sum_{T_2=B_2+1}^{M-1}f(B_1,T_1,B_2,T_2)
 k^{B_1,T_1}_{j}k^{B_2,T_2}_j
\nonumber\\
&\qquad -\sum_{j=1}^{N-2}\sum_{B_1,B_2=1}^{M-2}\sum_{T_1=B_1+1}^{M-1}\sum_{T_2=B_2+1}^{M-1}g(B_1,T_1,B_2,T_2)k^{B_1,T_1}_j k^{B_2,T_2}_{j+1}~,
\end{align}
where
\begin{align}
f(B_1,T_1,B_2,T_2) &\equiv o(B_1,T_1,B_2,T_2) - h(B_1,T_1,B_2,T_2)~,
\\[3mm]
o(B_1,T_1,B_2,T_2) &\equiv \left\{
\begin{array}{l}
\text{min}(T_1,T_2) - \text{max}(B_1,B_2) + 1 \qquad \text{if}\qquad
 \text{max}(B_1,B_2)\leq \text{min}(T_1,T_2)
 \\[2mm]
0 \qquad \text{otherwise}
\end{array}
\right.~,
\\[2mm]
 h(B_1,T_1,B_2,T_2) &\equiv \left\{
\begin{array}{l} 
 \frac{1}{2}\Big(1+(-1)^{o(B_1,T_1,B_2,T_2)}\Big)\quad \text{if}\quad
 B_1<B_2\leq T_1<T_2 \quad \text{or} \quad B_2<B_1\leq T_2<T_1
\\[2mm]
1 \qquad \text{if}\qquad B_1=B_2 \quad \text{or}\quad T_1=T_2
\\[2mm]
\frac{1}{2}\Big(1 - (-1)^{o(B_1,T_1,B_2,T_2)}\Big)\quad \text{if}\quad
B_1<B_2<T_2<T_1\quad \text{or}\quad B_2<B_1<T_1<T_2
\\[3mm]
0\qquad \text{otherwise}
\end{array}
\right.~,
\\[2mm]
g(B_1,T_1,B_2,T_2) &\equiv \text{floor}\left(\frac{o(B_1,T_1,B_2,T_2)}{2}\right)~.
\end{align}
Note that $o(B_1,T_1,B_2,T_2)$ is the number of integers shared by two
intervals $[B_1,T_1]$ and $[B_2,T_2]$. We have checked the conjectural
expression \eqref{eq:S-AMAN} for various coprime $M$ and $N$ with Mathematica.

Based on the above conjecture, the Schur index of the $(A_{M-1},
A_{N-1})$ theory is expressed as
\begin{equation}
 \mathcal{I}^\text{4d} = \sum_{a_{i,j}=0}^\infty\sum_{k^{B,T}_j,\tilde k^{B,T}_j=0}^\infty
  \frac{q^{\frac{A}{2}}q^{\sum_{i=1}^{M-1}\sum_{j=1}^{N-1}a_{i,j}}(-q^{-\frac{1}{2}})^{\sum_{j=1}^{N-1}\sum_{1\leq
  B< T\leq
  M-1}(T-B)(k^{B,T}_j + \tilde k^{B,T}_j)}}{
  \Big(\prod_{i=1}^{M-1}\prod_{j=1}^{N-1}(q)_{\hat{a}_{i,j}}(q)_{\tilde{\hat a}_{i,j}} \Big) \Big(\prod_{j=1}^{N-1}\prod_{1\leq B< T\leq
  M-1}(q)_{k^{B,T}_j}(q)_{\tilde k^{B,T}_j}\Big)}~,
\label{eq:index-general}
\end{equation}
where
\begin{align}
 \hat{a}_{i,j} &= a_{i,j}-
  \sum_{\substack{1\leq B<T\leq M-1 \\ (B\leq i\leq
 T)}}k_j^{B,T}~,\qquad  \tilde{\hat{a}}_{i,j} = a_{i,j}-
  \sum_{\substack{1\leq B<T\leq M-1 \\ (B\leq i\leq T)}}\tilde k_j^{B,T}~,
\\[3mm]
 A+\tilde A &= -2\sum_{i=1}^{M-2}\sum_{j=1}^{N-1}a_{i,j}a_{i+1,j}
 +2\sum_{i=1}^{N-1}\sum_{j=1}^{N-2}a_{i,j}a_{i,j+1} 
\nonumber\\
&\qquad
 +\sum_{j=1}^{N-1}\sum_{B_1,B_2=1}^{M-2}\sum_{T_1=B_1+1}^{M-1}\sum_{T_2=B_2+1}^{M-1}f(B_1,T_1,B_2,T_2)\Big(
 k^{B_1,T_1}_{j}k^{B_2,T_2}_j+  \tilde k^{B_1,T_1}_{j}\tilde k^{B_2,T_2}_j  \Big)
\nonumber\\
&\qquad
 -\sum_{j=1}^{N-2}\sum_{B_1,B_2=1}^{M-2}\sum_{T_1=B_1+1}^{M-1}\sum_{T_2=B_2+1}^{M-1}g(B_1,T_1,B_2,T_2)\Big(k^{B_1,T_1}_j
 k^{B_2,T_2}_{j+1} +  \tilde k^{B_1,T_1}_{j}\tilde k^{B_2,T_2}_j \Big)~.
\end{align}
From this expression, one can read off a 3d CS matter theory
describing the twisted compactification of the $(A_{M-1},A_{N-1})$
theory. This CS matter theory involves $U(1)^{(M-1)^2(N-1)}$ gauge symmetry and $M(M-1)(N-1)$
chiral multiplets. The CS levels and the gauge charges of the chiral
multiplets can be read off from \eqref{eq:index-general}.

\section{Summary and discussions}
\label{sec:summary}

In this paper, we have studied a series of 3d $\mathcal{N}=2$ CS matter
theories describing the $U(1)_r$-twisted $S^1$-reduction of the 4d AD
theories of $(A_{M-1},A_{N-1})$ type with $\text{gcd}(M,N)=1$, by using
the recent proposal of \cite{Gaiotto:2024ioj}. In
particular, for the $(A_2,A_{N-1})$ theories with $N$ coprime to three
and for the $(A_3,A_{N-1})$ theories with $N$ coprime to four, we have
identified the gauge group, matter content, mixed CS levels and monopole
superpotentials of the CS matter theories that are expected to give rise to an $\mathcal{N}=4$
supersymmetry enhancement in the infrared. For the $(A_{M-1},A_{N-1})$
theories with coprime $M\geq 5$ and $N$, we have conjectured in
Sec.~\ref{subsec:conjecture} an expression for the trace of the quantum
monodromy, from which one can read off the gauge group, matter content,
mixed CS levels of the corresponding CS matter theory. 

As a by-product, even independently of the above discussions, we have
also found a novel Nahm sum formula
for the vacuum character of the $(3,8)$ $W_3$ minimal model as discussed
in Sec.~\ref{subsec:Nahm}, which can
be regarded as a natural generalization of a similar formula
for the $(3,7)$ $W_3$ minimal model discovered in
\cite{Feigin:2005gz}. This novel formula has then led us to another
CS matter theory describing the $U(1)_r$-twisted $S^1$-reduction of
the $(A_2,A_4)$ theory.

Very recently, the authors of \cite{Kim:2025rog} have studied R-twisted
circle compactifications of AD theories of $(G,G')$ type, where $G$ and
$G'$ are Lie algebras of ADE type. They used the same proposal of
\cite{Gaiotto:2024ioj} and then found $\mathcal{N}=2$ CS matter theories
for some of the $(A_2,A_{N-1})$ and $(A_3,A_{N-1})$ theories. While
their CS matter theories and ours involve different gauge groups and matter
content, they are expected to be IR dual to each other in the sense that
they flow to the same fixed point in the infrared. Indeed, there are
generally many different UV $\mathcal{N}=2$ CS matter theories that flow
in the infrared to a single $\mathcal{N}=4$ SCFT, as pointed out and
demonstrated in \cite{Gaiotto:2024ioj}. It would be interesting to study
in detail
the duality between the CS matter theories discussed in \cite{Kim:2025rog} and those
discussed in this paper.

Another possible future direction is to generalize our work to the
$(A_{M-1},A_{N-1})$ theories for $M$ and $N$ that are not coprime to
each other. In this case, the 4d theory has a flavor symmetry and
therefore the trace of the quantum monodromy is evaluated in a slightly
different way \cite{Cordova:2015nma}. Even in that case, one can use the
proposal of \cite{Gaiotto:2024ioj} to identify an $\mathcal{N}=2$ CS
matter theory. In particular, when $N=nM$ for an integer $n$,
the half index of the resulting CS matter theories is expected to be
identical to the Schur index of the $(A_{M-1}, A_{nM-1})$ theories,
whose 
closed-form expression  was conjectured in \cite{Buican:2017uka}.

\section*{Acknowledgements}

T.~N. thanks Shou Tanigawa for helpful
discussions in a separate but related collaboration.
T.~N.'s research is partially supported by JSPS KAKENHI Grant
Numbers JP21H04993, JP23K03394 and JP23K03393.  Y.~Y.'s research is partially supported by
JSPS KAKENHI Grant Numbers 21K03382.

\begin{appendix}

\section{Three dimensional localization formulas }
\label{app:local}
In this appendix we summarize the localization formulas for the half index, the superconformal index, and the ellipsoid partition function of 3d $\mathcal{N}=2$ abelian CS matter theories.
Let us consider an abelian CS matter theory with gauge group $U(1)^N$ coupled to 
$L$ chiral multiplets with the gauge charges   $Q_{a, i}$ for $a = 1,\ldots,N$ and $i = 1,\ldots,L$ and with  the R-charge $r_i$ for $i = 1,\ldots,L$.
We denote the effective gauge CS level\footnote{
Here we use the term {\it effective} in the sense of the half index. 
In flat space,  the level shift usually depends on the sign of the fermion mass, whereas for the  half index,  the sign of level shift 
depends on the choice of boundary conditions for the vector and chiral multiplets. 
This shift was first pointed out in \cite{Yoshida:2014ssa} as a regularization factor of the divergence of the one-loop determinants. 
An interpretation  in terms of edge modes was given in \cite{Dimofte:2017tpi}. 
}
by $K_{ab}$ with $a,b = 1,\ldots,N$, and the effective gauge-R-symmetry CS level 
by $K_{a R}$ with $a,b = 1,\ldots,N$.
 For our purposes, we may set the gauge–flavor mixed CS levels to zero without loss of generality. 
We will also omit background CS levels, as they  contribute an overall multiplicative factor.

\subsubsection*{Half index}
First, we summarize the basic properties of the half index \cite{Beem:2012mb,Yoshida:2014ssa,Dimofte:2017tpi}, which is formally defined on the spacetime $S^1 \times D^2$ as
\begin{align}
  I\! \! I
  = \mathrm{Tr}_{\mathcal{H}(D^2)} (-1)^{F} \,
    q^{J_3+\frac{R}{2}} \prod_{i} x_i^{F_i} \,.
    \label{eq:halfindex}
\end{align}
Here $F$ is the fermion number, $R$ is the $U(1)$ R-charge, $J_3$ is the generator of rotations on the hemisphere $D^2$,
and $F_i$ denotes a $U(1)$ global symmetry charge, which is a topological symmetry  acting on monopole operators or a
flavor symmetry acting on  chiral multiplets. 
The  $\mathcal{H}(D^2)$ is the space of BPS operators on $D^2$.

The half index depends on the choice of boundary conditions for both vector and chiral multiplets at the boundary of $S^1 \times D^2$. 
For a vector multiplet, there are two standard types of boundary conditions: the Neumann boundary condition ($\mathcal{N}$) \cite{Yoshida:2014ssa} and the Dirichlet boundary condition ($\mathcal{D}$)\cite{Dimofte:2017tpi}. Since the Dirichlet boundary condition for the vector multiplet is directly related to the Schur index, we will focus on this choice in what follows. 
For a chiral multiplet, there are again Neumann ($N$) and Dirichlet ($D$) boundary conditions, and in addition one can impose a deformed Dirichlet boundary condition ($D_c$). 
Since a chiral multiplet with the deformed Dirichlet boundary condition acquires a non-zero boundary vev, only those flavor symmetries that leave this vev invariant remain unbroken.
Among these, it is the deformed Dirichlet boundary condition that is relevant for the Schur index, and this will be the choice we adopt in the following.
Then the half index is written as follows:
\begin{align}
I\! \! I_{(\mathcal{D}, D_c)}= \frac{1}{(q;q)_{\infty}^N} \sum_{a=1}^N\sum_{n_a \in \mathbb{Z}} 
q^{\frac{1}{2} \sum_{a,b=1}^N K_{ab} n_a n_b+\frac{1}{2}\sum_{a=1}^N K_{a R} n_a } 
\Bigl(\prod_{a=1}^{N} x_a^{n_a} \Bigr)
\prod_{i=1}^L (q^{1-\sum_a n_a Q_{a, i}};q)_{\infty}\,.
\label{eq:DDhalfindex}
\end{align}
Here $x_a$ is the fugacity for the $U(1)$ topological symmetry associated with the $a$-th $U(1)$ gauge group.
For the deformed boundary condition, the boundary value of chiral multiplet imposes  the condition $r_i=0$.

\subsubsection*{Superconformal index}
The superconformal index \cite{Bhattacharya:2008zy} for 3d $\mathcal{N}=2$ theory  is defined by 
\begin{align}
I_{\rm SCI} = \mathrm{Tr}_{\mathcal{H}(S^2)} (-1)^{F}\,
    q^{J_3+\frac{R}{2}} \prod_{i} x_i^{F_i} \,.
    \label{eq:3dSCI}
\end{align}
Here  $\mathcal{H}(S^2)$ is the space of BPS operators on defined $S^2$.
The localization formula \cite{Kim:2009wb, Imamura:2011su, Kapustin:2011jm} for the superconformal index is given by
\begin{align}
I_{\rm SCI}&= \sum_{a=1}^N \sum_{n_a \in \mathbb{Z}}\oint \prod_{i=1}^N \frac{dz_a}{2 \pi i z_{a}} 
\Bigl(\prod_{a,b=1}^N z_a^{K_{ab} n_b} \Bigr)
\Bigl( \prod_{a=1}^{N}  x^{n_a}_a \Bigr)
\nonumber \\
&\times
\prod_{a=1}^N q^{\frac{1}{2}K_{a R}  n_a } \prod_{i=1}^L  (q^{-\frac{1}{2}} z_a^{Q_{a i}})^{-\frac{Q_i \cdot n+|Q_i \cdot n|}{2}}
\frac{(z^{-Q_{a i}}_a w_i^{-F_i} q^{1-\frac{r_i}{2} +\frac{|Q_i \cdot n|}{2}};q)_{\infty}}{(z^{Q_{a i}}_a w_i^{F_i} q^{\frac{r_i}{2} +\frac{|Q_i \cdot n|}{2}};q)_{\infty}}\,,
\label{eq:SCI3}
\end{align}
where $Q_i \cdot n =\sum_{a=1}^{N} Q_{a, i} n_a$. $F_i$ (resp. $w_i$) is the $U(1)$ flavor charge (resp. fugacity)
for the $i$-th chiral multiplet. $x_a$, $r_i$ are same as the above case.

\subsubsection*{Ellipsoid partition function}

The localization formula \cite{Hama:2011ea, Imamura:2011wg} 
for the partition function on the 3d ellipsoid $S^3_{\sf b}$ including the gauge-R-symmetry mixed CS term \cite{Closset:2012ru} is given by
\begin{align}
Z_{S^3_{\sf b}} &= C \int \prod_{a=1}^N  d\sigma_{a} 
e^{ \pi i \sum_{a,b=1}^N K^{\prime}_{a b} \sigma_a \sigma_b +2 \pi i  \sum_{a=1}^N \xi_a \sigma_a - \sum_{a=1}^N \pi ({\sf b}+{\sf b}^{-1}) K^{\prime}_{a R} \sigma_a    } \nonumber \\
& \qquad \quad \times\prod_{i=1}^L
 s_{\sf b} \Bigl( \frac{i}{2}({\sf b}+{\sf b}^{-1})(1-r_i)- \sum_a Q_{a, i} \sigma_a - F_i  m_i \Bigr)\,.
 \label{eq:3dsquashed}
\end{align}
Here an overall $\sigma_a$-independent constant $C$  comes from the background CS terms.   
$K^{\prime}_{ab}$ and $K^{\prime}_{g R}$ are the bare gauge  CS level and gauge-R-symmetry mixed CS level, related to the effective CS levels as
\begin{align}
    K_{ab}&=K^{\prime}_{ab}+\frac{1}{2} \sum_{i=1}^{L} Q_{a, i} Q_{b, i}\,, \\
    K_{aR}&=K^{\prime}_{aR}+\frac{1}{2} \sum_{i=1}^{L} Q_{a, i} (r_i-1)\,.
\end{align}
$\xi_a$ are the Fayet--Iliopoulos parameters.  $m_i$ is  the real mass for the $i$-th chiral multiplet. 
The function $s_{\sf b}(x)$ is the double sine function, given by
\begin{align}
s_{\sf b}(x) = e^{-  \frac{\pi {i}}{2} x^2 } \prod_{k=1}^{\infty} \frac{ 1+e^{2 \pi {\sf b} x} e^{2 \pi {i} {\sf b}^2 (k-\frac{1}{2}) } }
{ 1+e^{2 \pi  {\sf b}^{-1} x} e^{-2 \pi { i} {\sf b}^{-2} (k-\frac{1}{2}) } }\,.
\label{eq:doublesine}
\end{align}

\section{Derivation of formula for $\mathcal{I}^{\rm 4d}$ and $\mathcal{S}^{\rm 3d}_{\sf b}$ }
\label{app:derivationIR}
In this appendix we present  the derivation of  \eqref{eq:Schur4} and \eqref{eq:S3b4} from \eqref{eq:4Dindex} \eqref{eq:S3part}, respectively.
First,  using the relations \eqref{eq:torus}, \eqref{eq:factorE},   \eqref{eq:xcom} and \eqref{eq:Fourier},  we rewrite \eqref{eq:halfmd} and \eqref{eq:smalls}  as
\begin{align}
 S(q) &= \sum_{\ell=1}^{L^{\prime}} \sum_{m_{\ell}=0}^{\infty}
 \frac{ q^{\frac{1}{2}\sum_{\ell <\ell^{\prime}} A^{\prime}_{\ell \ell^{\prime}}m_{\ell} m_{\ell^{\prime}}} (-q^{\frac{1}{2}})^{\sum_{\ell} m_{\ell}}}{\prod_{\ell=1}^{L^{\prime}}(q)_{m_{\ell}}}
  X_{\sum_{\ell=1}^{L^{\prime}} \sum_{k=1}^{2r}  P_{ k \ell} m_{\ell} \gamma_k}\,, \\
 {\sf s}_{\sf b} &= \int  \prod_{\ell=1}^{L^{\prime}} d p_i
 e^{ \pi i \sum_{\ell < \ell^{\prime}}A^{\prime}_{\ell \ell^{\prime}}p_{\ell} p_{\ell^{\prime}} }
 \prod_{\ell=1}^{L^{\prime}} \widehat{\Phi}_{\sf b}(p_{\ell})
  e^{ 2 \pi i \sum_{\ell=1}^{L^{\prime}} \sum_{k=1}^{2r} P_{ k \ell} p_{\ell}  \gamma_k}\,,
 \end{align}
where $A^{\prime}_{\ell \ell^{\prime}}$ is defined  by the following relation:
\begin{align}
\sum_{1 \le \ell < \ell \le L^{\prime}}A^{\prime}_{\ell \ell^{\prime}}m_{\ell} m_{\ell^{\prime}}
\equiv \sum_{\ell=1}^{\ell^{\prime}-1} \sum_{\ell^{\prime}=2}^{L^{\prime}} P_{ k \ell} P_{ j \ell}  \langle \gamma_{k}, \gamma_{j} \rangle    m_{\ell}  m_{\ell^{\prime}} \,.
\end{align}
When the flavor symmetry is absent, it follows from \eqref{eq:trace} and \eqref{eq:deltafunc} that the traces for the Schur index and $\mathcal{S}^{3d}_{\sf b}$ 
are taken as
\begin{align}
  \mathcal{I}^\text{4d}  
&
=(q)^{2r}_{\infty}\sum_{\ell=1}^{L^{\prime}} 
\sum_{m_{\ell}, \tilde{m}_{\ell}=0}^{\infty}
 \frac{ q^{\frac{1}{2}\sum_{\ell < \ell^{\prime}}A^{\prime}_{\ell \ell^{\prime}} (m_{\ell} m_{\ell^{\prime}}+ 
 \tilde{m}_{\ell} \tilde{m}_{\ell^{\prime}})} (-q^{\frac{1}{2}})^{\sum_{\ell=1}^{L^{\prime}} (m_{\ell}+ \tilde{m}_{\ell})}}{\prod_{\ell=1}^{L^{\prime}}(q)_{m_{\ell}}
 (q)_{\tilde{m}_{\ell}}}
\prod_{k=1}^{2r }  \delta_{\sum_{\ell=1}^{L^{\prime}} P_{ k \ell} (m_{\ell} -\tilde{m}_{\ell} ) , 0}\,,
\label{eq:schurindx}
\\
\mathcal{S}^{\rm 3d}_{\sf b}&=
\int  \prod_{\ell=1}^{L^{\prime}} d p_i d \tilde{p}_i 
 e^{\pi i \sum_{\ell <\ell^{\prime}} A^{\prime}_{\ell \ell^{\prime}}(p_{\ell} p_{\ell^{\prime}}+\tilde{p}_{\ell} \tilde{p}_{\ell^{\prime}}) }
 \prod_{\ell=1}^{L^{\prime}} \widehat{\Phi}_{\sf b}(p_{\ell}) \widehat{\Phi}_{\sf b}(\tilde{p}_{\ell})
\prod_{k=1}^{2r}  \delta \Bigl(\sum_{\ell=1}^{L^{\prime}} P_{ k \ell} (  p_{\ell} -\tilde{p}_{\ell} ) \Bigr)
\label{eq:ellip}
\end{align}
The Kronecker delta in \eqref{eq:schurindx} and delta function constraints in \eqref{eq:ellip} can be solved as follows. 
Since the matrix $P=(P_{k\ell})_{1 \le k \le 2r,\; 1 \le \ell \le L'}$ is a $2r \times L^{\prime}$ integer matrix, i.e.  
$P \in {\rm Mat}_{2r \times L^{\prime}}(\mathbb{Z})$, it can be brought to Smith normal form:
\begin{align}
U P V = 
\begin{pmatrix}
D & {\bf 0}_{2r \times (L^{\prime}-2r)} \\
\end{pmatrix}\,,
\end{align}
where $D$ is the following $2r \times 2r$ diagonal matrix:
\begin{align}
 D= {\rm diag}(d_1, d_2, \cdots, d_{{2r}})\,,      
\end{align}
where $d_i \in \mathbb{Z}$ for $i=1,\cdots, 2r$ satisfy the condition $d_{i+1}/d_i \in \mathbb{Z}$. 
 ${\bf 0}_{M\times N}$ denotes the $M \times N$ zero  matrix.
$U$ and $V$ are unimodular matrices: $U \in GL_{2r}(\mathbb{Z})$, $V \in GL_{L^{\prime}}(\mathbb{Z})$, ${\rm det}(U), {\rm det}(V)  \in \{ \pm 1 \}$ .

Using the Smith normal form,  the the Kronecker delta and delta function constraints are solved as
\begin{align}
&\sum_{\ell=1}^{L^{\prime}}
\sum_{m_{\ell}, \tilde{m}_{\ell}=0}^{\infty}
\prod_{k=1}^{2r }  \delta_{\sum_{\ell=1}^{L^{\prime}} P_{ k \ell} (m_{\ell} -\tilde{m}_{\ell} ) , 0}
 f(m_{1}, \cdots, m_{L^{\prime}}, \tilde{m}_{1}, \cdots, \tilde{m}_{L^{\prime}})  \nonumber \\
&=
\sum_{i=1}^{L^{\prime}} \sum_{j=2r+1 }^{ L^{\prime}} \sum_{s_{i}, \tilde{s}_j \in \mathbb{Z} \atop (V \cdot {s})_i, (V \cdot \tilde{s})_j \ge 0 } 
f((V \cdot s )_1, \cdots, (V \cdot \tilde{s} )_{L^{\prime}})
\Big|_{\tilde{s}_i=s_{i} \text{ for  } i=1  \cdots,2r}
\,, 
\\
&\int  \prod_{\ell=1}^{L^{\prime}} d p_i d \tilde{p}_i 
\prod_{k=1}^{2r}   \delta \Bigl(\sum_{\ell=1}^{L^{\prime}} P_{ k \ell} (  p_{\ell} -\tilde{p}_{\ell} ) \Bigr)
f(p_{1}, \cdots, p_{L^{\prime}}, \tilde{p}_{1}, \cdots, \tilde{p}_{L^{\prime}}) \nonumber \\&
=
\frac{1}{\left| \prod_{k=1}^{2r} d_k \right|}  \int  \prod_{i=1}^{ L^{\prime}} d x_i \prod_{j=2r+1}^{ L^{\prime}} d \tilde{x}_j 
f( (V\cdot x )_1,  \cdots, (V \cdot \tilde{x} )_{L^{\prime}})
\Big|_{\tilde{x}_i=x_{i} \text{ for  } i=1  \cdots,2r}
\end{align}
Here $(V \cdot p)_{\ell^{\prime}} \equiv  \sum_{\ell=1}^{L^{\prime}} V_{\ell^{\prime} \ell}  p_{ \ell}$. 
$s_{\ell},\tilde{s}_{\ell}, x_{\ell}$ and $\tilde{x}_{\ell}$ for $\ell=1, \cdots, L^{\prime}$ are defined by
\begin{align}
s_{\ell^{\prime}}\equiv \sum_{\ell=1}^{L^{\prime}} (V^{-1})_{\ell^{\prime} \ell}  m_{ \ell}\,, \quad
\tilde{s}_{\ell^{\prime}}\equiv \sum_{\ell=1}^{L^{\prime}} (V^{-1})_{\ell^{\prime} \ell}  \tilde{m}_{ \ell}\,, 
\\
x_{\ell^{\prime}}\equiv \sum_{\ell=1}^{L^{\prime}} (V^{-1})_{\ell^{\prime} \ell}  p_{ \ell}\,, \quad
\tilde{x}_{\ell^{\prime}}\equiv \sum_{\ell=1}^{L^{\prime}} (V^{-1})_{\ell^{\prime} \ell}  \tilde{p}_{ \ell} \,.
\end{align}
We define $\sigma_a$ and  $Q_{a, i}$ for $a=1,\cdots, 2L^{\prime} -2r$ and $i=1,\cdots, 2L^{\prime} $ by
\begin{align}
\sigma_a \equiv
\begin{cases}
   x_a& \text{for } a = 1,\cdots, L^{\prime},\\[2pt]
  x_{a-L^{\prime}} & \text{for } a = L^{\prime}+1, \cdots, L^{\prime}+2r ,\\[2pt]
   \tilde{x}_{a-L^{\prime}-2r} & \text{if } a = L^{\prime}+2r + 1, \cdots, 2L^{\prime}-2r\,.
\end{cases}
\end{align}
and
\begin{align}
n_a \equiv
\begin{cases}
   s_a& \text{for } a = 1,\cdots, L^{\prime},\\[2pt]
  s_{a-L^{\prime}} & \text{for } a = L^{\prime}+1, \cdots, L^{\prime}+2r ,\\[2pt]
   \tilde{s}_{a-L^{\prime}-2r} & \text{if } a = L^{\prime}+2r + 1, \cdots, 2L^{\prime}-2r\,.
\end{cases}
\end{align}
and
\begin{align}
Q_{a,i } \equiv 
\begin{cases}
  V_{i, a} & \text{for } i =1, \cdots L^{\prime}, a =1, \cdots L^{\prime},\\[2pt]
 V_{i-L^{\prime}, a} & \text{for } i = L^{\prime}+1, \cdots, 2 L^{\prime} , a =1, \cdots, 2r\,,\\[2pt] 
  V_{a- L^{\prime}, i -L^{\prime} +2r} & \text{for } i = L^{\prime}+1, \cdots, 2L^{\prime}, a =L^{\prime}+1, \cdots 2L^{\prime}-2r\,,\\[2pt]
  0 & \text{the others}\,.
\end{cases}
\end{align}
Then  we obtain the following expressions for the Schur index $\mathcal{I}^{\rm 4d}$ and $\mathcal{S}^{\rm 3d}_{\sf b}$: 
\begin{align}
  \mathcal{I}^\text{4d}  
&
=(q)^{L-N}_{\infty} \sum_{n_1, \cdots, n_N \in \mathbb{Z} \atop (n \cdot Q)_1 \ge 0, \cdots (n \cdot Q)_{L} \ge 0} 
 \frac{ q^{\frac{1}{2}\sum_{a,b=1}^{N} K_{a b} n_{a} n_{b}}(-q^{\frac{1}{2}})^{ 
 \sum_{a=1}^N \sum_{i=1}^L Q_{a, i} n_{a}  }}{\prod_{i=1}^{L}(q)_{\sum_{a=1}^N n_a Q_{a, i} }
 }\,,
 \label{eq:Schur4}
\\
\mathcal{S}^{\rm 3d}_{\sf b}&= \int  \prod_{a=1}^{N} d \sigma_a  
 e^{ \pi i \sum_{a, b=1}^{N} K_{a b} \sigma_{a} \sigma_{b} }
 \prod_{i=1}^{L} \widehat{\Phi}_{\sf b}(\sum_{a=1}^{N} Q_{a, i} \sigma_{a}) \,.
 \label{eq:S3b4}
\end{align}
Here we omit an overall $\sigma_a$-independent prefactor in $\mathcal{S}^{\rm 3d}_{\sf b}$.
$N$ and $L$ are defined by
\begin{align}
L \equiv 2 L^{\prime}\,, \quad N \equiv  2 L^{\prime} -2r\,,
\end{align}
 and $(n \cdot Q)_i \equiv  \sum_{a=1}^{N} n_a Q_{a i} $.
 $K_{a b}$ for $a,b=1,\cdots, N$ is defined by the following relation:
\begin{align}
\sum_{a, b=1}^{N}    K_{a b} n_a n_b \equiv 
 \sum_{\ell=1}^{\ell^{\prime}-1} \sum_{\ell^{\prime}=2}^{L^{\prime}} P_{ k \ell} P_{ j {\ell}^{\prime}}  \langle \gamma_{k}, \gamma_{j} \rangle\Bigl(  
 (n \cdot Q)_{\ell}  (n \cdot Q)_{\ell^{\prime}} +(n \cdot Q)_{\ell+L^{\prime}}  (n \cdot Q)_{\ell^{\prime}+L^{\prime}} \Bigr) \,.
\end{align}

\section{Gauge and R-charge for dressed monopole operator}
\label{app:monopole}

Here we summarize the formulas, for example see \cite{Cremonesi:2016nbo}, for the gauge charge and $R$-charges of monopole operators, 
dressed by chiral multiplets.
Again let us consider an abelian $\mathcal{N}=2$ CS matter theory with the gauge group $U(1)^N$ coupled to 
$L$ chiral multiplets $\{ \phi_i \}_{i=1}^L$ with the R-charge $r_i$. We denote the gauge CS level by $K_{ab}$ with $a, b=1,\cdots, N$, and  denote
the gauge-R-symmetry mixed CS level by $K_{a R}$. We also denote the gauge charge of 
the chiral multiplets by $Q_{a, i}$ with $a=1,\cdots, N$ and $i=1,\cdots, L$.

We consider a  BPS dressed monopole operator $\mathcal{O}_{(n,m)}=(\prod_{i=1}^{L} \phi^{n_i}_{i} )V_{m}$ 
with magnetic charge $m=(m_1,\cdots, m_N) \in \mathbb{Z}^{N}$. Here  $n_i$ has to satisfy $n_i (m \cdot Q_i )=0$ 
for $i=1,\cdots, L$, and   $m \cdot Q_i = \sum_{a=1}^N m_a Q_{a, i}$.  

Then the gauge charge of dressed monopole operator $\mathcal{O}_{(n,m)} $ is given by
\begin{align}
Q_a[\mathcal{O}_{(n,m)}]= \sum_{b=1}^{N} K_{a b} m_b +\sum_{i =1}^L n_{i} Q_{a, i}-\sum_{i=1}^L\frac{1}{2} Q_{a, i}(| m \cdot Q_i |+ {m} \cdot Q_i ),
\label{eq:gaugeV}
\end{align}
and a reference R-charge is given by 
\begin{align}
R[\mathcal{O}_{(n,m)}]=\sum_{a=1}^N K_{ a R} m_a +\sum_{i=1}^N n_i r_i  -\sum_{i=1}^L\frac{1}{2} ( r_i-1)(| m \cdot Q_i |+ {m} \cdot Q_i ).
\label{eq:RchargeV}
\end{align}

\end{appendix}

\bibliography{VOA}

\end{document}